\documentclass[final,5p,times,twocolumn]{elsarticle}

\renewcommand{\bibname}{References}%

\let\ElseVierBibliography\bibliography%
\renewcommand{\bibliography}[1]{%
\section*{\bibname}%
\ElseVierBibliography{#1}%
}%

\usepackage{amsthm}
\usepackage{amsmath}
\usepackage{amssymb}
\usepackage{graphicx}
\usepackage{xspace}

\usepackage{url}
\usepackage{algorithmic}
\usepackage{algorithm} % For counting chapters
\algsetup{linenosize=\scriptsize}
\xspaceaddexceptions{=}
\xspaceaddexceptions{\}}
\xspaceaddexceptions{\in}
\algsetup{linenosize=\scriptsize}
\usepackage{algorithm} % For counting chapters
\usepackage[bf]{subfigure}
\usepackage[pdftex,table,fixpdftex]{xcolor}
\usepackage{longtable}
\usepackage{threeparttable}
\usepackage{multirow}
\usepackage{tabularx}
\usepackage{hhline}

\newcommand{\eg}{e.g.,\xspace}

\newcommand{\randFunct}{\mathop{\mathrm{rand}}}
\newcommand{\meanFunct}[1]{\ensuremath{\mu(#1)}\xspace}
\newcommand{\stdevFunct}[1]{\ensuremath{\sigma(#1)}\xspace}
\newcommand{\reverseFunct}[1]{\ensuremath{r(#1)}\xspace}
\newcommand{\bruteForceFunct}[1]{\ensuremath{b(#1)}\xspace}
\newcommand{\normalizeFunct}[1]{\ensuremath{n(#1)}\xspace}
\newcommand{\diversityFunct}[1]{\ensuremath{d(#1)}\xspace}
\newcommand{\minFunct}[1]{\ensuremath{\min(#1)}\xspace}
\newcommand{\maxFunct}[1]{\ensuremath{\max(#1)}\xspace}

\newcommand{\response}{\ensuremath{R}\xspace}
\newcommand{\savings}{\ensuremath{P}\xspace}
\newcommand{\error}[1]{\ensuremath{\epsilon_{#1}}\xspace}

\newcommand{\agent}{\ensuremath{i}\xspace}
\newcommand{\child}{\ensuremath{u}\xspace}
\newcommand{\plan}{\ensuremath{j}\xspace}
\newcommand{\selectedPlan}{\ensuremath{s}\xspace}
\newcommand{\timePoint}{\ensuremath{t}\xspace}
\newcommand{\newTimePoint}{\ensuremath{\hat{t}}\xspace}

\newcommand{\numOfAgents}{\ensuremath{n}\xspace}
\newcommand{\numOfPlans}{\ensuremath{p}\xspace}
\newcommand{\numOfChildren}{\ensuremath{k}\xspace}
\newcommand{\horizon}{\ensuremath{T}\xspace}

\newcommand{\heterogeneity}{\ensuremath{\varepsilon}\xspace}

\newcommand{\avgLoad}{\ensuremath{\meanFunct{\signal{\eTFSIndex}}}\xspace}
\newcommand{\minAvgLoad}{\ensuremath{\minFunct{\signal{\eTFSIndex}}}\xspace}
\newcommand{\maxAvgLoad}{\ensuremath{\maxFunct{\signal{\eTFSIndex}}}\xspace}
\newcommand{\randNum}{\ensuremath{x}\xspace}

\newcommand{\signalType}{\textsc{\tiny TS}\xspace}
\newcommand{\iTFSIndex}{\textsc{\tiny iTFS}\xspace}
\newcommand{\eTFSIndex}{\textsc{\tiny eTFS}\xspace}
\newcommand{\TISIndex}{\textsc{\tiny TIS}\xspace}

\newcommand{\upperBoundIndex}{\textsc{\tiny UB}\xspace}
\newcommand{\upperBoundAIndex}{\textsc{\tiny UB1}\xspace}
\newcommand{\upperBoundBIndex}{\textsc{\tiny UB2}\xspace}

\newcommand{\TS}{\textsc{\scriptsize TS}\xspace}
\newcommand{\iTFS}{\textsc{\scriptsize iTFS}\xspace}
\newcommand{\eTFS}{\textsc{\scriptsize eTFS}\xspace}
\newcommand{\TIS}{\textsc{\scriptsize TIS}\xspace}
\newcommand{\TFS}{\textsc{\scriptsize TFS}\xspace}
\newcommand{\upperBound}{\textsc{\scriptsize UB}\xspace}
\newcommand{\upperBoundA}{\textsc{\scriptsize UB1}\xspace}
\newcommand{\upperBoundB}{\textsc{\scriptsize UB2}\xspace}

\newcommand{\signal}[1]{\ensuremath{S_{#1}}\xspace}
\newcommand{\signalValue}[2]{\ensuremath{s_{#1,#2}}\xspace}

\newcommand{\demand}[3]{\ensuremath{d_{#1,#2,#3}}\xspace} % agent, plan and time index
\newcommand{\combinationalDemand}[3]{\ensuremath{c_{#1,#2,#3}}\xspace} % agent, plan and time index
\newcommand{\aggregateDemand}[3]{\ensuremath{a_{#1,#2,#3}}\xspace} % agent, plan and time index
\newcommand{\demandPlan}[2]{\ensuremath{d_{#1,#2}}\xspace}
\newcommand{\demandPlans}[1]{\ensuremath{D_{#1}}\xspace}
\newcommand{\aggregatePlan}[2]{\ensuremath{a_{#1,#2}}\xspace}
\newcommand{\aggregatePlans}[1]{\ensuremath{A_{#1}}\xspace}
\newcommand{\combinationalDemandPlan}[2]{\ensuremath{c_{#1,#2}}\xspace}
\newcommand{\combinationalDemandPlans}[1]{\ensuremath{C_{#1}}\xspace}

\newcommand{\shuffle}{\textsc{\scriptsize SHUFFLE}\xspace}
\newcommand{\shift}{\textsc{\scriptsize SHIFT}\xspace}
\newcommand{\shiftParam}[1]{\textsc{\scriptsize SHIFT(#1)}\xspace}
\newcommand{\swap}{\textsc{\scriptsize SWAP}\xspace}
\newcommand{\swapParam}[1]{\textsc{\scriptsize SWAP(#1)}\xspace}

\newcommand{\minCost}{\textsc{\scriptsize MIN-COST}\xspace}
\newcommand{\minRmseA}{\textsc{\scriptsize MIN-RMSE-UB1}\xspace}
\newcommand{\minRmseB}{\textsc{\scriptsize MIN-RMSE-UB2}\xspace}

\newcommand{\SIM}{\textsc{\scriptsize SIM}\xspace}
\newcommand{\EDF}{\textsc{\scriptsize EDF}\xspace}
\newcommand{\PNW}{\textsc{\scriptsize PNW}\xspace}
\newcommand{\PNWMorning}{\textsc{\scriptsize PNW-MORNING}\xspace}
\newcommand{\PNWEvening}{\textsc{\scriptsize PNW-EVENING}\xspace}

\DeclareMathOperator*{\argmin}{arg\,min}

\newtheorem{bounds}{Upper Bound}

\journal{Future Generation Computer Systems}

\begin{document}

\begin{frontmatter}



\title{Self-regulating Supply-Demand Systems}


\author[ethz]{Evangelos~Pournaras}
\author[ibm]{Mark~Yao}
\author[ethz]{Dirk~Helbing}

\address[ethz]{Professorship of Computational Social Science\\
ETH Zurich, Zurich, Switzerland\\
\{epournaras,dhelbing\}@ethz.ch
}

\address[ibm]{Smart Energy\\
IBM T.J Watson Research Center\\
Yorktown Heights, NY USA\\
markyao@us.ibm.com
}

\begin{abstract}
Supply-demand systems in Smart City sectors such as energy, transportation, telecommunication, are subject of unprecedented technological transformations by the Internet of Things. Usually, supply-demand systems involve actors that produce and consume resources, e.g. energy, and they are regulated such that supply meets demand, or demand meets available supply. Mismatches of supply and demand may increase operational costs, can cause catastrophic damage in infrastructure, for instance power blackouts, and may even lead to social unrest and security threats. Long-term, operationally offline and top-down regulatory decision-making by governmental officers, policy makers or system operators may turn out to be ineffective for matching supply-demand under new dynamics and opportunities that Internet of Things technologies bring to supply-demand systems, for instance, interactive cyber-physical systems and software agents running locally in physical assets to monitor and apply automated control actions in real-time. e.g. power flow redistributions by smart transformers to improve the Smart Grid reliability. Existing work on online regulatory mechanisms of matching supply-demand either focuses on game-theoretic solutions with assumptions that cannot be easily met in real-world systems or assume centralized management entities and local access to global information. This paper contributes a generic decentralized self-regulatory framework, which, in contrast to related work, is shaped around standardized control system concepts and Internet of Things technologies for an easier adoption and applicability. The framework involves a decentralized combinatorial optimization mechanism that matches supply-demand under different regulatory scenarios. An evaluation methodology, integrated within this framework, is introduced that allows the systematic assessment of optimality and system constraints, resulting in more informative and meaningful comparisons of self-regulatory settings. Evidence using real-world datasets of energy supply-demand systems confirms the effectiveness and applicability of the self-regulatory framework. It is shown that a higher informational diversity in the options, from which agents make local selections, results in a higher system-wide performance. Several strategies with which agents make selections come along with measurable performance trade-offs creating a vast potential for online adjustments incentivized by utilities, system operators and policy makers. 
\end{abstract}


\end{frontmatter}

\section{Introduction}\label{sec:introduction}

Regulating supply-demand systems in techno-socio-economic sectors of Smart Cities, such as energy, transportation, telecommunication, financial markets and others, is a complex, yet critical endeavor for societal development and sustainability. Regulation usually implies government-imposed controls on business activities and infrastructural operations~\cite{Brown2006}, in which decison-making by the involved actors is often centralized and operationally offline. However, the introduction of Internet of Things technologies~\cite{Gubbi2013} to tackle the inter-temporal nature of supply and demand~\cite{Obizhaeva2013} creates a vast potential for an automated, online and decentralized self-regulation using remotely communicating sensors, controllers and actuators. For example, the regulation of electricity prices used to be a result of institutional and legal policy designs strictly imposed in a top-down fashion. In contrast, Smart Grid technologies enable regimes within which bottom-up price formation can be automated via real-time pricing schemes and tariffs that continuously balance supply and demand~\cite{MohsenianRad2010,Conejo2010,Samadi2010,Hopper2006a,Akasiadis2016}. 

This new type of online bottom-up self-regulation imposes two grand scientific challenges: (i) How to effectively balance supply and demand in different regulatory scenarios by steering the price elasticity of demand in a fully decentralized fashion? (ii) How to continuously measure and evaluate this self-regulatory capability in different regulatory scenarios of supply-demand systems?  Regulatory scenarios refer to the mitigation of imbalances in supply-demand originated from varying generation and/or consumption, system failures, system reforms, etc. This paper studies the balance of supply and demand with a fundamental, yet highly applicable, approach by introducing a minimal and standardized information exchange. This approach does not require major system interventions and can make a self-regulation mechanism applicable in various Smart City sectors.

How to balance supply-demand or how to design efficient computational markets remains a fascinating challenge for a very broad range of scientific communities. Most remarkably, game-theoretical approaches in combinatorial games are relevant in systems in which agents have a finite number of options to choose from and each choice influences the evolution and outcome of the game. Although mathematical tools exist for solutions to particular problems~\cite{Bewersdorff2005}, there is no unified theory or framework addressing combinatorial elements in these games. This observation usually characterizes matching mechanisms~\cite{Sandholm1999} and learning techniques that have at least the following limitations in applied contexts: (i) long convergence times~\cite{Azevedo2014,Gale1962} that can make them inapplicable in practice due to technological limitations, e.g. communication bandwidth or computational capacity. (ii) Solutions have often a limited scope, meaning that equilibrium is not guaranteed with a minor problem modification or when adding/removing a constraint~\cite{Saad2012}. For example, the literature on certain game-theoretical approaches in Smart Grids reveals the limited applicability of cooperative games and dynamic models to tackle the pervasive presence of time-varying parameters such as generation and demand~\cite{Saad2012}. Moreover, it is also shown that network delays may lead to incorrect price inquiries. A similar conclusion is shared within other related work~\cite{Callaway2011} claiming that a highly responsive and non-disruptive regulatory system for load requires direct centralized control by utilities~\cite{Callaway2011}. There is recent ongoing work to tackle these challenges~\cite{Akasiadis2016}, however, it remains an open question how fully decentralized coordination mechanisms can operate without relying on central broker entities or assuming global/aggregate information available locally. 

This paper contributes a novel and generic self-regulatory framework for supply-demand systems. A single-commodity resource-oriented equilibrium market~\cite{Sandholm1999} resembles the operations of the proposed self-regulated supply-demand system that interacts with a loop of (i) incentive signals that communicate supply costs and (ii) feedback signals that communicate the demand acquired. Agent-based bilateral interactions are performed in the background of a bottom-up computational mechanism advanced to operate in this generic context of distributed supply-demand systems. This mechanism self-regulates the balance of supply and demand in a similar fashion with the intuitive paradigm of Adams Smith's `invisible hand'~\cite{Smith1776}. Regulation becomes an emergent system property~\cite{Veetil2015}, originated from reconfigurable interactions and adaptive decision-making structured over a self-organized tree topology. Earlier work motivates the adoption of tree structures in computational markets and supply-demand systems~\cite{Carlsson2005,Andersson1998,Ygge1998}, however, the proposed framework does not exclude the adoption of other mechanisms and structures~\cite{Jacyno2013}. 

%\footnote{Resource-oriented and price-oriented market models can be equivalent as shown in earlier work~\cite{Andersson1998}.}

Evidence using real-world datasets of energy supply-demand systems confirms the effectiveness and applicability of the framework in a broad spectrum of regulatory scenarios. The implications of informational diversity in consumers' options but also the actual consumers' selections are studied by quantifying the regulatory capability with two metrics referred to as `response' and `savings'. A novel characteristic of these metrics is that they are relative to different upper bounds. This allows the systematic evaluation of optimality and constraints resulting in more informative and meaningful comparisons of different self-regulatory settings. Findings show measurable trade-offs between response and savings creating a vast potential for online adjustments incentivized by utilities, system operators and policy makers.

\section{Self-regulatory Framework}\label{sec:framework}

In this paper, a \emph{self-regulated supply-demand system} is defined by the inner operational system capability of distributed producers and consumers to collectively coordinate their actions remotely based on a fully decentralized computational design such that supply and demand are met under different regulatory scenarios. This section introduces a self-regulatory framework for supply-demand systems. It is shown how this framework can be realized as a decentralized transactive control system that is a well-established standardized concept and implemented technology in real-world supply-demand systems of Smart Cities. Table~\ref{table:math-symbols} in~\ref{app:math} outlines all mathematical symbols used in this paper in the order they appear. 

\subsection{Overview}\label{subsec:overview}

\begin{figure}[!htb]
\centering
\includegraphics[width=0.85\columnwidth]{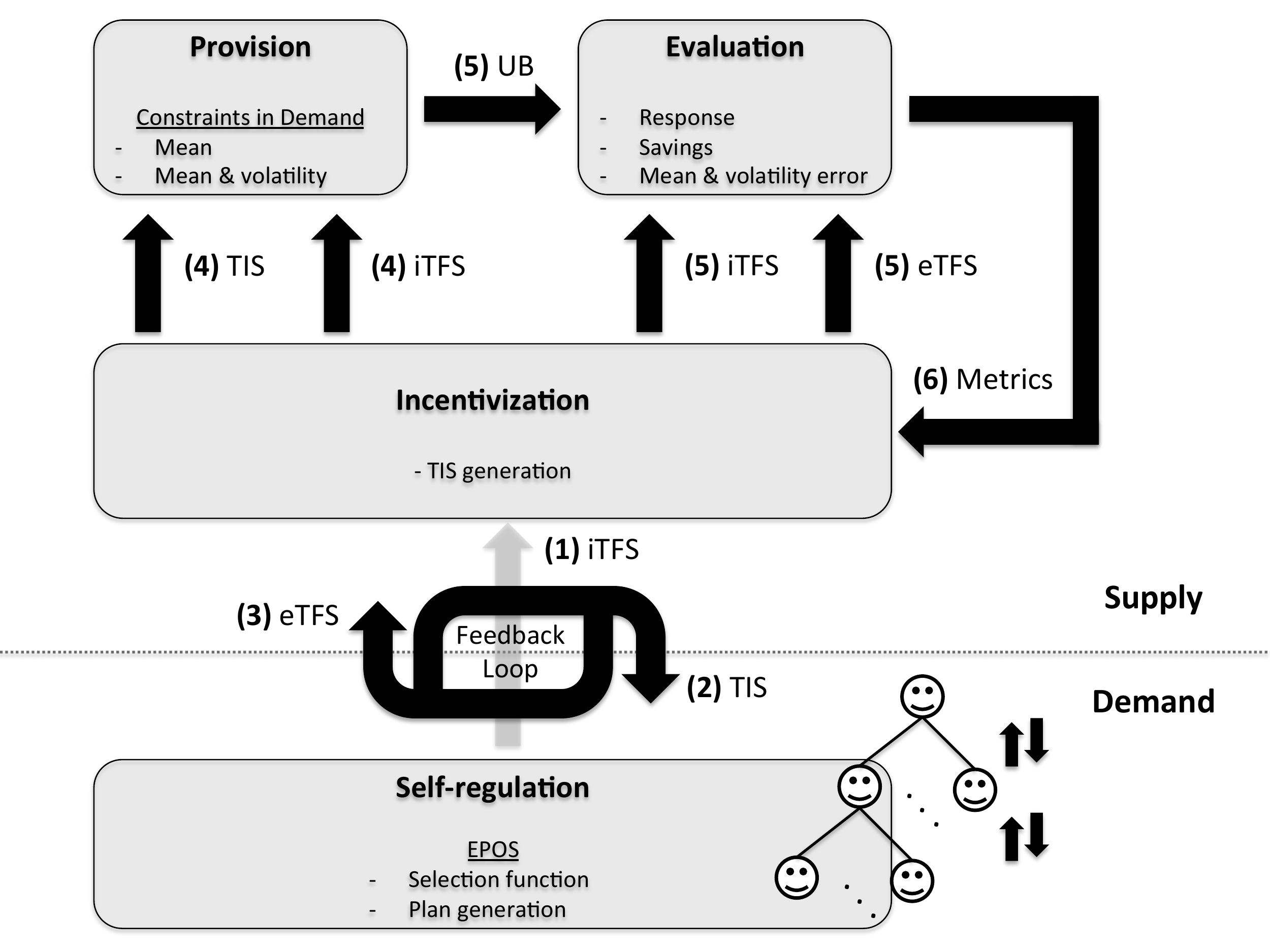}
\caption{A self-regulatory framework for supply-demand systems.}\label{fig:framework}
\end{figure}

The studied supply-demand system consists of communicating (software) agents that represent or control the actions of producers and consumers, for instance, smart meters in smart grids or vehicles in transportation systems. Communication between agents can be performed over the Internet or via other dedicated communication infrastructures of certain supply-demand systems~\cite{Gao2012}, for instance, a SCADA system in smart grids. The agent facilitates the self-regulation logic and represents social or operational requirements such as the type/amount of resources supplied and demanded. Depending on how a supply-demand system is institutionalized, three modes of self-regulation can be applied: (i) \emph{demand-side}, (ii) \emph{supply-side} and (iii) \emph{supply-demand-side}. The demand-side mode assumes the production of resources is given by system operators and consumers need to adapt to the available supply. Respectively, the supply-side mode assumes the demand of consumers is given by system operators and production needs the adapt to the formed demand. Finally, in the supply-demand mode, none of supply or demand are given and they are both formed interactively. Due to space limitations, this paper mainly focuses on the first two self-regulation modes. For consistency and an intuitive understanding of the framework, the rest of this paper illustrates the framework from the viewpoint of self-regulation in demand-side mode. Figure~\ref{fig:framework} introduces an overview of the framework. 

Self-regulation aims at balancing supply and demand under various operational scenarios and constraints, such as keeping demand within a certain range, distributing demand fairly among consumers, or supplying a minimum amount of resources to each consumer. Self-regulation is based on two types of information signals: (i) \emph{incentive signals} and (ii) \emph{feedback signals}. The incentive signal is the price for each unit of resource supplied. It is provided by supply to demand. The feedback signal is the amount of demand requested based on the incentive signal provided. It is provided by demand to supply. Self-regulation is governed by a feedback loop: agents coordinate demand in a fully decentralized fashion to collectively form a feedback signal as a response to a given incentive signal. A new incentive signal is potentially formed to further steer an improved feedback signal. 

\subsection{Transactive control}\label{subsec:TCS}

This framework is far beyond conceptual. Regulating supply and demand with incentive and feedback signals is the basis of \emph{transactive control}, a well-established standardized concept and implemented technology in the real world~\cite{Huang2010,Subbarao2013}. The term `transactive' refers to the use of economic and control techniques to improve system reliability and efficiency. Transactive control defines the minimal interaction types and information exchange in the domain of Smart Grids. However, it does not define any computational logic or formal methodology for the regulation of supply and demand, or the evaluation of the regulatory capability. The proposed framework fills this gap by introducing a decentralized regulation mechanism and an evaluation methodology that motivates the further adoption of transactive control in other Smart City applications such as gas networks, water networks, traffic systems, supply chain systems, etc. 

A \emph{transactive signal} is a sequence\footnote{This paper assumes positive signals for simplifying the illustration of the framework and results. Negative signals are possible as well and correspond to supply-demand systems in which an agent may act as both, consumer and producer.} $\signal{\signalType}=(\signalValue{\signalType}{\timePoint})_{\timePoint=1}^{\horizon}$ of numerical values, where $\signalValue{\signalType}{\timePoint} \in \mathbb{R}_{>0}$. This paper assumes that all signals are time series, and therefore, \horizon is a future time horizon. However, there is no limitation to other signal types applicable to several computational problems~\cite{Gale1962,Azevedo2014}. A sequence provides a computational degree of freedom in self-regulation: in the case of time series, temporal adjustments scheduled in the transactive signals can be used as regulatory tools, e.g. load-shifting in power grids, load-balancing between vehicles in transportation systems, etc. 

The types of transactive signals are defined as follows:

\begin{itemize}
\item \emph{Transactive Incentive Signal} (\TIS): A time series with the \emph{price} in currency per unit of resources supplied.  
\item \emph{Transactive Feedback Signal} (\TFS): A time series with the units of \emph{resources} supplied or demanded. This paper additionally distinguishes the following \TFS signals:
\begin{itemize}
\item \emph{Inelastic Transactive Feedback Signal} (\iTFS): A non-regulated historic or forecast \TFS signal. 
\item \emph{Elastic Transactive Feedback Signal} (\eTFS): A regulated \TFS signal.
\item \emph{Upper Bound Signal} (\upperBound): A reference \TFS signal for the evaluation of self-regulation. 
\end{itemize}
\end{itemize}

\noindent All interactions within the proposed framework are entirely based on the above transactive signals as shown in Figure~\ref{fig:framework}.

\subsection{Operational lifecycle}\label{subsec:operation}

The self-regulatory framework of Figure~\ref{fig:framework} performs the following four operations: (i) \emph{incentivization}, (ii) \emph{self-regulation}, (iii) \emph{provision} and (iv) \emph{evaluation}. Six transactive signalings are performed to complete an operational cycle. 

In the first signaling, depicted by Arrow 1 in Figure~\ref{fig:framework}, an \iTFS signal is provided by demand to supply. This signal is used as a historic or forecast base information by the other operations of the framework. For instance, the historic/forecasting information about the energy demand based on which energy prices are formed plays the role of an \iTFS signal. 

In the second signaling, depicted by Arrow 2 in Figure~\ref{fig:framework}, a \TIS signal is provided by supply to demand. A \TIS signal may be the result of incentivization performed based on the outcome of previous operational cycles and the current system requirements. Incentivization is performed by system operators, policy makers or even automated software systems that choose how to optimize and regulate a supply-demand system~\cite{Joskow2007}. The price information in the \TIS signal is computed by using information about the available supply or constraints in the characteristics of future demand, e.g. the mean and/or volatility. For instance, the costs involved in the scheduling of power generators form the \TIS signal.

In the third signaling, depicted by Arrow 3 in Figure~\ref{fig:framework}, an \eTFS signal is provided by demand to supply as a response to the received \TIS signal. The \eTFS signal is computed by the adopted self-regulation mechanism. For instance, the energy demand formed under the influence of a demand-response program is an \eTFS signal. At this stage, supply has all information to provision and evaluate the regulatory capability achieved in this operational cycle as outlines below. 

In the fourth signaling, depicted by Arrow 4 in Figure~\ref{fig:framework}, a hypothetical optimum \TFS signal is computed, the upper bound, given the \TIS and \iTFS signals, along with constraints in characteristics of demand. Such constraints may concern the computation of an upper bound signal with the same mean or, mean and volatility, with the \iTFS signal. For instance, if system operators plan to turn off a power generator for repair or maintenance, the hypothetical optimum energy demand is reduced to fully encounter the missing power generator. 

In the fifth signaling, depicted by Arrow 5 in Figure~\ref{fig:framework}, the evaluation of the \eTFS signal is triggered in the light of the \iTFS signal and the \upperBound signal. Evaluation measures the response, savings and the mean/volatility error. For instance these metrics can measure energy cost reductions and power peak shavings. 

Finally, in the sixth signaling, depicted by Arrow 6 in Figure~\ref{fig:framework}, an operational cycle completes by providing the values of the measured metrics as input to incentivization. The next operational cycle initiates the generation of an improved \TIS signals that can potentially achieve a higher response and savings but lower mean and volatility error.  

The rest of this paper illustrates the self-regulation framework in detail. 

\section{Incentivization}\label{sec:incentivization}

Incentivization concerns the construction of \TIS signals that accurately reflect the costs of supply, costs of infrastructure, operational constraints and certain regulatory actions required. For instance, when a power generator fails, costs increase as the available supply is lower, infrastructure may require maintenance and complex operational constraints should be met, i.e. ramp up times for the activation of operational power reserves~\cite{Joskow2007}. Moreover, system operators and utilities may initiate costly demand-response events so that consumers lower demand. The costs of all these actions sum up and form the prices of a \TIS signal. Forecasting is a critical component of incentivization as cost information is not always available~\cite{Timmermann2004,Carbonneau2008}. For example, the availability of energy supply generated by renewables is highly stochastic. Therefore the accuracy of weather forecasting techniques influence the effectiveness of a \TIS signal. These observations suggest that incentivization is a highly domain-dependent process. 

A \TIS incentive signal results in a \eTFS response signal evaluated by the metrics at the stage of evaluation. A \TIS signal can be modified to trigger a new \eTFS signal that optimizes in a evolutionary fashion the evaluation metrics according to a a weighting scheme or an elasticity function~\cite{Kirschen2000}. This paper focuses on evaluating a single operational cycle by using real-world \TIS signals computed via toolkit functions based on advanced predictive analytics. Section~\ref{sec:experimental-evaluation} illustrates the \TIS signals studied in this paper. The evaluation of an evolutionary operation of the self-regulatory framework is part of future work. 

%Implementations of the proposed self-regulatory framework can make use of existing techniques adopted by various application domains for constructing \TIS signals. However, the framework introduces the option to construct incentive signals in an evolutionary fashion.

\section{Provision}\label{sec:provision}

This section shows how to construct hypothetical \eTFS signals representing an optimally regulated supply-demand given the incentive signal \signal{\TISIndex} and a set of demand constraints. These hypothetical signals are referred to as \emph{upper bounds}. The incentive and feedback signals are assumed to follow the law of demand, meaning that higher prices in supply shall result in lower demand and the other way around. For the mathematical formulations of this paper, a linear reverse relationship is assumed that can be relaxed with empirical data or domain-specific models of supply-demand systems. The incentive signal \signal{\TISIndex} is used to construct a lower bound signal by computing its reflections as follows:

\begin{equation}
\reverseFunct{\timePoint,\signal{\TISIndex}} = 2\meanFunct{\signal{\TISIndex}}-\signalValue{\TISIndex}{\timePoint}
\end{equation}

%\begin{equation}
%\reverseFunct{\timePoint,\signal{\TISIndex}} = \begin{cases}
%\meanFunct{\signal{\TISIndex}}-\signalValue{\TISIndex}{\timePoint}, &\text{if $\signalValue{\TISIndex}{\timePoint}\geq\meanFunct{\signal{\TISIndex}}$}\\
%\meanFunct{\signal{\TISIndex}}+\signalValue{\TISIndex}{\timePoint}, &\text{if $\signalValue{\TISIndex}{\timePoint}<\meanFunct{\signal{\TISIndex}}$}
%\end{cases}
%\end{equation}

\noindent where the mean over the values of the incentive signal \meanFunct{\signal{\TISIndex}} is used as the hyperplane for the flip. The isometry of the reflection \reverseFunct{\timePoint,\signal{\TISIndex}} and the selection of the mean \meanFunct{\signal{\TISIndex}} as hyperplane result in a price signal with the same total cost for one unit of resources compared to the initial incentive signal \signal{\TISIndex}. An undesired effect of a reflection is that it may result in a signal with negative values. A normalization is illustrated in~\ref{app:reflection} that guarantees a signal \normalizeFunct{\timePoint,\signal{\TISIndex}} with positive values. It is used for the computation of the \upperBoundA.

This paper introduces two upper bounds distinguished by two demand constraints: 

\begin{itemize}
\item The mean \meanFunct{\signal{\upperBoundIndex}} of the \upperBound signal equals the mean \meanFunct{\signal{\iTFSIndex}} of the \iTFS signal. This constraint keeps the total amount of demand equal to the one of the \iTFS signal, however demand is distributed more effectively as defined by the reflected \iTFS signal \reverseFunct{\timePoint,\signal{\TISIndex}}.
\item The mean \meanFunct{\signal{\upperBoundIndex}} and volatility \stdevFunct{\signal{\upperBoundIndex}} of the \upperBound signal equal the mean \meanFunct{\signal{\iTFSIndex}} and volatility \stdevFunct{\signal{\iTFSIndex}} of the \iTFS signal. This constraint keeps the total amount of demand equal to the one of the \iTFS signal and, additionally, it limits the standard deviation of the values to the level of the \iTFS signal. 
\end{itemize}

\begin{bounds}
Given a \TIS signal \signal{\TISIndex} and an \iTFS signal \signal{\iTFSIndex}, an upper bound \signal{\upperBoundAIndex} is given as follows:

\begin{equation}\label{eq:upperBoundA}
\begin{split}
&\signalValue{\upperBoundAIndex}{\timePoint}=\frac{\meanFunct{\signal{\iTFSIndex}}}{\meanFunct{\signal{\TISIndex}}}\normalizeFunct{\timePoint,\signal{\TISIndex}}, \forall \timePoint \in [1,\horizon]\\
&\text{subject to }\meanFunct{\signal{\upperBoundAIndex}}=\meanFunct{\signal{\iTFSIndex}}
\end{split}
\end{equation}

\end{bounds}

\begin{proof}
The average cost of one demand unit during the time horizon \horizon is given by \meanFunct{\signal{\TISIndex}}. Subject to $\meanFunct{\signal{\upperBoundAIndex}}=\meanFunct{\signal{\iTFSIndex}}$, the average demand acquired for one cost unit of demand is \meanFunct{\signal{\iTFSIndex}}/\meanFunct{\signal{\TISIndex}}. Then, by proportionality, for \normalizeFunct{\timePoint,\signal{\TISIndex}} cost units of demand, $\frac{\meanFunct{\signal{\iTFSIndex}}}{\meanFunct{\signal{\TISIndex}}}\normalizeFunct{\timePoint,\signal{\TISIndex}}$ units of demand are acquired for the upper bound \signalValue{\upperBoundAIndex}{\timePoint}.
\end{proof}

\begin{bounds}
Given a \TIS signal \signal{\TISIndex} and an \iTFS signal \signal{\iTFSIndex}, an upper bound \signal{\upperBoundBIndex} is given as follows:

\begin{equation}\label{eq:upperBoundB}
\begin{split}
&\signalValue{\upperBoundBIndex}{\timePoint}=\stdevFunct{\signal{\iTFSIndex}}\frac{\reverseFunct{\timePoint,\signal{\TISIndex}}-\meanFunct{\signal{\TISIndex}}}{\stdevFunct{\signal{\TISIndex}}}+\meanFunct{\signal{\iTFSIndex}}, \forall \timePoint \in [1,\horizon]\\
&\text{subject to }\meanFunct{\signal{\upperBoundBIndex}}=\meanFunct{\signal{\iTFSIndex}} \text{ and }\stdevFunct{\signal{\upperBoundBIndex}}=\stdevFunct{\signal{\iTFSIndex}}
\end{split}
\end{equation}

\end{bounds}

\begin{proof}

Let $\reverseFunct{\timePoint,\signal{\TISIndex}}-\meanFunct{\signal{\TISIndex}}$ represent the price difference at time \timePoint from the average price \meanFunct{\signal{\TISIndex}}. This difference, measured in units of price, can be measured in units of price volatility of the \TIS signal as $\frac{\reverseFunct{\timePoint,\signal{\TISIndex}}-\meanFunct{\signal{\TISIndex}}}{\stdevFunct{\signal{\TISIndex}}}$. The demand difference from the average demand \meanFunct{\signal{\TISIndex}} at time \timePoint can be constructed using the demand volatility \stdevFunct{\signal{\iTFSIndex}} as $\stdevFunct{\signal{\iTFSIndex}}\frac{\reverseFunct{\timePoint,\signal{\TISIndex}}-\meanFunct{\signal{\TISIndex}}}{\stdevFunct{\signal{\TISIndex}}}$ subject to $\stdevFunct{\signal{\upperBoundBIndex}}=\stdevFunct{\signal{\iTFSIndex}}$. Given that $\meanFunct{\signal{\upperBoundBIndex}}=\meanFunct{\signal{\iTFSIndex}}$, the final upper bound signal is constructed by summing the average demand $\meanFunct{\signal{\iTFSIndex}}$ as $\stdevFunct{\signal{\iTFSIndex}}\frac{\reverseFunct{\timePoint,\signal{\TISIndex}}-\meanFunct{\signal{\TISIndex}}}{\stdevFunct{\signal{\TISIndex}}}+\meanFunct{\signal{\iTFSIndex3}}$. 

\end{proof}

Although these hypothetical upper bounds are constrained to aggregate demand characteristics of the \iTFS signal, other more complex operational constraints can be captured as well, depending on the application domain considered. For example, in Smart Grids, operational constraints of the individual household devices or the availability of renewable energy resources can be considered~\cite{Koutitas2012,Ramchurn2011,Brandstatt2011}.

\section{Evaluation}\label{sec:evaluation}

This paper studies three factors that measure the effectiveness of self-regulation: (i) \emph{response}, (ii) \emph{savings} and (iii) \emph{mean/volatility error}. Given an initial feedback signal \signal{\iTFSIndex} and a feedback signal \signal{\eTFSIndex} computed by a regulation mechanism after applying an incentive signal \signal{\TISIndex}, response and savings are computed as follows: 

\begin{equation}\label{eq:response}
\response=\frac{\sum_{\timePoint=1}^{\horizon}|\signalValue{\iTFSIndex}{\timePoint}-\signalValue{\eTFSIndex}{\timePoint}|}{\sum_{\timePoint=1}^{\horizon}|\signalValue{\iTFSIndex}{\timePoint}-\signalValue{\upperBoundIndex}{\timePoint}|}
\end{equation}

\begin{equation}\label{eq:savings}
\savings=\frac{\sum_{\timePoint=1}^{\horizon}\signalValue{\TISIndex}{\timePoint}*\signalValue{\iTFSIndex}{\timePoint}-\sum_{\timePoint=1}^{\horizon}\signalValue{\TISIndex}{\timePoint}*\signalValue{\eTFSIndex}{\timePoint}}{\sum_{\timePoint=1}^{\horizon}\signalValue{\TISIndex}{\timePoint}*\signalValue{\iTFSIndex}{\timePoint}-\sum_{\timePoint=1}^{\horizon}\signalValue{\TISIndex}{\timePoint}*\signalValue{\upperBoundIndex}{\timePoint}}
\end{equation}

\noindent where \signal{\upperBoundIndex} is an upper bound signal. 

Response represents the similarity of the self-regulated demand with the optimal demand computed by the upper bound. Response measures the distance $|\signalValue{\iTFSIndex}{\timePoint}-\signalValue{\eTFSIndex}{\timePoint}|$ of the two feedback signals \signal{\iTFSIndex} and \signal{\eTFSIndex} in relation to the maximum distance that these two signals can hypothetically have. This maximum distance is computed using the upper bound signal as $|\signalValue{\iTFSIndex}{\timePoint}-\signalValue{\upperBoundIndex}{\timePoint}|$. 

Savings payoff measures the cost reduction between the feedback signals \signal{\iTFSIndex} and \signal{\eTFSIndex} in relation to the maximum hypothetical cost reduction between the feedback signal \signal{\eTFSIndex} and the upper bound signal \signal{\upperBoundIndex}. In other words, savings measures the response in economic terms. Cost is computed by multiplying the incentive signal (price) with the feedback signal (quantity), for example, $\signalValue{\TISIndex}{\timePoint}*\signalValue{\eTFSIndex}{\timePoint}$ provides the cost of \signalValue{\eTFSIndex}{\timePoint} amount of resources with price \signalValue{\TISIndex}{\timePoint} at time \timePoint. 

An upper bound \signal{\upperBoundIndex} is required to compute both response and savings. With these two metrics based on an upper bound, different self-regulation strategies can be meaningfully compared. For example, a self-regulation strategy `A' may decrease the demand cost 5\% more than another strategy 'B'. This finding is not so informative compared to knowing a maximum possible cost reduction of demand that indicates the actual effectiveness of these strategies compared to a hypothetical optimal case. This evaluation methodology is especially useful in decentralized self-regulation mechanisms that capture objectives with optimum solutions being computationally intractable.  

The capability of a regulatory mechanism to satisfy the constraints of the upper bounds can be measured by the mean and volatility errors as follows:

\begin{equation}
\error{\mu}=|1-\frac{\meanFunct{\signal{\eTFSIndex}}}{\meanFunct{\signal{\iTFSIndex}}}|\text{, }\error{\sigma}=|1-\frac{\stdevFunct{\signal{\eTFSIndex}}}{\stdevFunct{\signal{\iTFSIndex}}}|,
\end{equation}

\noindent where \error{\mu} is the relative error of the mean in the \eTFS signal relative to the mean of the \iTFS signal, and \error{\sigma} is the respective relative error of demand volatility. The mean and volatility errors represent violations of aggregate characteristics in demand. For instance, a minimal mean error means that suppliers do not need to supply a higher aggregated amount of resources that the one allocated over the future horizon.

\section{Self-regulation}\label{sec:self-regulation}

This paper advances the EPOS mechanism~\cite{Pournaras2010b} to employ it as a highly generic self-regulation mechanism for supply-demand systems. EPOS is earlier studied in the Smart Grid domain as a fully decentralized mechanism to improve the robustness of Smart grids by stabilizing energy demand, lowering power peaks and other regulatory actions. Loads, representing electrical household devices or even the consumption of a household as a whole~\cite{Pournaras2014c,Pournaras2014d}, are regulated by software agents self-organized~\cite{Pournaras2014a} in a tree topology\footnote{A tree overlay network is the logical structuring of the network communication at the application-level.} for structuring their interactions. A tree overlay network provides a cost-effective aggregation of the demand level required for coordinating the decision-making: each agent generates a number of transactive signals \numOfPlans referred to as \emph{possible plans} $\demandPlans{\agent}=(\demandPlan{\agent}{\plan})_{\plan=1}^{\numOfPlans}$ and collectively selects the plan \demandPlan{\agent}{\selectedPlan} that fits best to an objective. Decision-making is performed with selection functions~\cite{Pournaras2014c} that evaluate characteristics of the demand plan, e.g. how uniformly distributed demand is over time, how large the demand peaks are, etc. Starting from the leaf agents, parents select the selected plans of their children using the possible plans of their children and the aggregated selections of the agents in the branch underneath. The process repeats level-by-level and completes with a propagation of the final \eTFS signal by the root agent to all agents in the tree. 

The main limitation of EPOS is that it does not interact with supply and, therefore, it cannot be used for balancing supply and demand under a broader range of operational scenarios. This paper fills this gap by introducing two new selection functions without any change in the algorithm and architecture of EPOS.  

The plans of EPOS and the transactive signals are both mathematically equivalent sequences. This makes a transactive control system fully interoperable with the EPOS mechanism. The supply-side provides to EPOS a \TIS signal\footnote{The \TIS signal can be distributed to all agents of the tree overlay network via application-level broadcasting supported in EPOS for distributing the global plan~\cite{Pournaras2010b}.} and receives back an \eTFS signal computed in EPOS as follows:

\begin{equation}\label{eq:eTFS-EPOS}
\signalValue{\eTFSIndex}{\timePoint}=\sum_{\agent=1}^{\numOfAgents}\demand{\agent}{\selectedPlan}{\timePoint}
\end{equation}

\noindent where \numOfAgents is the total number of agents connected in the tree overlay network of EPOS and \demand{\agent}{\selectedPlan}{\timePoint} is the selected demand of agent \agent at time \timePoint. 

Plan generation computes the possible plans. Each one can be executed by the agent locally if selected. The possible plans represent a degree of freedom in the operation of the physical assets that the agent regulates. Therefore, the physical characteristics and the parameterization of these physical assets can make the computation of possible plans feasible in various application domains. For example, possible plans can be operationally equivalent, and therefore alternatives, when household appliances permit this~\cite{Pournaras2010b,Callaway2011}. Moreover, possible plans may represent social preferences, such as comfort levels~\cite{Pournaras2017,Hinrichs2013} or even recommendations extracted from historic data~\cite{Pournaras2014c,Pournaras2014d}. This paper studies three generation schemes of possible plans that can be applied in several application domains and scenarios: (i) \shuffle, (ii) \shift and (iii) \swap. In \shuffle, each possible plan is a random permutation of the other ones. This scheme represents scenarios in which agents have a significant controllability in the operation of the regulated assets. In \shift, the values of a plan are rotated by pushing values certain time steps ahead in time. This scheme represents load-shifting scenarios or future resource reallocation and rescheduling. In \swap, a number of values in a plan interchange their positions. This scheme represents scenarios in which scheduled operations can be interchanged within the future horizon. 

Each scheme generates plans that contain the same values and, therefore, they have the same mean, volatility and entropy that is useful for the interpretation of the measurements performed. In this way, the generation schemes are not highly disruptive for consumers~\cite{Callaway2011} and do not influence the total amount of supply scheduled within the period \horizon. Moreover, the elasticity of volatility is exclusively governed by EPOS and not by the generation schemes that would bias the measurements of response and savings.

The informational diversity of the possible plans is defined by the Euclidean distance between the positions of the demand values after applying a generation scheme. Let \demandPlan{\agent}{1} being a seed possible plan and \demandPlan{\agent}{2} the diversified possible plan with one of the proposed generation schemes. The informational diversity is measured as $\sum_{\timePoint=1}^{\horizon}|\timePoint-\newTimePoint|$, where \timePoint refers to the time point of demand \demand{\agent}{1}{\timePoint} and \newTimePoint=\diversityFunct{\timePoint} refers to the new time point of demand \demand{\agent}{2}{\newTimePoint} after diversification with a function \diversityFunct{\timePoint} such that $\demand{\agent}{1}{\timePoint}\equiv\demand{\agent}{2}{\newTimePoint}$.

\begin{figure}[!htb]
\centering
\includegraphics[width=0.49\columnwidth]{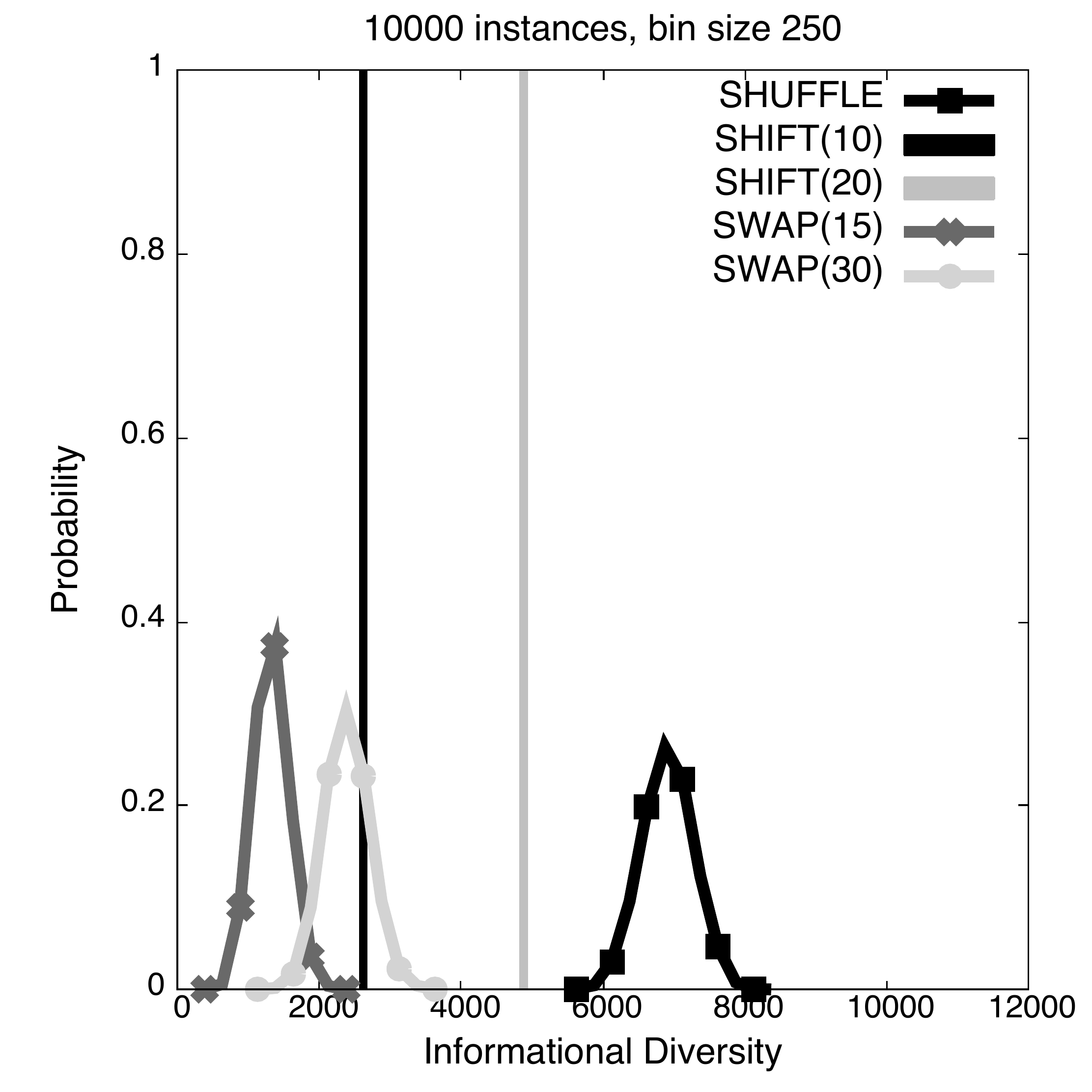}
\caption{Probability density function of informational diversity for the five generation schemes.}\label{fig:diversity}
\end{figure}

Figure~\ref{fig:diversity} illustrates the probability density functions of informational diversity for the five generation schemes\footnote{The test is implemented in Java using randomization methods of JDK 1.8.}. \shuffle has the highest informational diversity followed by \shiftParam{20}. \shiftParam{10}, \swapParam{30} follow very closely and \swapParam{15} has the lowest diversity. 

Decision-making is performed bottom-up and level-by-level: children interact with their parents and collectively decide which plan to execute based on (i) their own demand plans and (ii) the selected plans of the agents connected to each of their branch underneath. Selection is computationally performed by the parent of the children that computes the combinational plans as follows:  

\begin{equation}\label{eq:combinational-plan}
\combinationalDemandPlans{\agent}=(\combinationalDemandPlan{\agent}{\plan})_{\plan=1}^{\numOfPlans^{\numOfChildren}}=\bruteForceFunct{\aggregatePlans{\child},...,\aggregatePlans{\child+\numOfChildren}}
\end{equation}

\noindent where the parent agent \agent performs a brute force operation \bruteForceFunct{\aggregatePlans{\child},...,\aggregatePlans{\child+\numOfChildren}} to compute all combinations between the sequences of aggregates plans $\aggregatePlans{\child},...,\aggregatePlans{\child+\numOfChildren}$ received by its \numOfChildren children. The child \child of agent \agent is assumed to be the first one and the child \child+\numOfChildren the last one. An aggregate plan $\aggregatePlan{\child}{\plan} \in \aggregatePlans{\child}$ is computed by summing the demand plan \demandPlan{\child}{\plan} and all selected plans of the agents received through the branch underneath child \child. 

Three selection functions are introduced with the following self-regulation objective: match the actual demand to any desired demand curve. This is an especially challenging objective to fulfill given that decision-making is performed in a fully decentralized fashion. The three selection functions use different mathematical approaches to achieve the same self-regulation objective. 

The first two selection functions \minRmseA and \minRmseB are devised to maximize response\footnote{They are also expected to decrease the cost given that a reflection of the \TIS signal is used for the computation of the upper bounds.}. They compute the root mean square error between the combinational plans and one of the upper bounds illustrated in Section~\ref{sec:provision}:

\begin{equation}\label{eq:maxResponse}
\begin{split}
&\selectedPlan=\argmin_{\plan=1}^{\numOfPlans^{\numOfChildren}}\sqrt{\frac{\sum_{\timePoint=1}^{\horizon}(\combinationalDemand{\agent}{\plan}{\timePoint}-\signalValue{\upperBoundIndex}{\timePoint})^{2}}{\horizon}}\\
&\text{subject to }\signalValue{\iTFSIndex}{\timePoint}=\combinationalDemand{\agent}{\plan}{\timePoint}
\end{split}
\end{equation}

\noindent This relation gives the combinational plan, and therefore the selected plan of each child, with the minimum root mean square error. This is the combinational plan with the highest matching to the upper bound. The upper bound is recomputed for each combinational plan given that $\signalValue{\iTFSIndex}{\timePoint}=\combinationalDemand{\agent}{\plan}{\timePoint}$. This condition is necessary as it makes collective decision-making possible\footnote{Several other versions of selection functions are experimentally evaluated, however, it is this function that results in \eTFS signals with improved response and savings.}. It actually means that when a number of agents have a selected plan, meaning the agents in the branch underneath each child, the rest of the agents need to bind their decision-making to these existing choices. 

The third selection function \minCost is devised to minimize the cost of demand:

\begin{equation}\label{eq:1}
\selectedPlan=\argmin_{\plan=1}^{\numOfPlans^{\numOfChildren}}\sum_{\timePoint=1}^{\horizon}\signalValue{\TISIndex}{\timePoint}*\combinationalDemand{\agent}{\plan}{\timePoint}
\end{equation}

\noindent where the price of supply \signalValue{\TISIndex}{\timePoint} in the \TIS signal multiplied by the planned demand \combinationalDemand{\agent}{\plan}{\timePoint} in the combinational plan at time \timePoint provides the respective cost of demand. 

The generation schemes, selection functions and number of possible plans are all configurations that influence the performance of self-regulations. They can be incentivized by supply-demand service providers via some payoff.

\section{Experimental Framework Evaluation}\label{sec:experimental-evaluation}

This section illustrates the applicability and evaluation of the self-regulatory framework in four scenarios of balancing energy supply and demand: (i) \emph{ramp down}, (ii) \emph{generation failure}, (iii) \emph{maximum entropy} and (iv) \emph{minimum entropy}. Each regulatory scenario corresponds to a \TIS signal of size 144 as standardized in transactive control for Smart Grids\footnote{This corresponds to a 72 hours forecasting of variable granularity within a market operating every 5 minutes.}~\cite{Hoke2013,Subbarao2013}. Evaluation is based on both simulated and real-world Smart Grid data sources: (i) the simulated zonal power transmission in the Pacific Northwest\footnote{Available upon request at \url{http://www.pnwsmartgrid.org/participants.asp} (last accessed: November 2015)} (\SIM), (ii) the \'{E}lectricite\'{e} de France\footnote{Available at \url{https://archive.ics.uci.edu/ml/datasets/Individual+household+electric+power+consumption} (last accessed: November 2015)} utility (\EDF) and (iii) the Pacific Northwest Smart Grid Demonstration Project\footnotemark[8] (\PNW).

\begin{figure}[!htb]
\centering
\subfigure[\TIS signals from the \SIM dataset.]{\includegraphics[width=0.49\columnwidth]{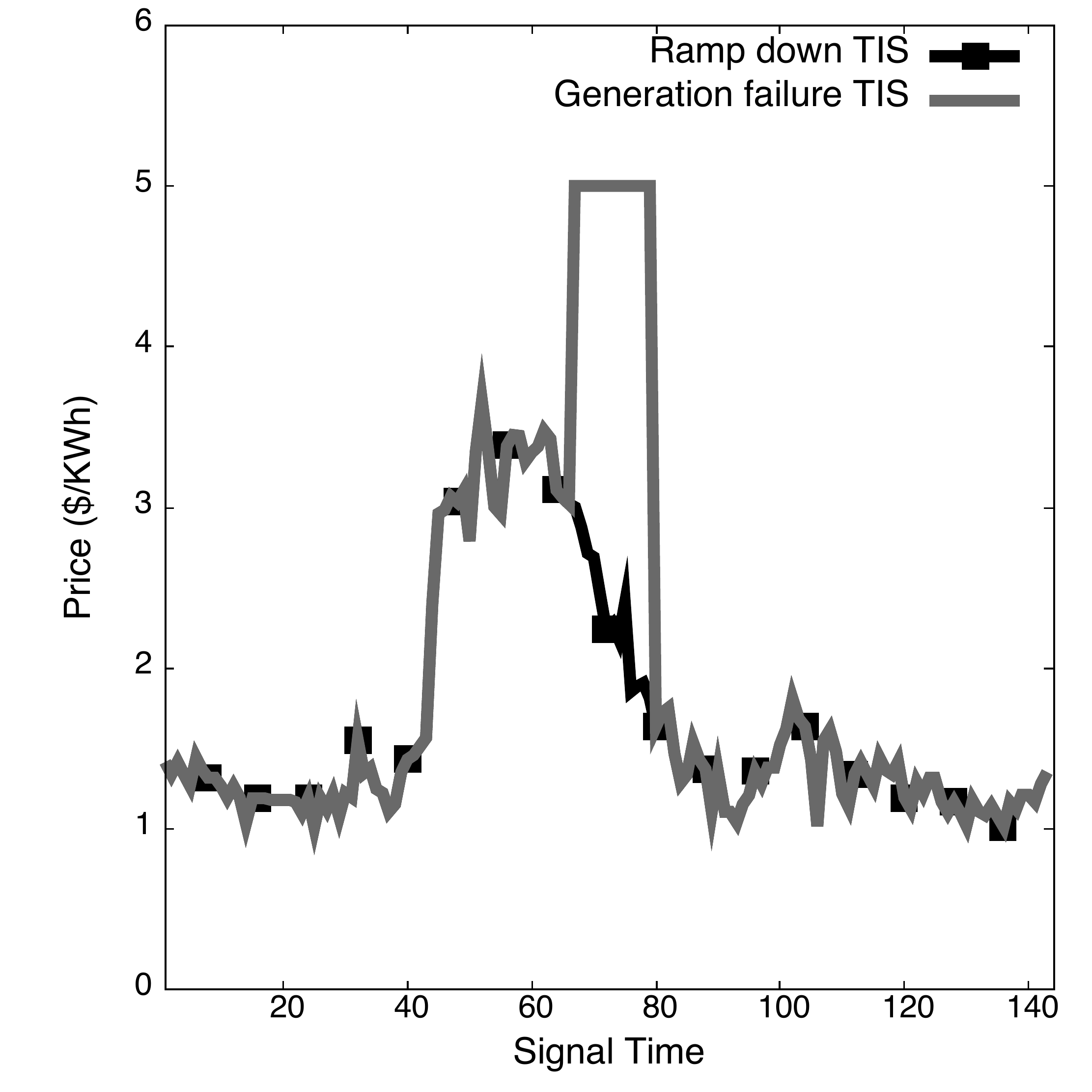}}
\subfigure[\TIS signals from the \PNW dataset.]{\includegraphics[width=0.49\columnwidth]{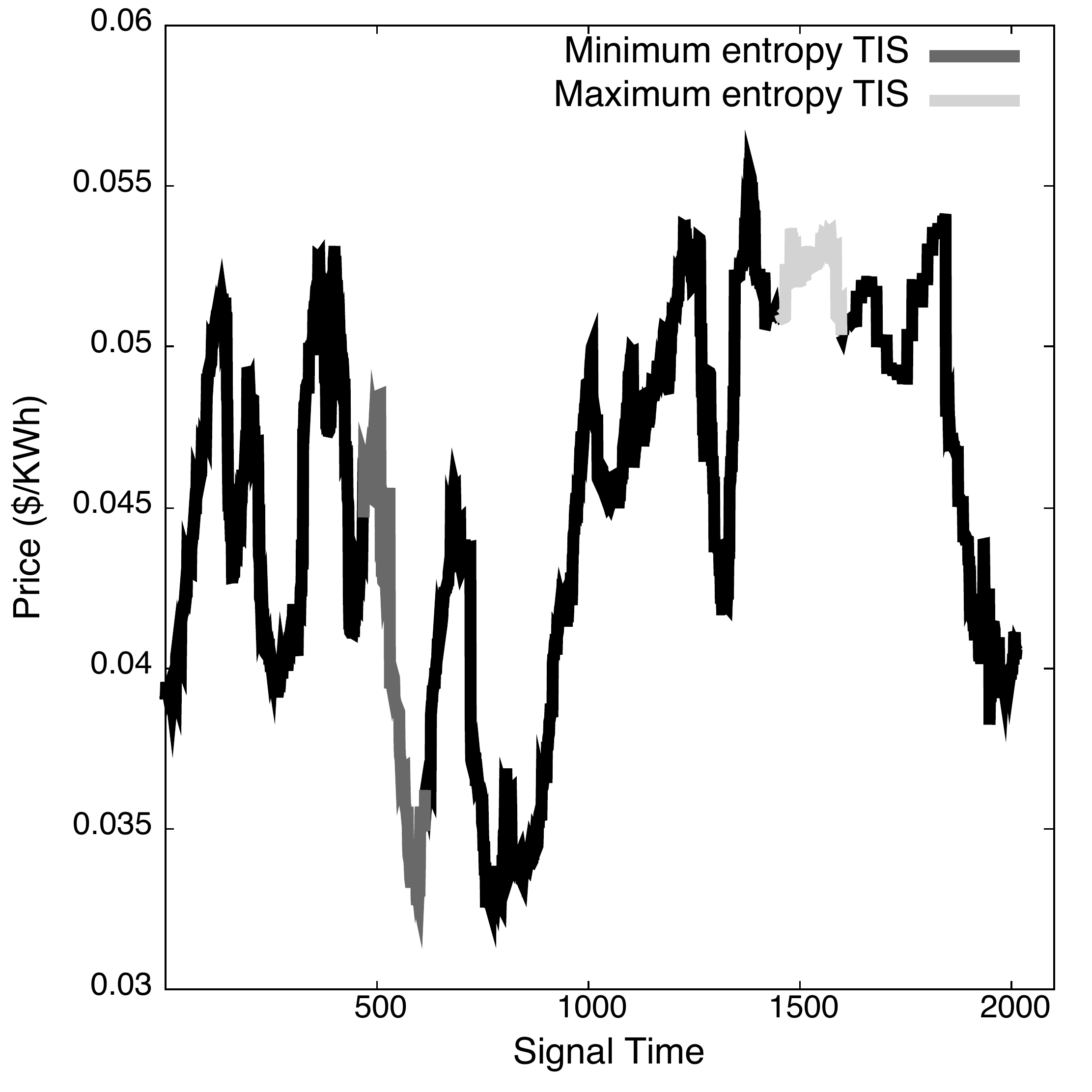}}
\caption{The \TIS signals corresponding to the four regulatory scenarios: (i) ramp down, (ii) generation failure, (iii) minimum entropy and (iv) maximum entropy.}\label{fig:tis-signals}
\end{figure}

A ramp down scenario concerns a situation in which power supply suddenly drops due to, for example, a low availability of renewable energy resources, \eg wind generation. In this case, operating reserves with high operational costs are activated such as backup diesel generators. Figure~\ref{fig:tis-signals}a shows that during the time period $[40,80]$ the generation costs in the \TIS signal of the \SIM dataset doubles reflecting the low availability of cheap renewable energy resources during this period. 

The generation failure scenario concerns the unexpected damage of an operational power generator at the same time in which the ramp down scenario occurs as well. Further expensive operating reserves need to be activated followed up by urgent repair costs to meet demand. This scenario is illustrated in Figure~\ref{fig:tis-signals}a with the \TIS signal from the \SIM dataset that shows the increase in the price of generation during the time period $[60,80]$. 

The minimum entropy scenario concerns a situation in which the availability of power generation is highly dispersed over time. This dispersion is measured with the entropy of Shannon's information theory~\cite{Jaynes82}. This scenario assumes that a low entropy in the incentive signal corresponds to a high dispersion in power generation. Therefore, matching demand to the available supply is more challenging. The opposite holds for the maximum-entropy scenario that assumes a low dispersion in the power generation over time. When demand is regulated to have a maximum entropy, the required adjustments in the generation are minimized as demand is more uniformly distributed compared to the minimum entropy scenario. Figure~\ref{fig:tis-signals}b illustrates a series of consecutive \TIS signals used from the \PNW dataset\footnote{The \TIS signals concern the transmission zone 12 at project site 14 for the period starting on 30/07/13 at 09:45 to 06/08/13 at 09:30.}. An exhaustive heuristic algorithm is applied over all consecutive \TIS signals of size 144 to discover the ones with the minimum and maximum entropy as shown in this figure.  

\begin{figure}[!htb]
\centering
\includegraphics[width=0.85\columnwidth]{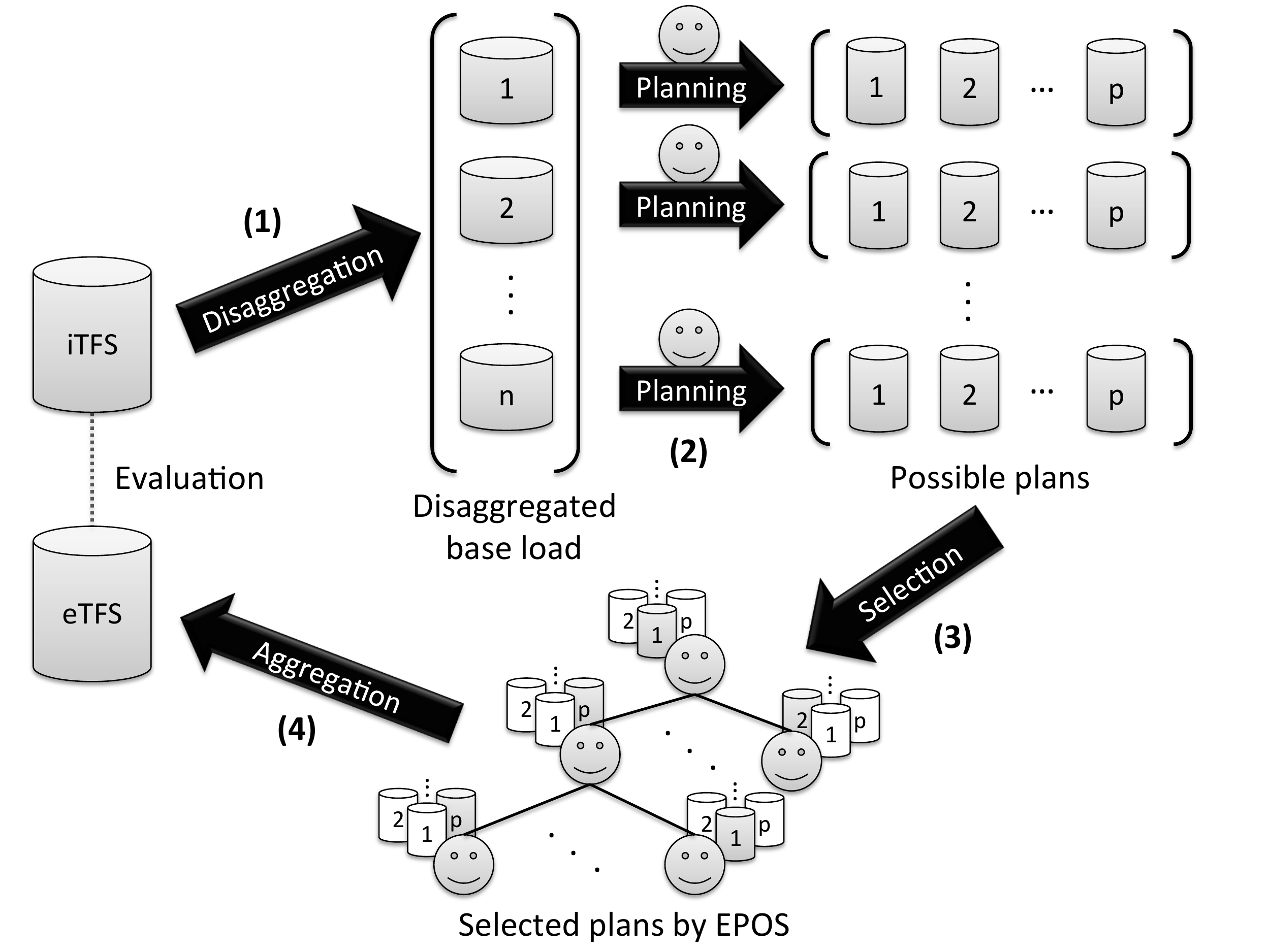}
\caption{The experimental setup for generating an \eTFS signal: (1) The aggregated base load from different real-world Smart Grid demonstration projects is disaggregated to \numOfAgents individual household base loads. (2) An agent in each household generates \numOfPlans number of alternative demand plans based on a generation scheme that receives as input the base load of a household. (3) The agents are self-organized in a tree topology over which the decentralized algorithm of EPOS runs to select the demand plan for execution. (4) The agents perform a tree aggregation that computes the \eTFS signal.}\label{fig:data-pipeline}
\end{figure}

Experimental evaluation is performed for each regulatory scenario by measuring and comparing the response, savings and volatility error\footnote{The mean error is not illustrated as the possible plans for each agent have the same mean and therefore, for any selection performed, $\meanFunct{\signal{\eTFSIndex}}=\meanFunct{\signal{\iTFSIndex}}$ is satisfied. Experimental evaluation confirms an average $\error{\mu}=0.09$ for all performed experiments.} of the \eTFS signals using the upper bound \upperBoundB\footnote{Response and savings are also measured with the upper bound \upperBoundA. These results are not shown in this paper given the space limitations and the fact that they indicate similar findings.}. The three following settings vary in each experiment: (i) generation scheme, (ii) selection function and (iii) dataset. Each \eTFS signal is generated by a process pipeline of four stages as illustrated in Figure~\ref{fig:data-pipeline}. The first stage in this pipeline, the disaggregation, is applied in the \SIM and \EDF datasets that only contain the aggregate base load of all residential consumers. In contrast, the \PNW dataset contains the base load for each individual household.

Each regulatory scenario is applied in two project sites\footnote{They are enumerated as project site 12 and 14.} that belong in a single transmission zone\footnote{It is enumerated as transmission zone 14}. The loads in each project site are regulated by the EPOS agents that are self-organized in an overlay network with a 3-ary balanced tree topology. Self-organization is performed with the AETOS overlay service~\cite{Pournaras2014a}. Each EPOS agent generates 4 possible plans using the following generation schemes: (i) \shuffle, (ii) \shiftParam{10}, (iii) \shiftParam{20}, (iv) \swapParam{15} and (v) \swapParam{30}. Plan selection is performed with the \minRmseA, \minRmseB and \minCost selection functions of EPOS. The \iTFS signal of the \SIM dataset is disaggregated in 5600 and 2374 loads for each project site respectively, emulating in this way the loads of the `Flathead, Libby' and `Flathead, Haskill' project sites. Disaggregation is performed as illustrated in~\ref{app:disaggregation} with parameter $\heterogeneity=0.20$. The \iTFS signal of the \EDF dataset is disaggregated, as illustrated in~\ref{app:disaggregation}, in 724 loads for each project site respectively. The \PNW dataset contains the residential base load of approximately 1000 residential consumers on the date 23.07.2014. It is split in two subsets referred to as \PNWMorning and \PNWEvening. The \iTFS signal for \PNWMorning concerns the duration 01:00-13:00, whereas the \iTFS for \PNWEvening concerns the duration 11:00-23:00.

\subsection{Response, savings and volatility error}\label{subsec:response-savings}

\begin{figure}[!htb]
\centering
\subfigure[Generation Schemes]{\includegraphics[width=0.49\columnwidth]{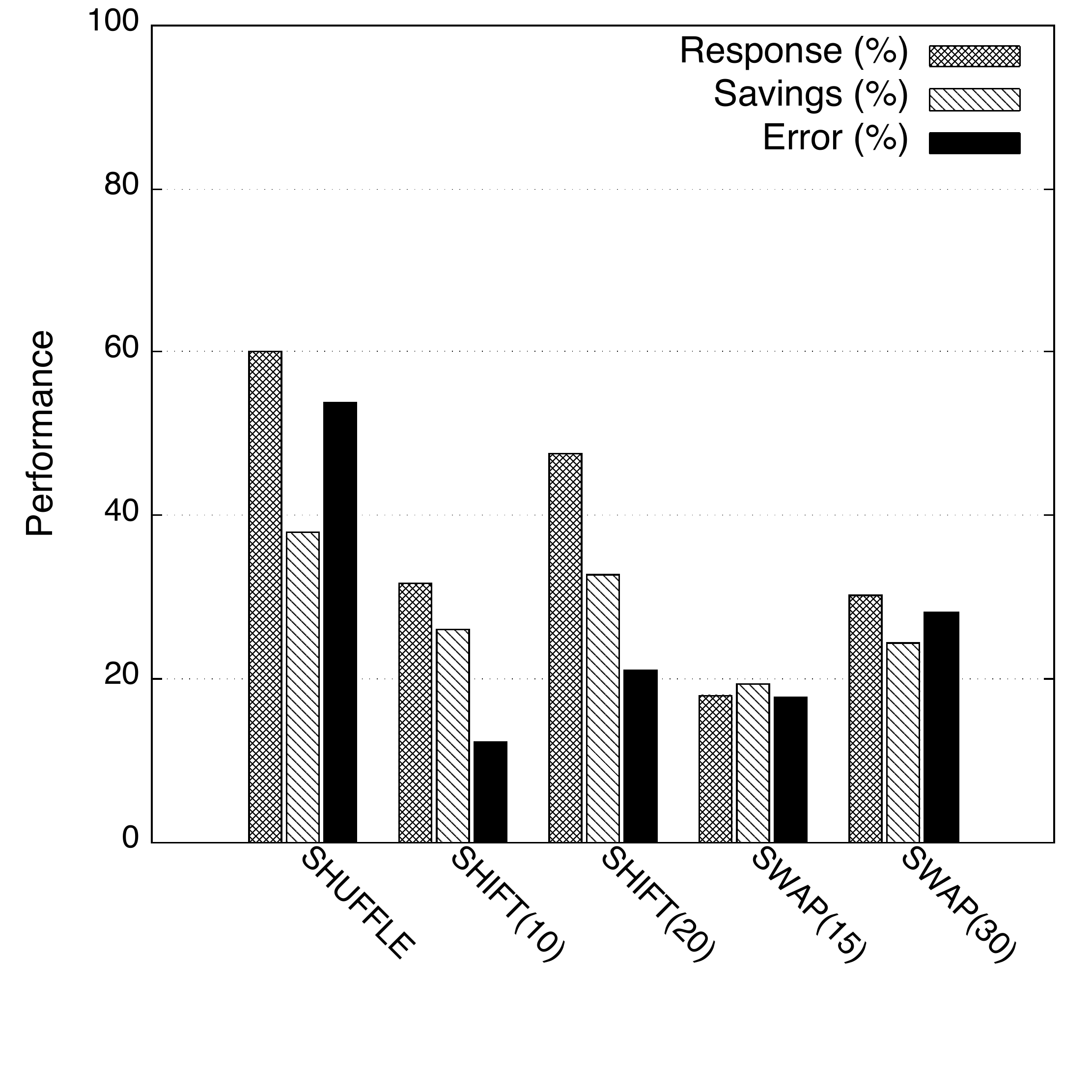}}
\subfigure[Selection Functions]{\includegraphics[width=0.49\columnwidth]{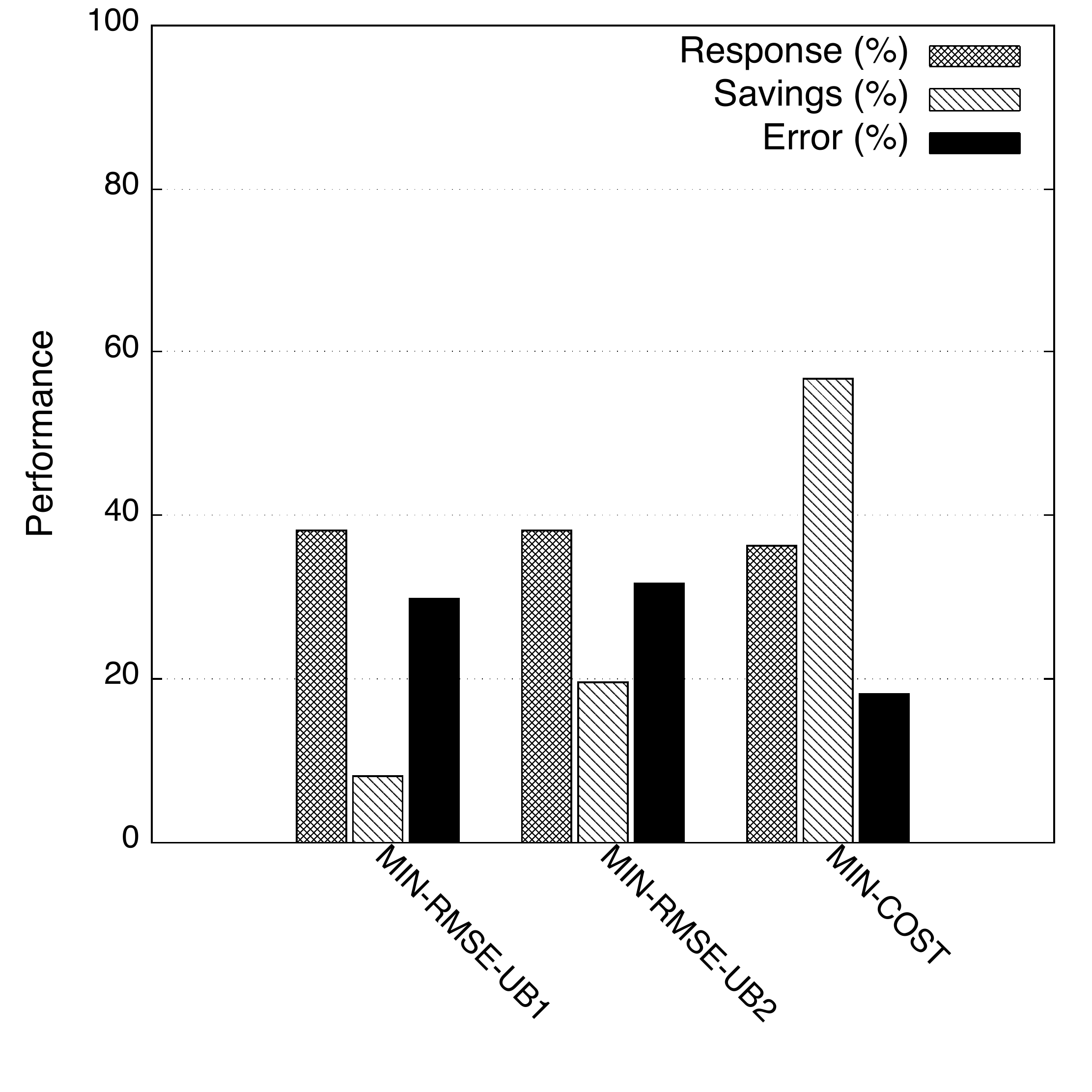}}
\subfigure[Datasets]{\includegraphics[width=0.49\columnwidth]{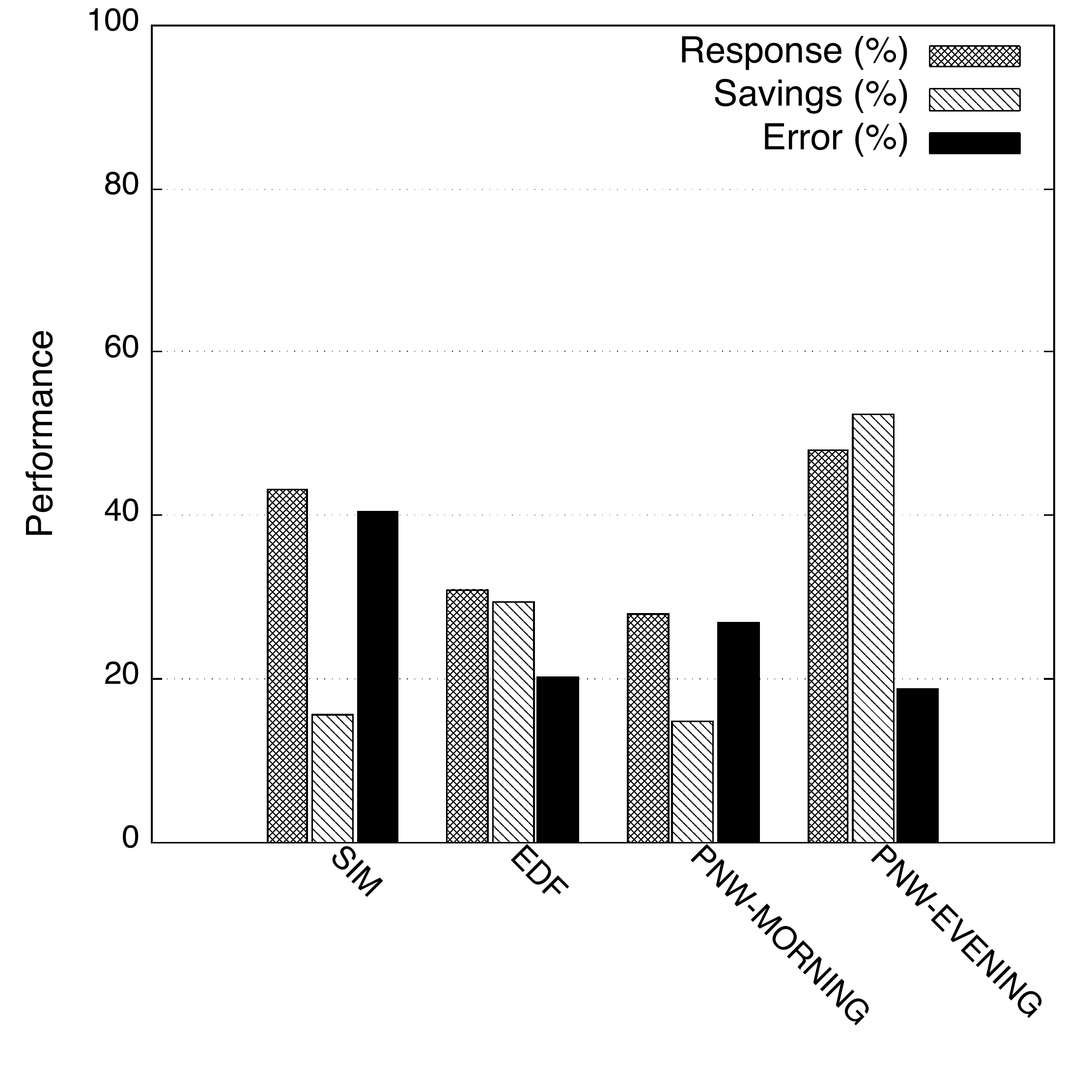}}
\subfigure[Regulatory Scenarios]{\includegraphics[width=0.49\columnwidth]{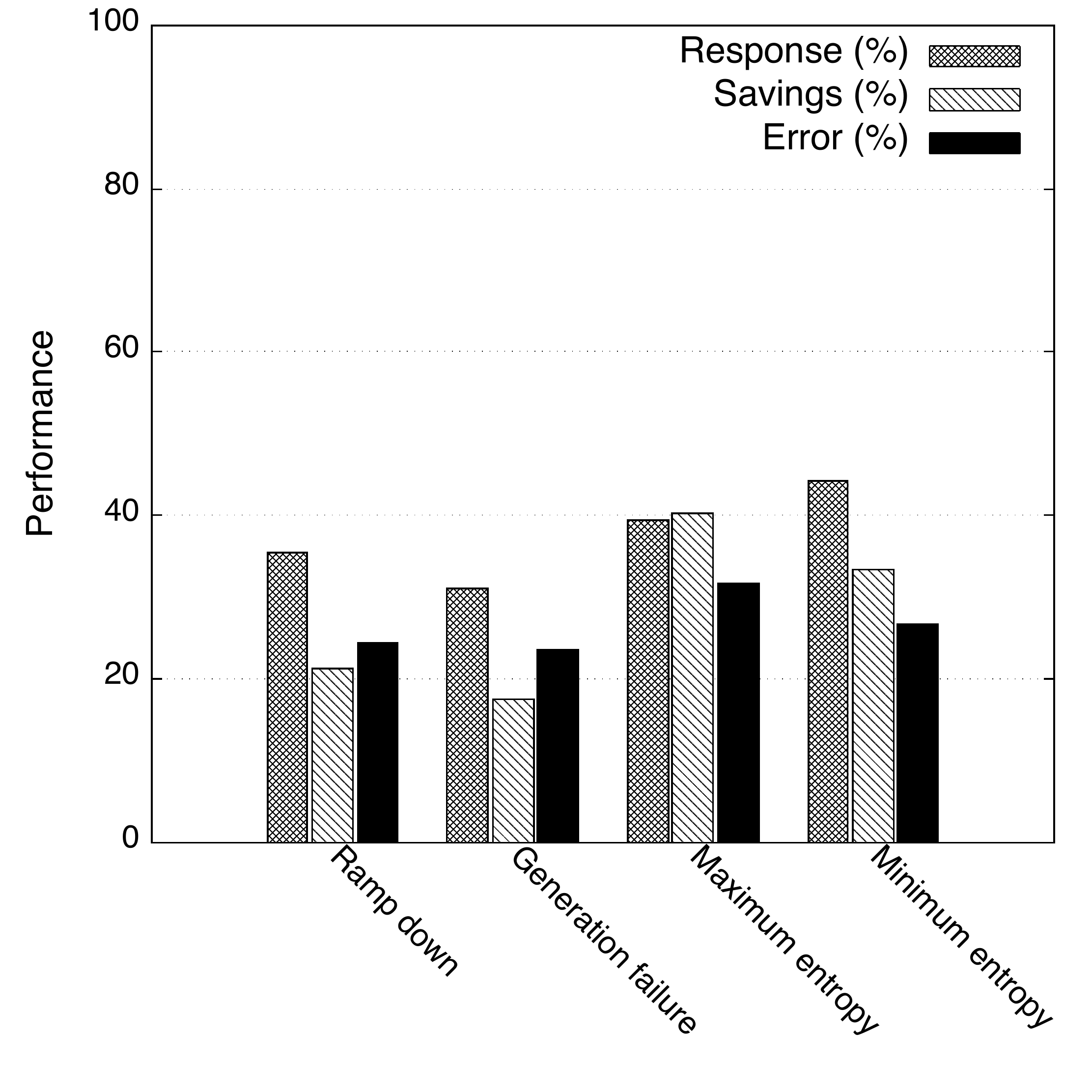}}
\caption{Response, savings and volatility error in each studied aspect of the self-regulatory framework.}\label{fig:performance}
\end{figure}

Figure~\ref{fig:performance} illustrates the performance of the self-regulatory framework in the light of four studied aspects: (i) generation schemes, (ii) selection functions, (iii) datasets and (iv) regulatory scenarios. Performance is evaluated by averaging the response, savings and volatility error in each experiment that involves the choice of a generation scheme, selection function, dataset and regulatory scenario. Figure~\ref{fig:response-detail},~\ref{fig:savings-detail} and~\ref{fig:error-detail} in~\ref{app:response-savings-detail} illustrate the individual results in each experiment performed. 

Figure~\ref{fig:performance}a shows that \shuffle and \shiftParam{20} have the highest response of 60.06\% and 38.02\% respectively. These schemes also have the highest savings of 47.63\% and 32.67\%. In contrast, the lowest performance is observed in \swapParam{15} with 18.01\% response and 19.3\% savings. \shuffle and \swapParam{30} cause the highest volatility error of 53.72\% and 28.15\%, while \shiftParam{10} the lowest one of 12.32\%. These results confirm that a higher informational diversity results in higher performance. 

Figure~\ref{fig:performance}b shows that the \minRmseA selection function has the highest response of 38.12\%, though \minRmseB also has a significant response of 38.09\%. In contrast, \minCost achieves the highest savings of 56.67\% and \minRmseA the lowest savings of 7.99\%. The highest volatility error is 29.79\% by \minRmseA and the lowest one is 18.2\% by \minCost. 

Figure~\ref{fig:performance}c illustrates that the highest response of 48.06\% is observed in the \PNWEvening dataset and the lowest response of 28.02\% at the \PNWMorning dataset. Similarly, the highest savings of 52.31\% are observed in \PNWEvening and the lowest ones of 14.75\% in \PNWMorning. The influence of a dataset to the performance concerns the complexity of the \iTFS signal in comparison to \upperBoundB. For example, when both signals are monotonously increasing, they may require a lower degree of adjustment compared to the case in which the one is monotonously increasing and the second monotonously decreasing\footnote{For example, Figure~\ref{fig:signals-minimum-entropy}d shows an \iTFS and an \upperBoundB with a very similar complexity.}. The complexity of the signals in each dataset is shown in the figures of Section~\ref{subsec:signals}. The highest volatility error of 40.37\% is observed in \SIM and the lowest one of 18.82\% in \PNWEvening. 

Figure~\ref{fig:performance}d illustrates the response and savings under the four different regulatory scenarios. The minimum entropy scenario has the highest response of 44.14\%, whereas generation failure has the lowest response of 31.05\%. The maximum entropy has the highest savings of 40.22\%, in contrast to the generation failure scenario with the lowest savings of 17.51\%. The highest volatility error is 31.61\% for maximum entropy and the lowest one is 23.55\% for generation failure. 

%\footnote{An upper bound signal is constructed by a \TIS signal that corresponds to a regulatory scenario. Therefore, a regulatory scenario influences performance on the same conceptual basis as explained earlier for the datasets.}

\begin{figure*}[!htb]
\centering
\subfigure[\shuffle, \minRmseB]{\includegraphics[width=0.3\textwidth]{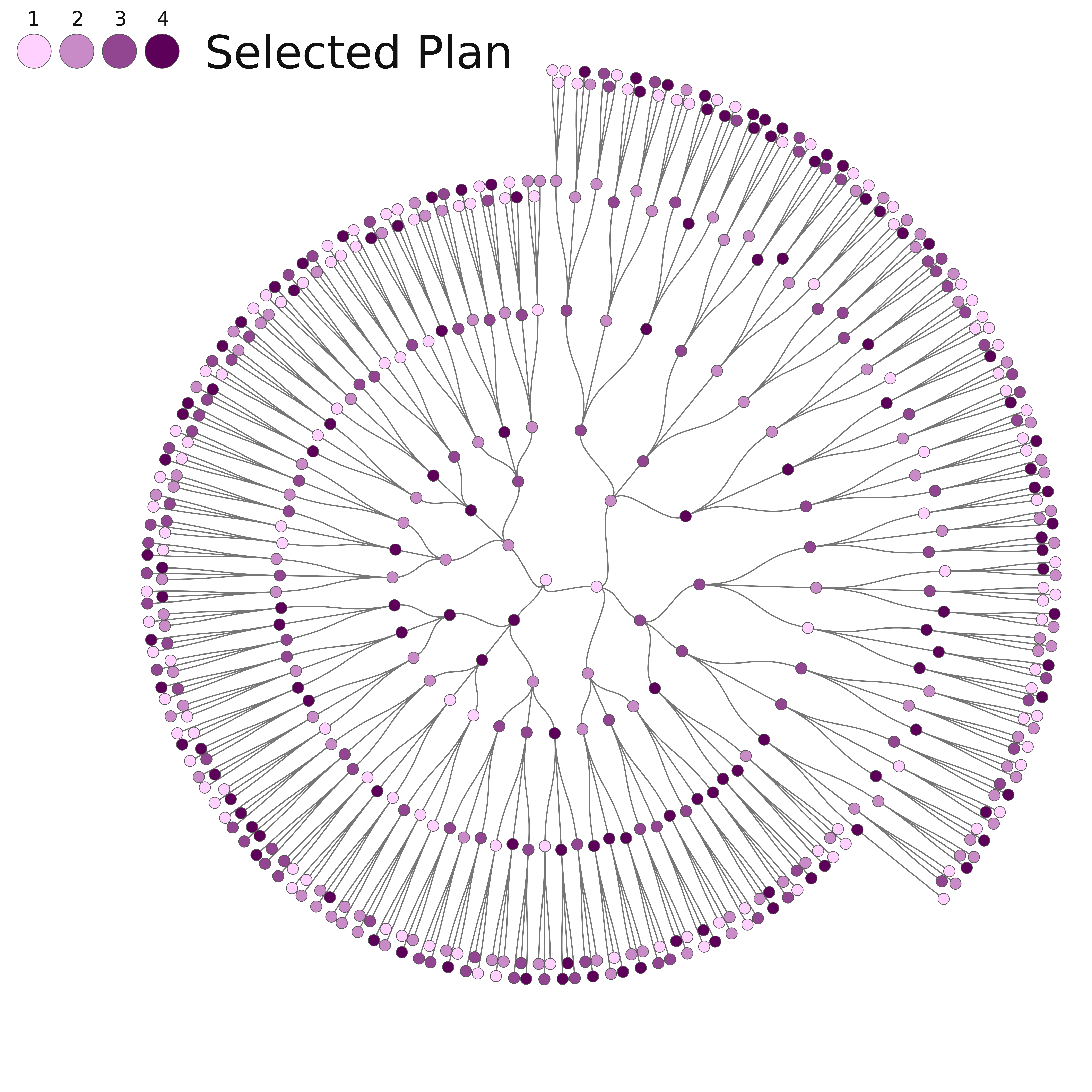}}
\subfigure[\shuffle, \minCost]{\includegraphics[width=0.3\textwidth]{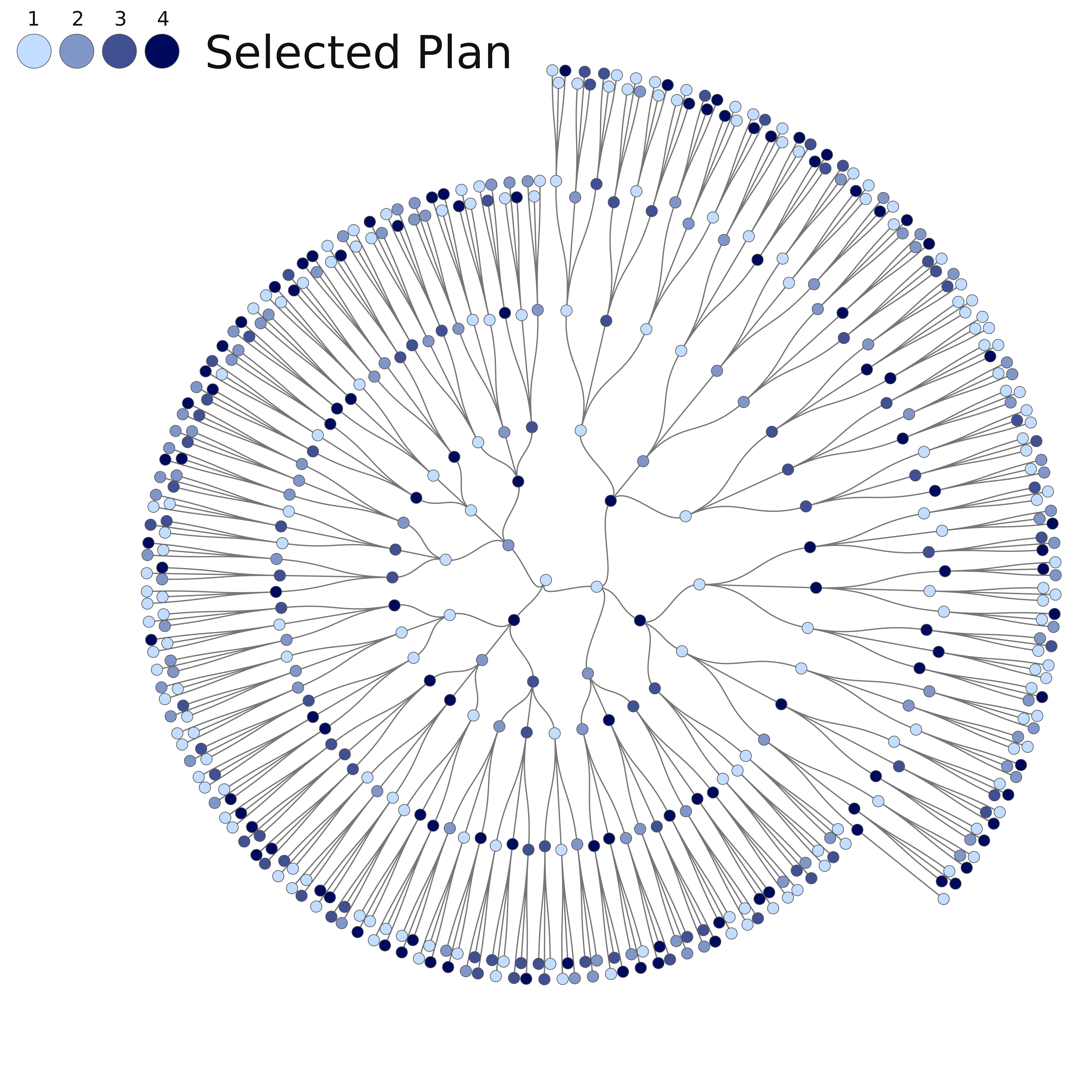}}
\subfigure[\shuffle, \minRmseB vs. \minCost]{\includegraphics[width=0.3\textwidth]{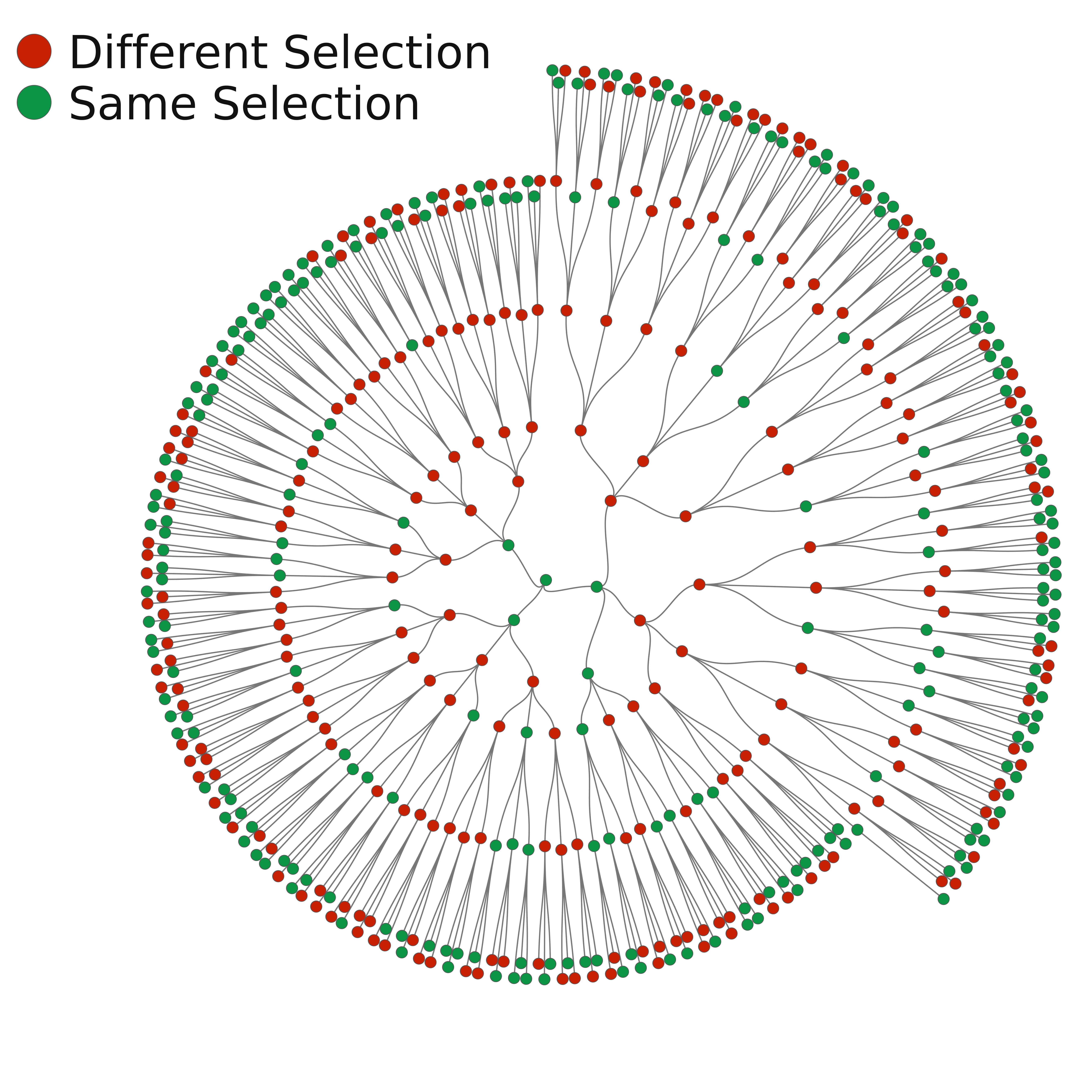}}
\subfigure[\swapParam{15}, \minRmseB]{\includegraphics[width=0.3\textwidth]{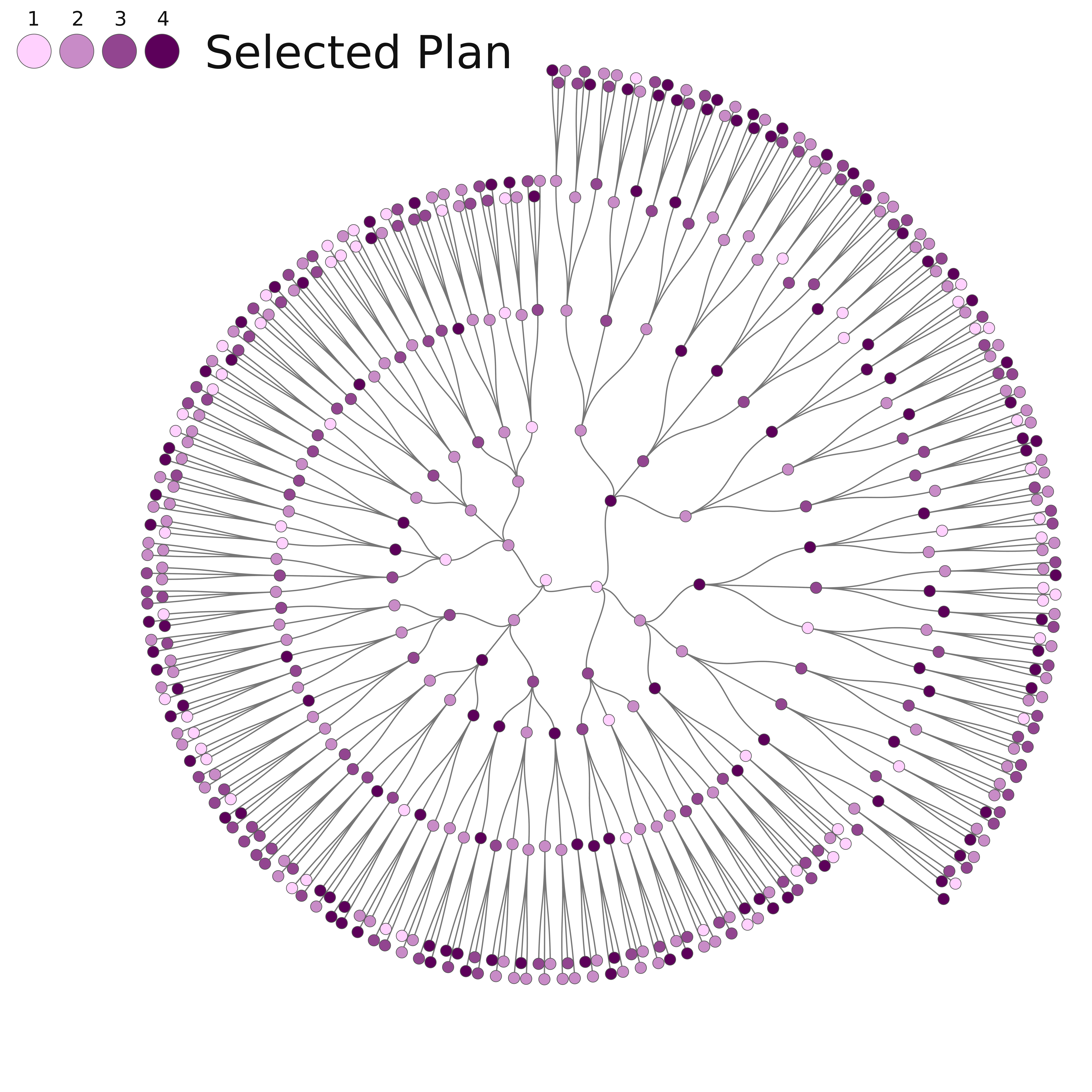}}
\subfigure[\swapParam{30}, \minRmseB]{\includegraphics[width=0.3\textwidth]{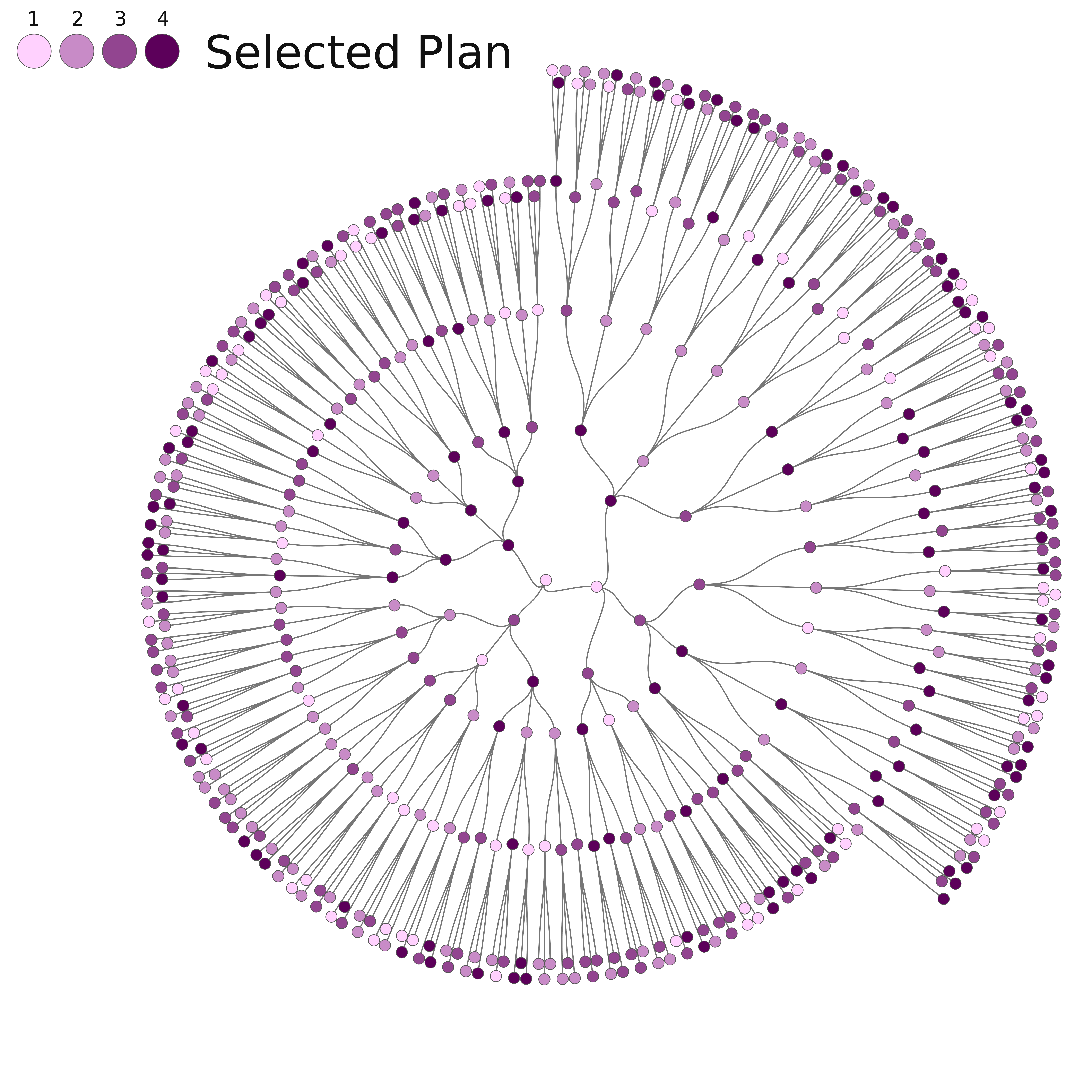}}
\subfigure[\swapParam{15} vs. \swapParam{30}, \minRmseB]{\includegraphics[width=0.3\textwidth]{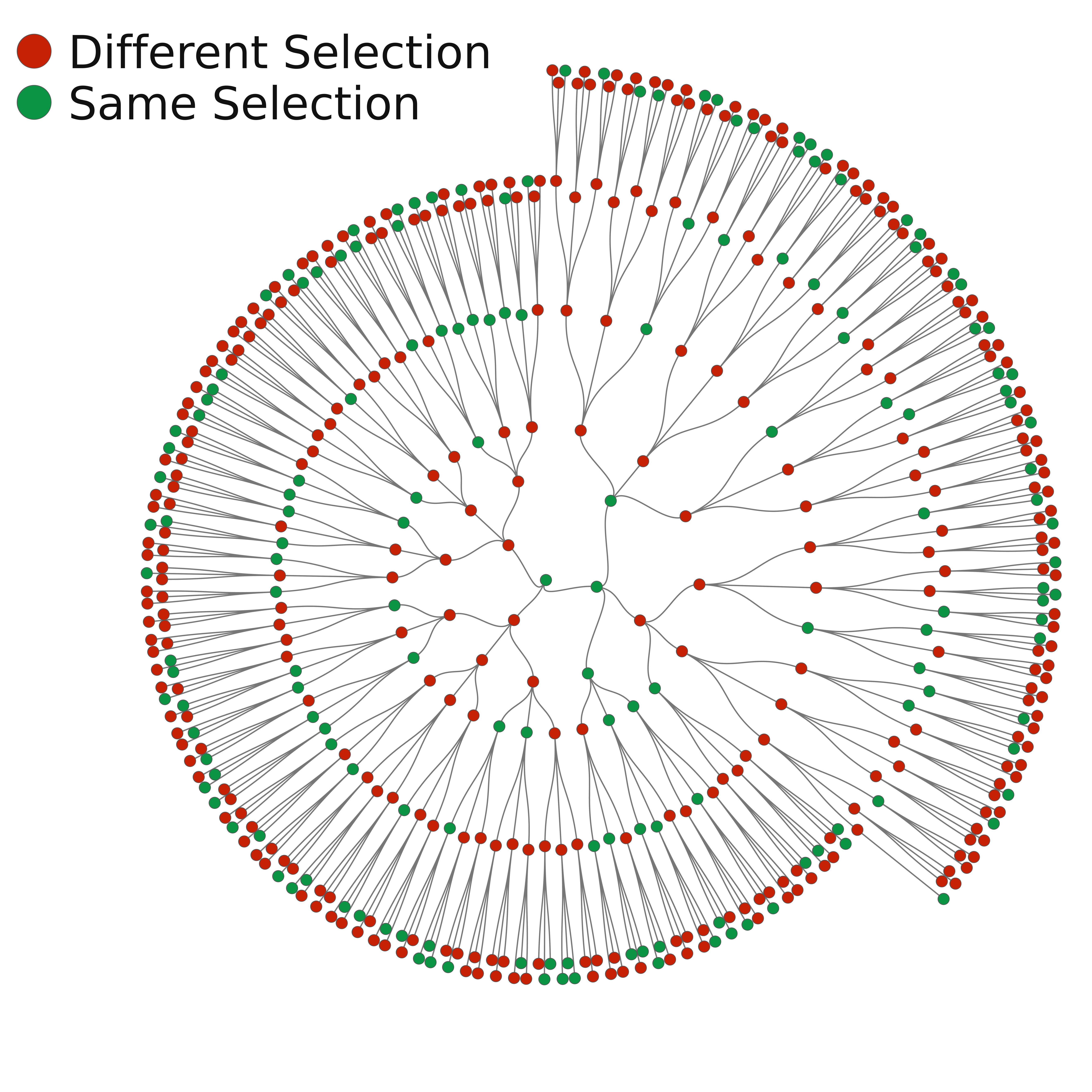}}
\subfigure[\swapParam{15}, \minCost]{\includegraphics[width=0.3\textwidth]{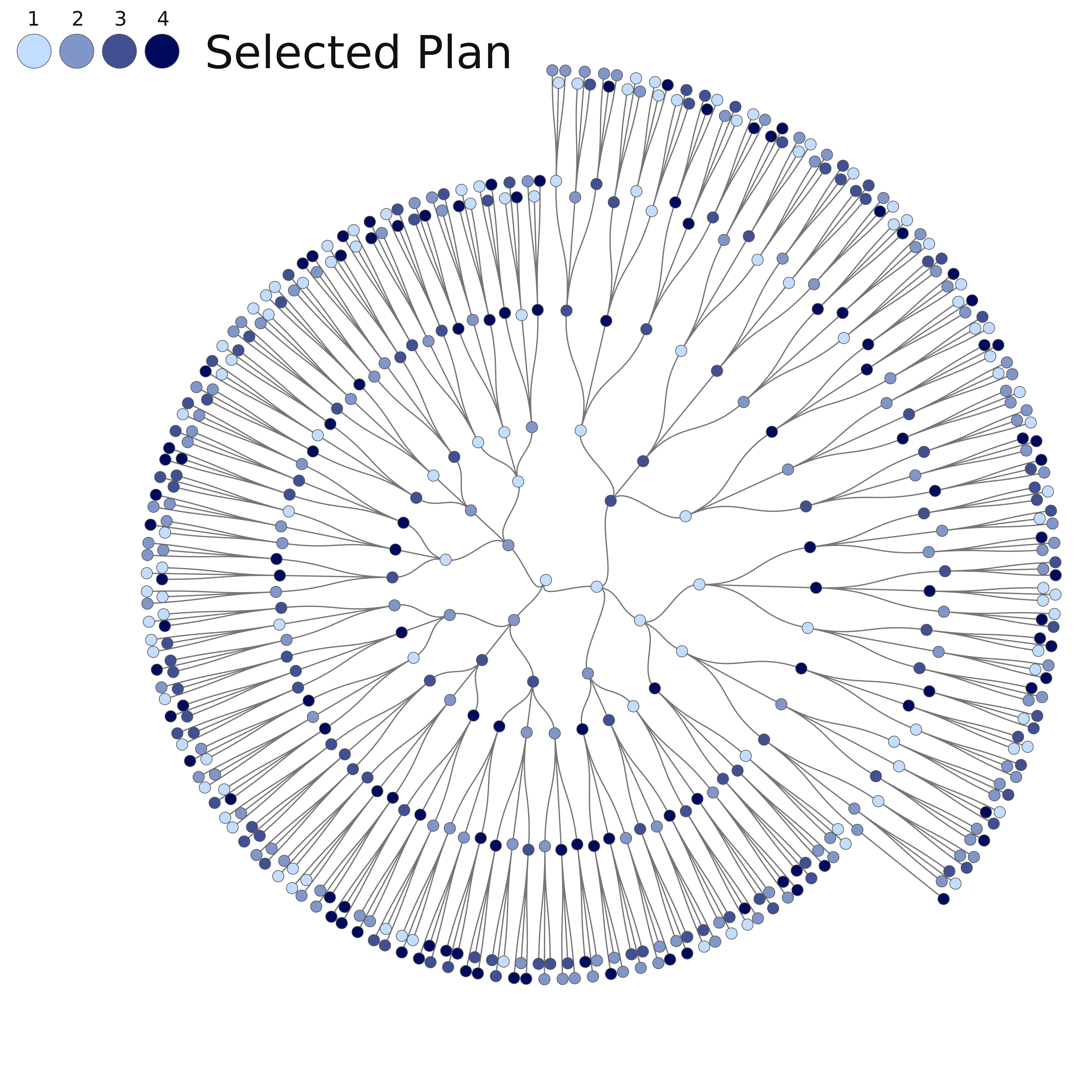}}
\subfigure[\swapParam{30}, \minCost]{\includegraphics[width=0.3\textwidth]{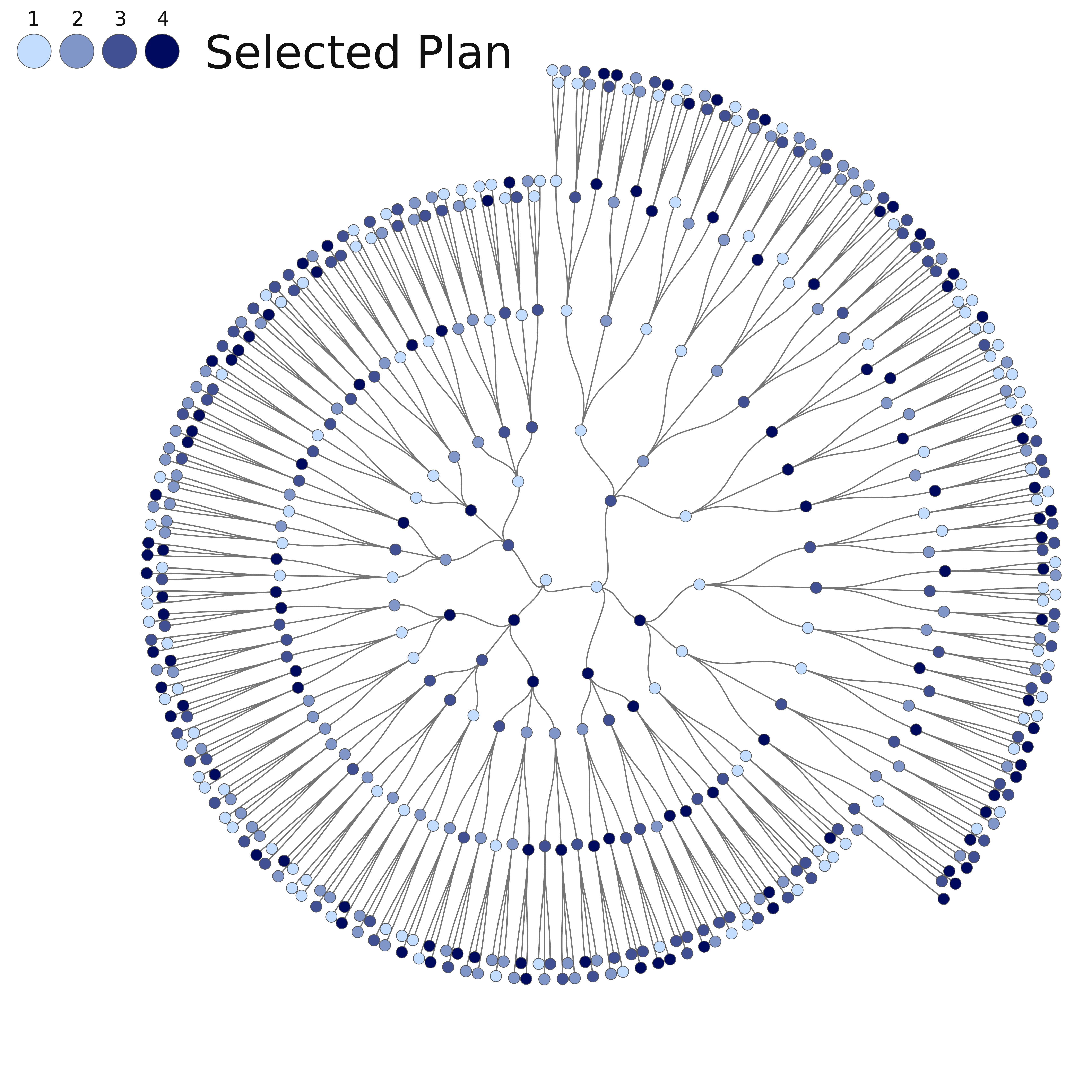}}
\subfigure[\swapParam{15} vs. \swapParam{30}, \minCost]{\includegraphics[width=0.3\textwidth]{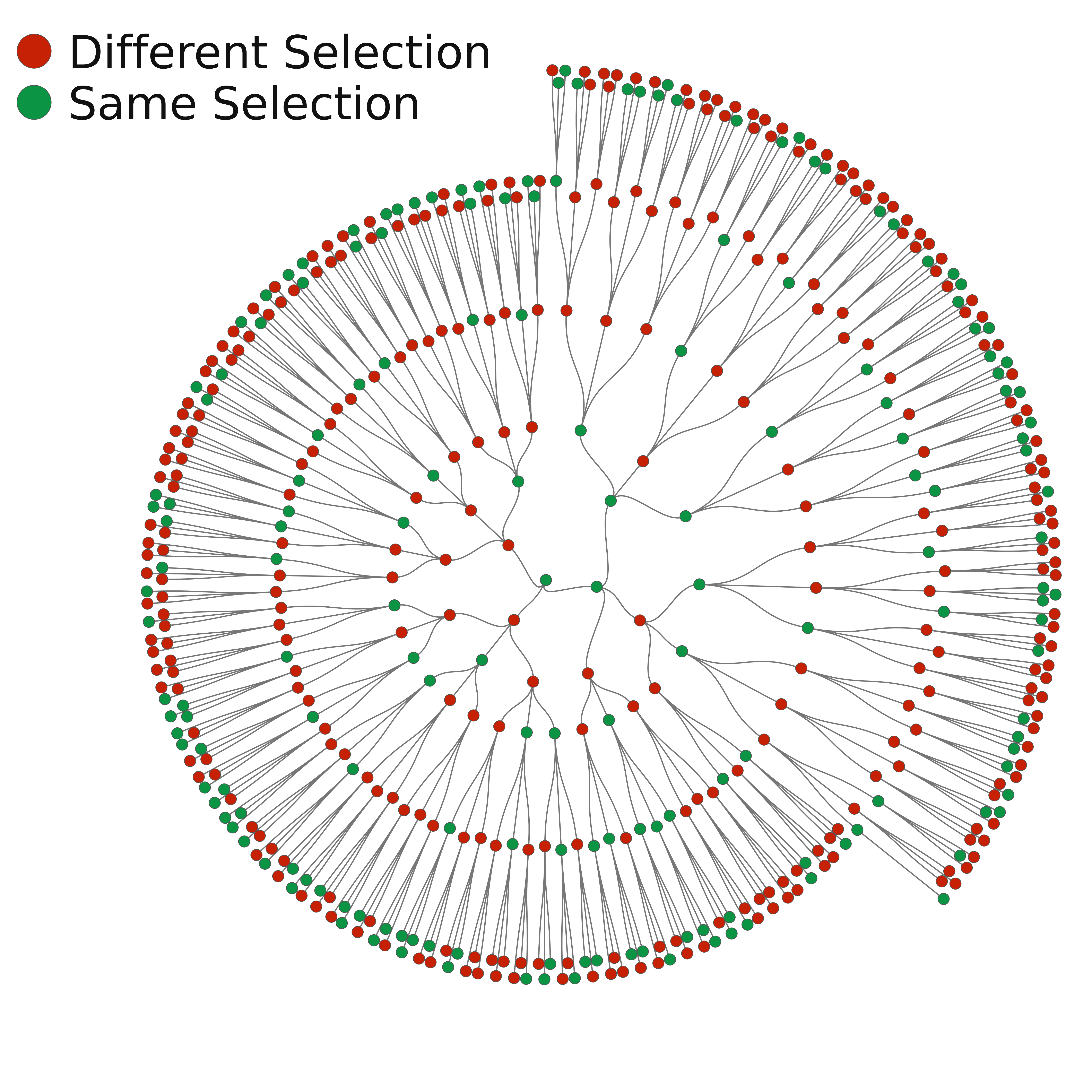}}
\caption{Agent selections under the generation failure scenario for the \PNWMorning dataset.}\label{fig:visualizations}
\end{figure*}

Figure~\ref{fig:visualizations} visualizes the agent selections under the generation failure scenario for the \PNWMorning dataset. Figure~\ref{fig:visualizations}a-c provide qualitative comparisons about the selection made with \minRmseB and \minCost using the generation scheme of \shuffle. Agents adopting the \shuffle generation scheme under the \minRmseB and \minCost selections functions make 50.91\% different selections. Figure~\ref{fig:visualizations}d-f show selections made with \minRmseB, though using the generation schemes \swapParam{15} and \swapParam{30}. Agents adopting the \minRmseB selection function under the \swapParam{15} and \swap{30} generation schemes make 67\% different selections. Finally, Figure~\ref{fig:visualizations}g-i show selections made with \minCost, though using the generation schemes \swapParam{15} and \swapParam{30}. Agents adopting the \minCost selection function under the \swapParam{15} and \swap{30} generation schemes make 64.59\% different selections. 

Given the performance results of response, savings and volatility error, a plausible question that arises is if these metrics are correlated in practice. Intuitively, a high response can be incentivized via reducing the demand cost. Similarly, high savings require a significant response. Finally a high response can originate from adjustment in demand volatility. How correlated are these metrics in the empirical findings illustrated? Which system parameter can be used to make trade-offs between response and savings, while minimizing volatility error? 

The total number of experiments performed are a population on which the correlation between response-savings, error-response and error-savings can be measured. In this case, the correlation coefficient is 0.3, 0.44 and -0.08. However, not all experiments vary the same parameter. Four different aspects as studied as shown in Figure~\ref{fig:performance}. For this reason, it is meaningful to restrict the correlation measurements for each studied aspect. 

\begin{figure}[!htb]
\centering
\includegraphics[width=0.49\columnwidth]{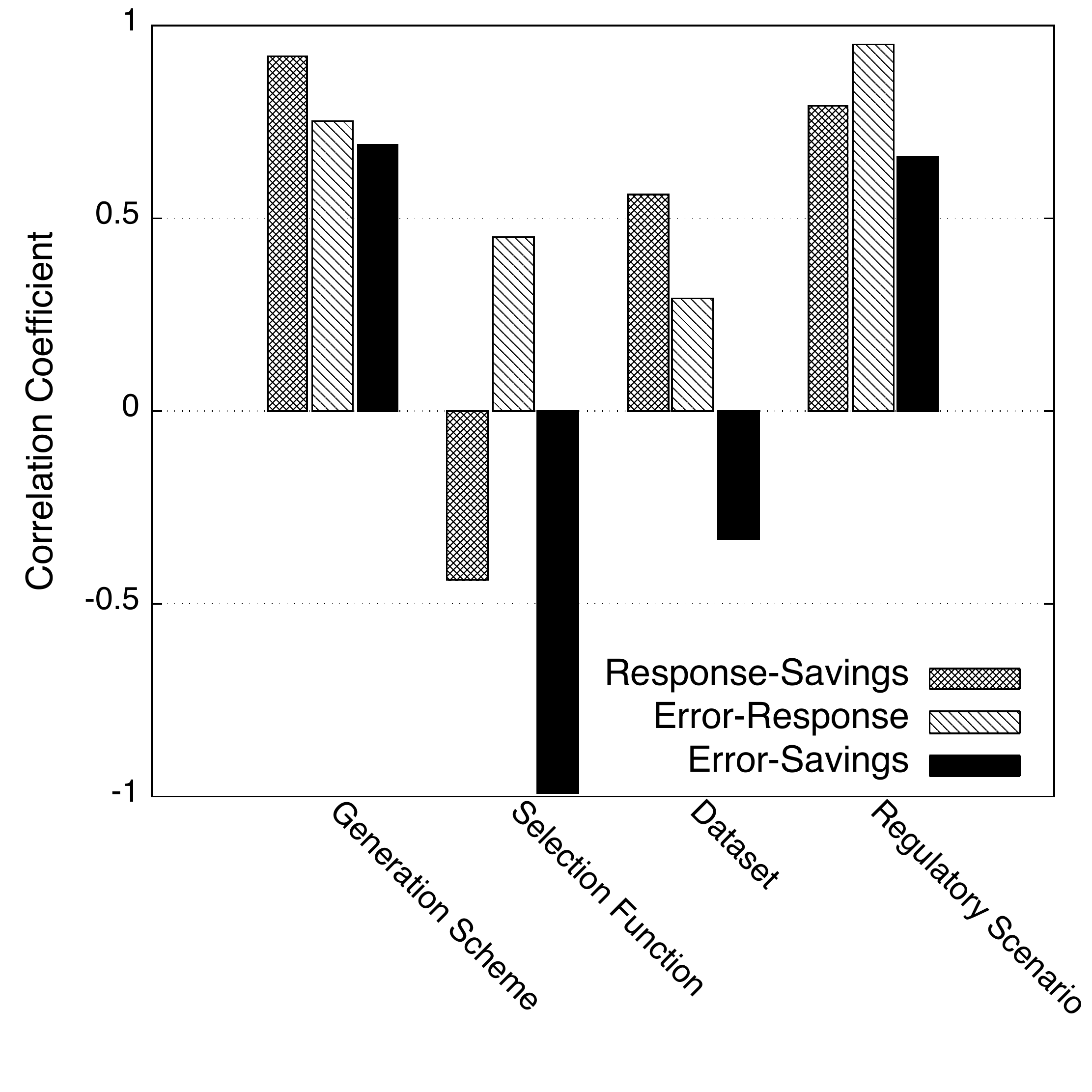}
\caption{Correlation coefficient between response, savings and volatility error for each studied aspect of the self-regulatory framework.}\label{fig:correlation-coefficient}
\end{figure}

Figure~\ref{fig:correlation-coefficient} illustrates the correlation coefficient for each studied aspect of the self-regulatory framework. The three metrics appear highly correlated when measurements are performed for the experiments that vary the generation scheme and regulatory scenario. For example, the correlation range [0.69,0.93] indicates that response, savings and volatility error are likely to increase or decrease simultaneously by only changing the generation scheme. This is a desired property of the framework as the choice of the generation scheme and regulatory scenario should not govern the dynamics between the three metrics. Under different datasets, response-savings have a correlation of 0.56, error-response have a correlation of 0.29 and error-savings have a negative correlation of -0.33. Finally, under different selection functions, response-savings have a correlation of -0.44, error-response have a correlation of 0.45 and error-savings have a significant negative correlation of -0.995. 

The correlation results under different selection functions have two striking implications. The first implication is that the design of selection functions is empirically validated: \minCost maximizes the savings while \minRmseA and \minRmseB maximize the response. Trade-offs between response and savings can be performed by choosing the respective selection function. In practice, these trade-offs can be incentivized by system operators or policy makers~\cite{Strbac2008a,Giannoccaro2009,Raghunathan2003}: each consumer of the supplied resources may change its selection function in exchange of a monetary payoff. The second implication is that when the volatility error should remain strictly minimal, it is more likely for the self-regulatory system to increase the savings rather than the response.

\subsection{Signals with maximum response}\label{subsec:signals}

This section illustrates the resulting \eTFS signals with the maximum response in each dataset and regulatory scenario. The \eTFS signal is compared to \iTFS and \upperBoundB.

\begin{figure}[!htb]
\centering
\subfigure[\SIM, \shiftParam{20}, \minCost]{\includegraphics[width=0.49\columnwidth]{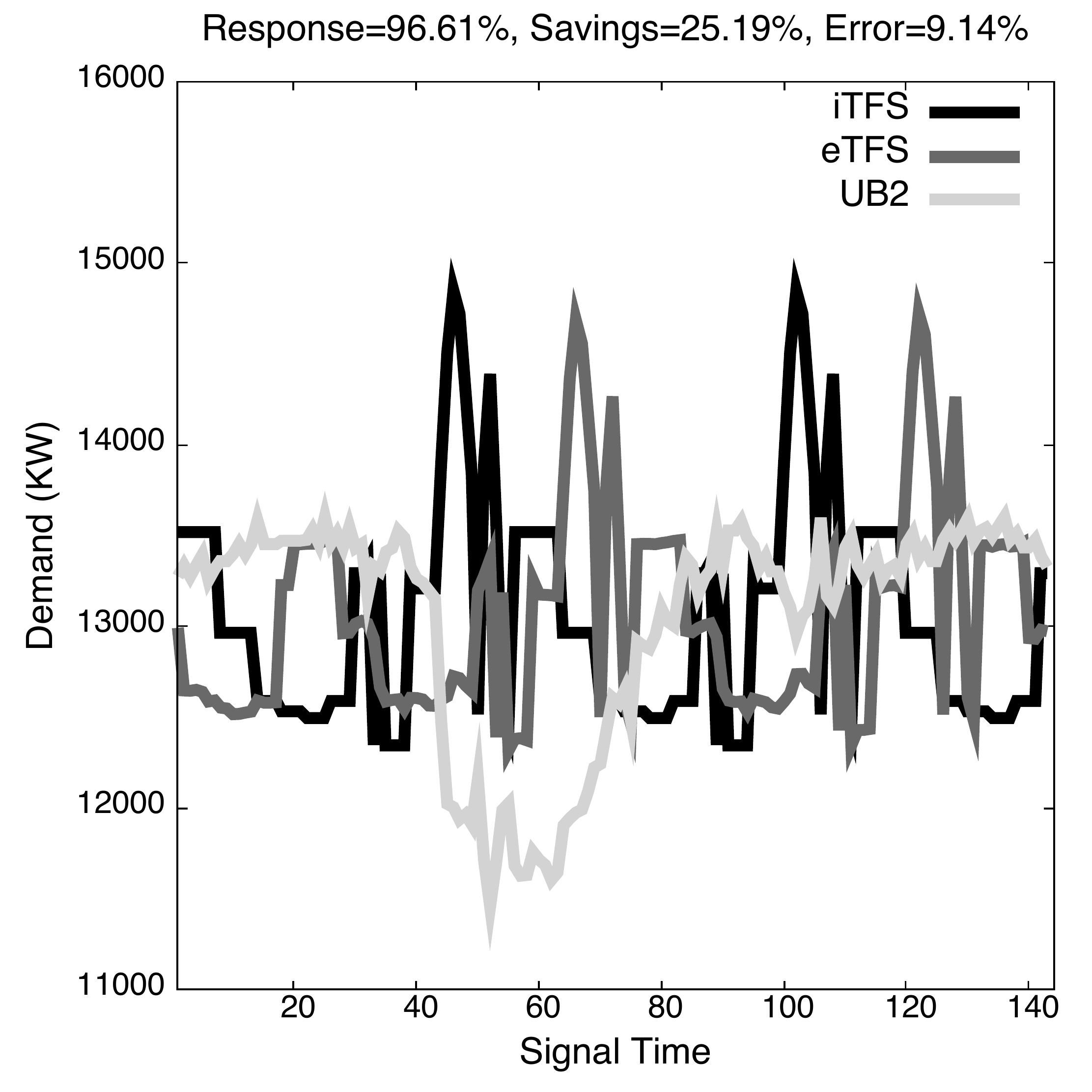}}
\subfigure[\EDF, \shuffle, \minRmseB]{\includegraphics[width=0.49\columnwidth]{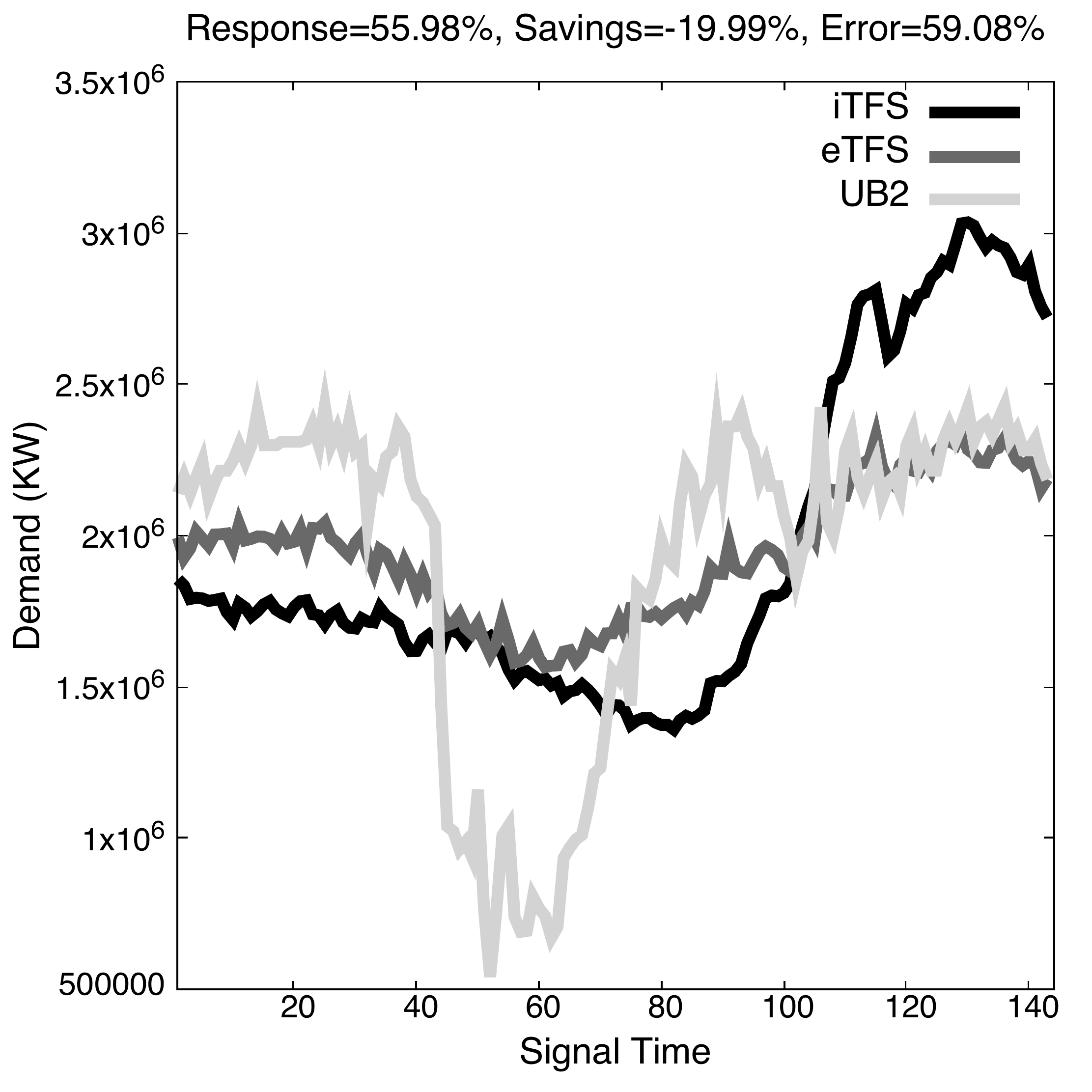}}
\subfigure[\PNWMorning, \shuffle, \minRmseB]{\includegraphics[width=0.49\columnwidth]{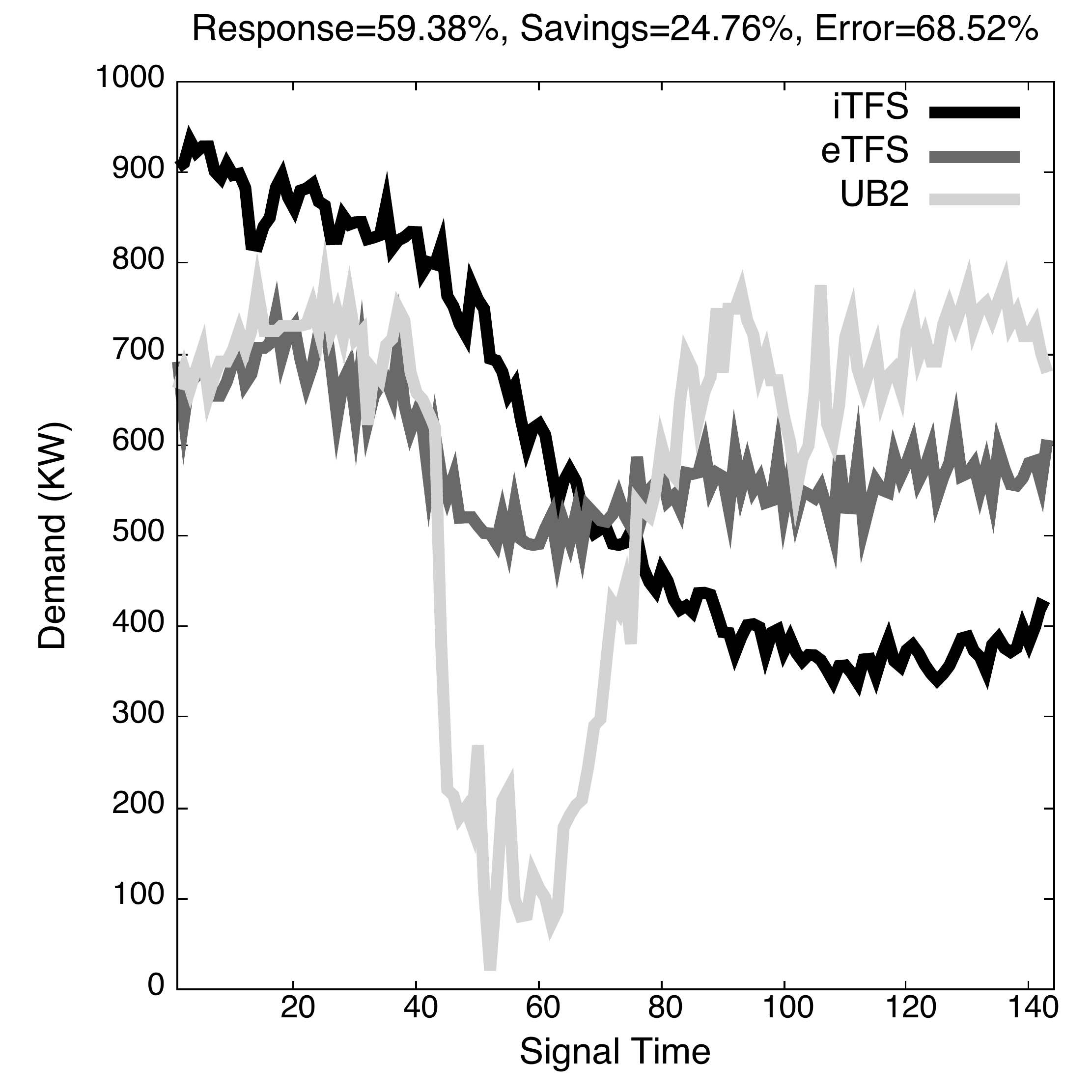}}
\subfigure[\PNWEvening, \shiftParam{20}, \minCost]{\includegraphics[width=0.49\columnwidth]{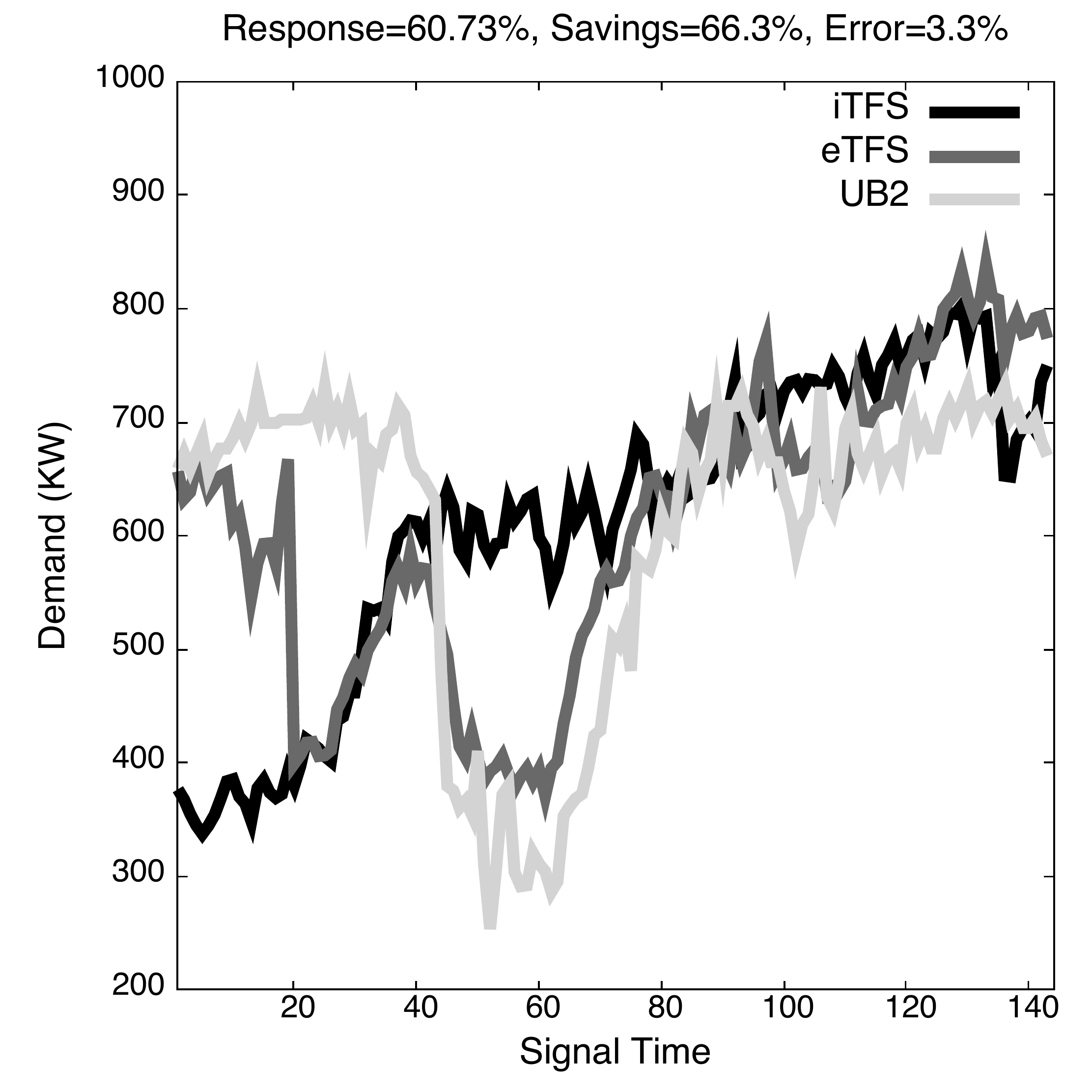}}
\caption{\eTFS signals with maximum response under the ramp down scenario. For comparison, the \iTFS and \upperBoundB signals are shown.}\label{fig:signals-ramp-down}
\end{figure}

Figure~\ref{fig:signals-ramp-down} illustrates the ramp down scenario. The maximum response is 96.61\%, 55.98\%, 59.38\% and 60.73\% for \SIM, \EDF, \PNWMorning and \PNWEvening respectively. It is shown that during the period $[40,80]$ in which the cost generation is maximized, the majority of the values in the \eTFS signals is decreased compared to the respective values of the \iTFS signal. Figure~\ref{fig:signals-ramp-down}d shows a significant match of the \eTFS signal with \upperBoundB. In the case of Figure~\ref{fig:signals-ramp-down}a and~\ref{fig:signals-ramp-down}d that maximum response is achieved with \shiftParam{20}, a significantly lower volatility error of 9.14\% and 3.3\% is observed. 

\begin{figure}[!htb]
\centering
\subfigure[\SIM, \shuffle, \minRmseB]{\includegraphics[width=0.49\columnwidth]{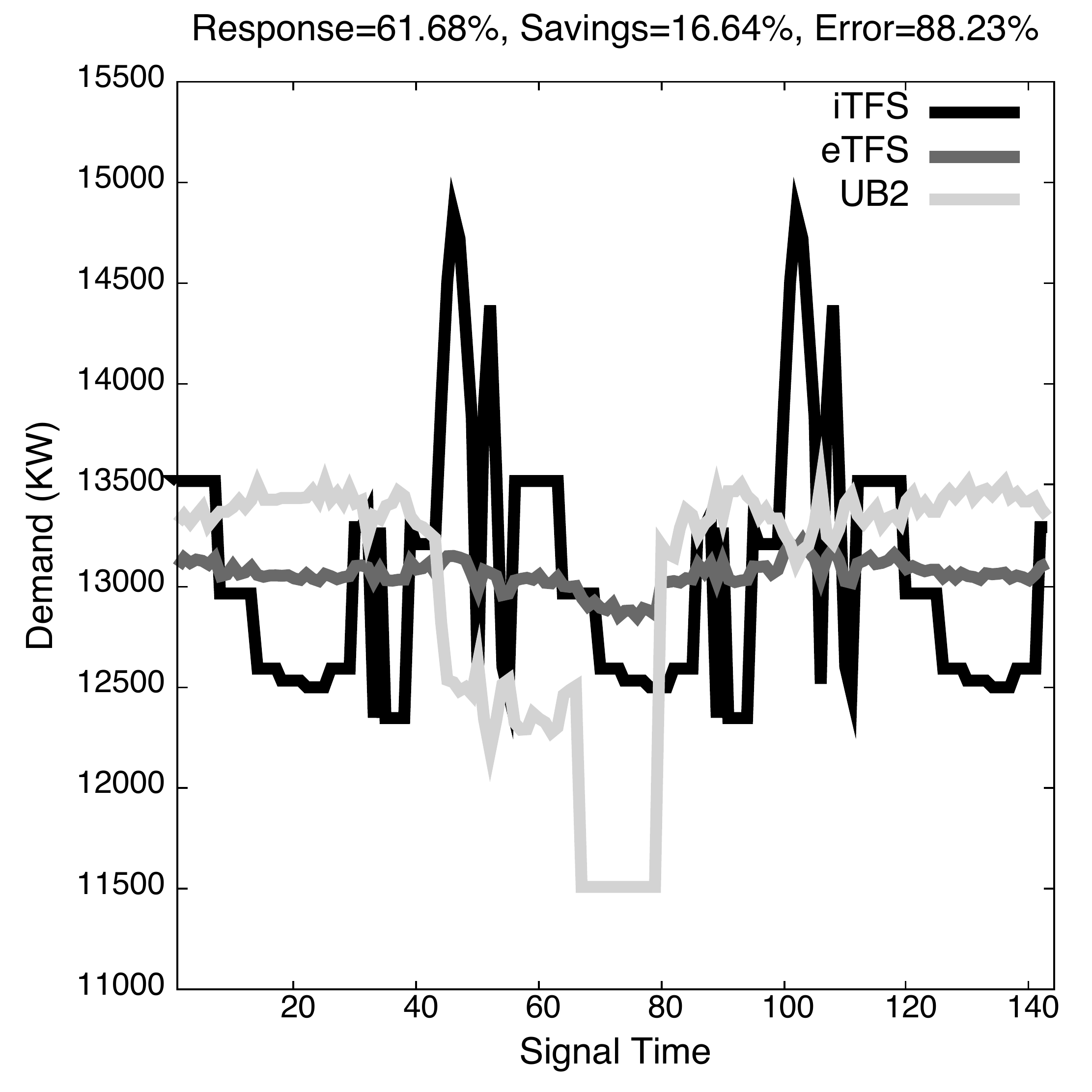}}
\subfigure[\EDF, \shuffle, \minRmseB]{\includegraphics[width=0.49\columnwidth]{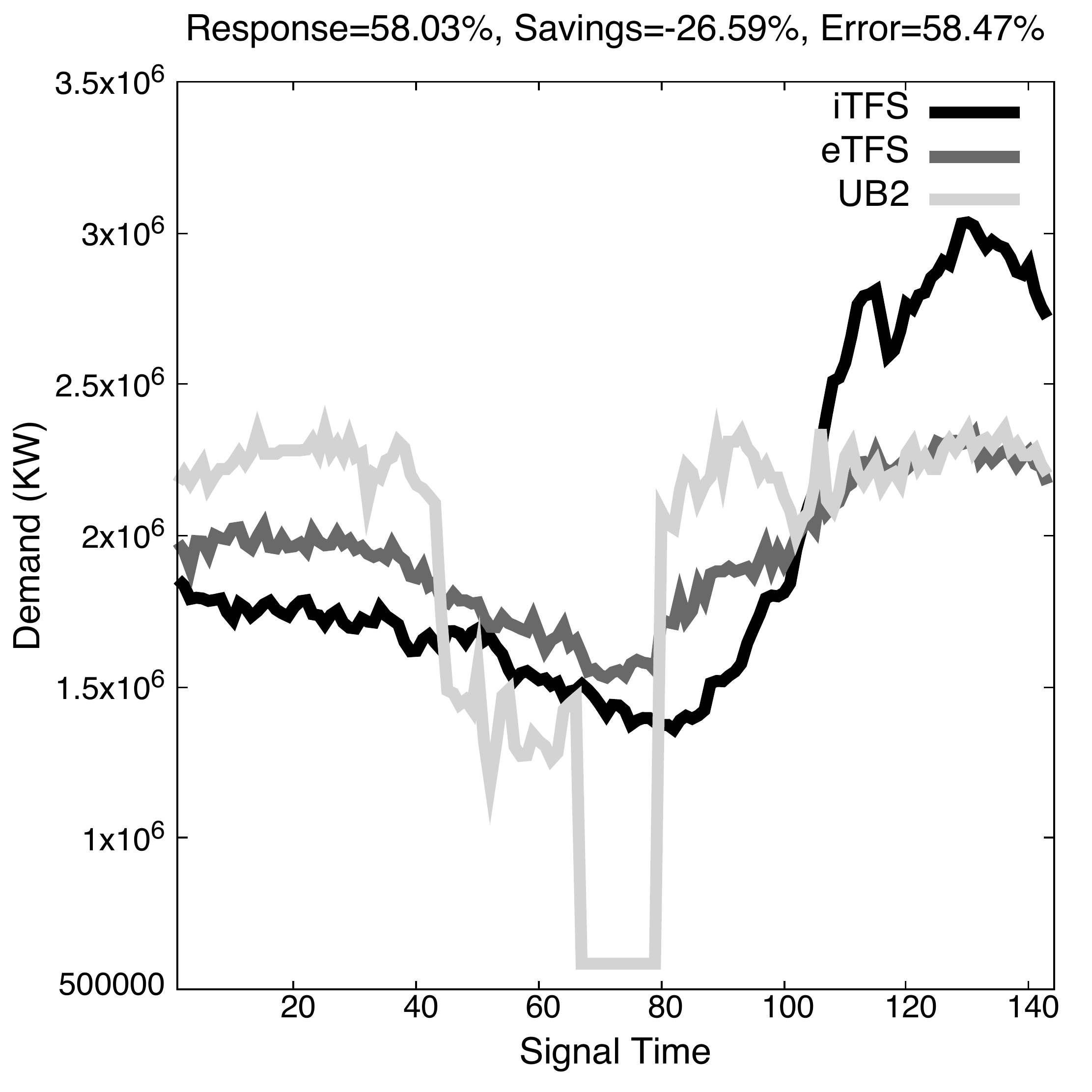}}
\subfigure[\PNWMorning, \shuffle, \minRmseB]{\includegraphics[width=0.49\columnwidth]{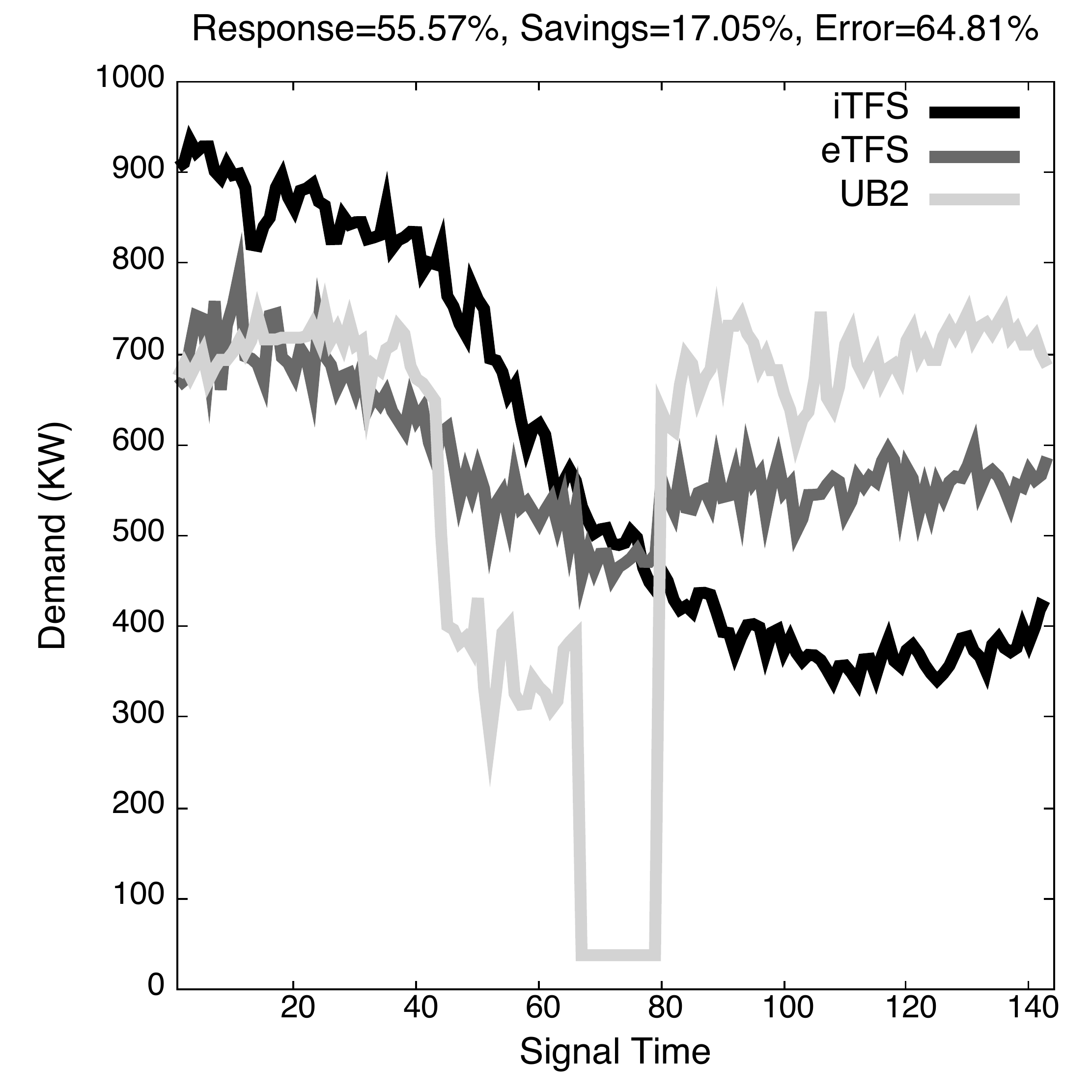}}
\subfigure[\PNWEvening, \shuffle, \minRmseB]{\includegraphics[width=0.49\columnwidth]{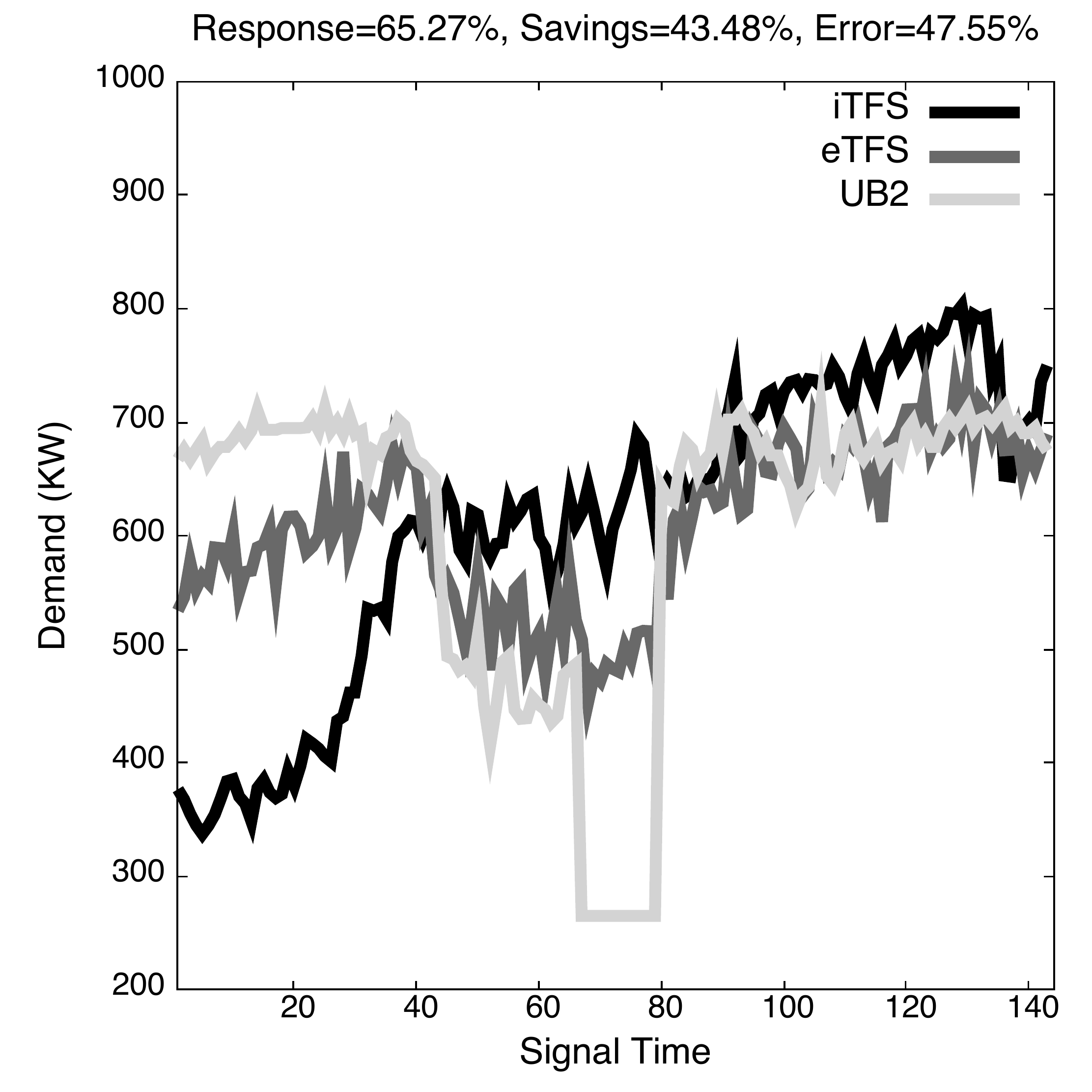}}
\caption{\eTFS signals with maximum response under the generation failure scenario. For comparison, the \iTFS and \upperBoundB signals are shown.}\label{fig:signals-generation-failure}
\end{figure}

Figure~\ref{fig:signals-generation-failure} illustrates the generation failure scenario. The maximum response is 61.68\%, 58.03\%, 55.57\% and 65.27\% for \SIM, \EDF, \PNWMorning and \PNWEvening respectively. Similarly to the ramp down scenario, the majority of the values in the \eTFS signals during the period $[60,80]$ is decreased compared to the respective values of the \iTFS signal. 

\begin{figure}[!htb]
\centering
\subfigure[\SIM, \shuffle, \minRmseB]{\includegraphics[width=0.49\columnwidth]{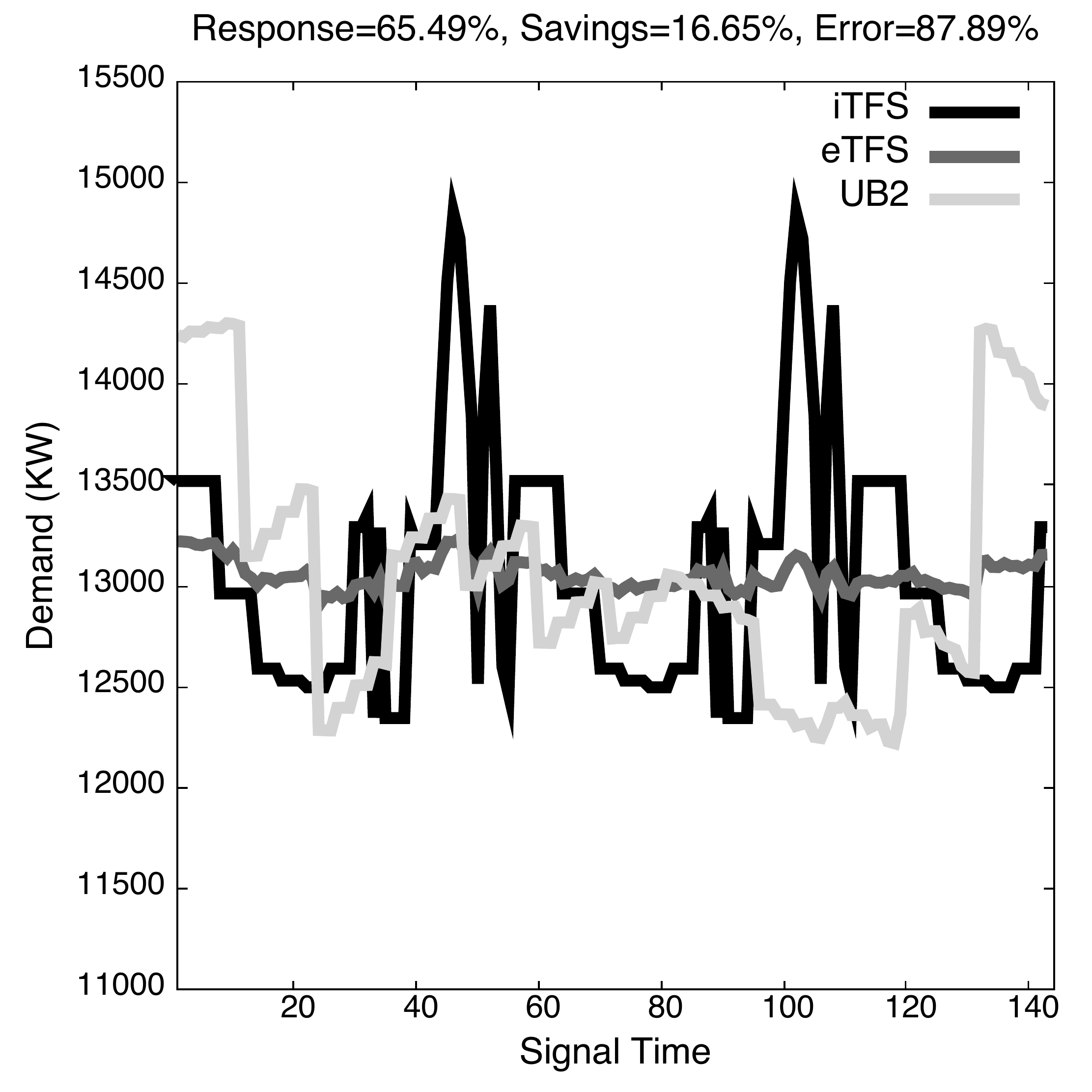}}
\subfigure[\EDF, \shuffle, \minRmseB]{\includegraphics[width=0.49\columnwidth]{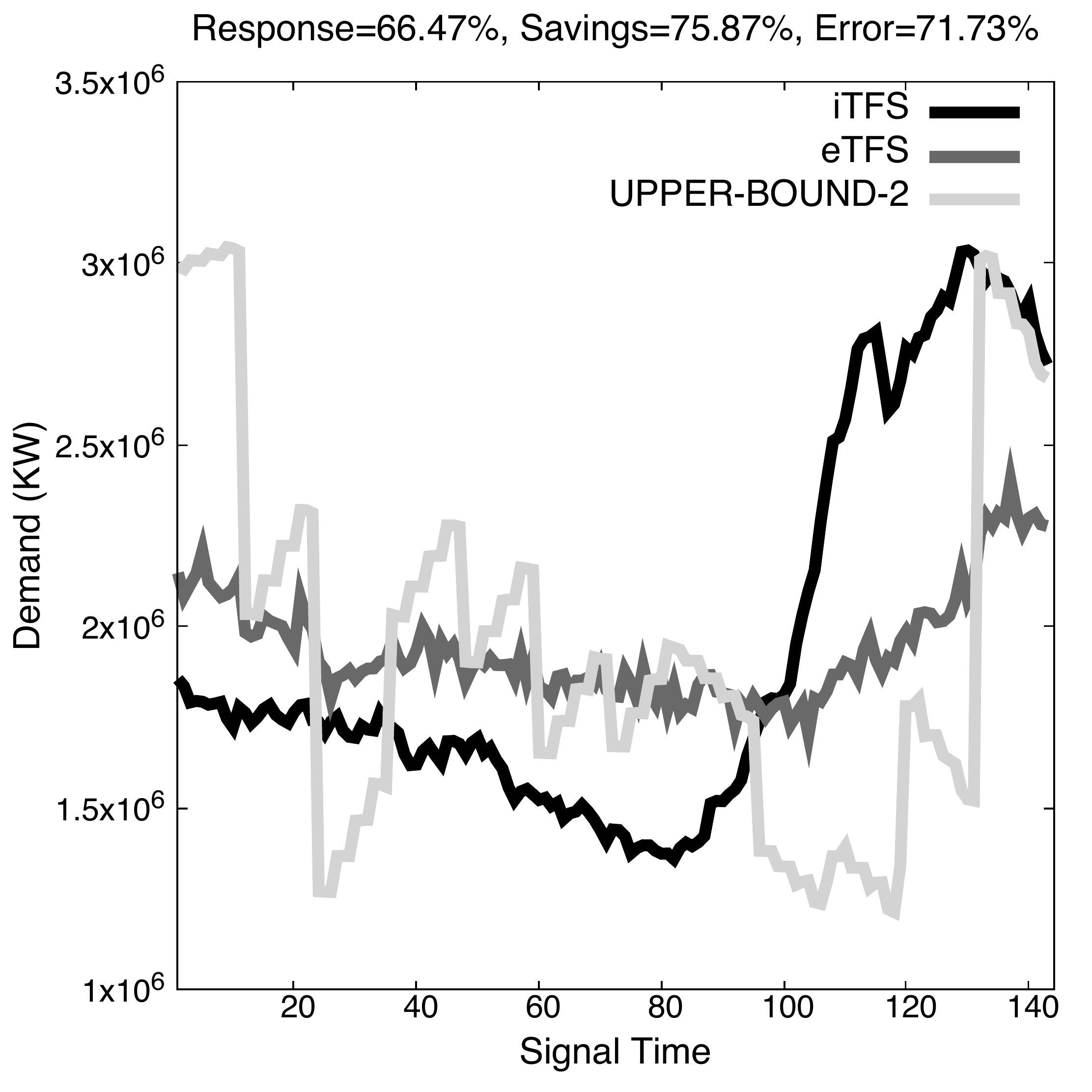}}
\subfigure[\PNWMorning, \shuffle, \minRmseA]{\includegraphics[width=0.49\columnwidth]{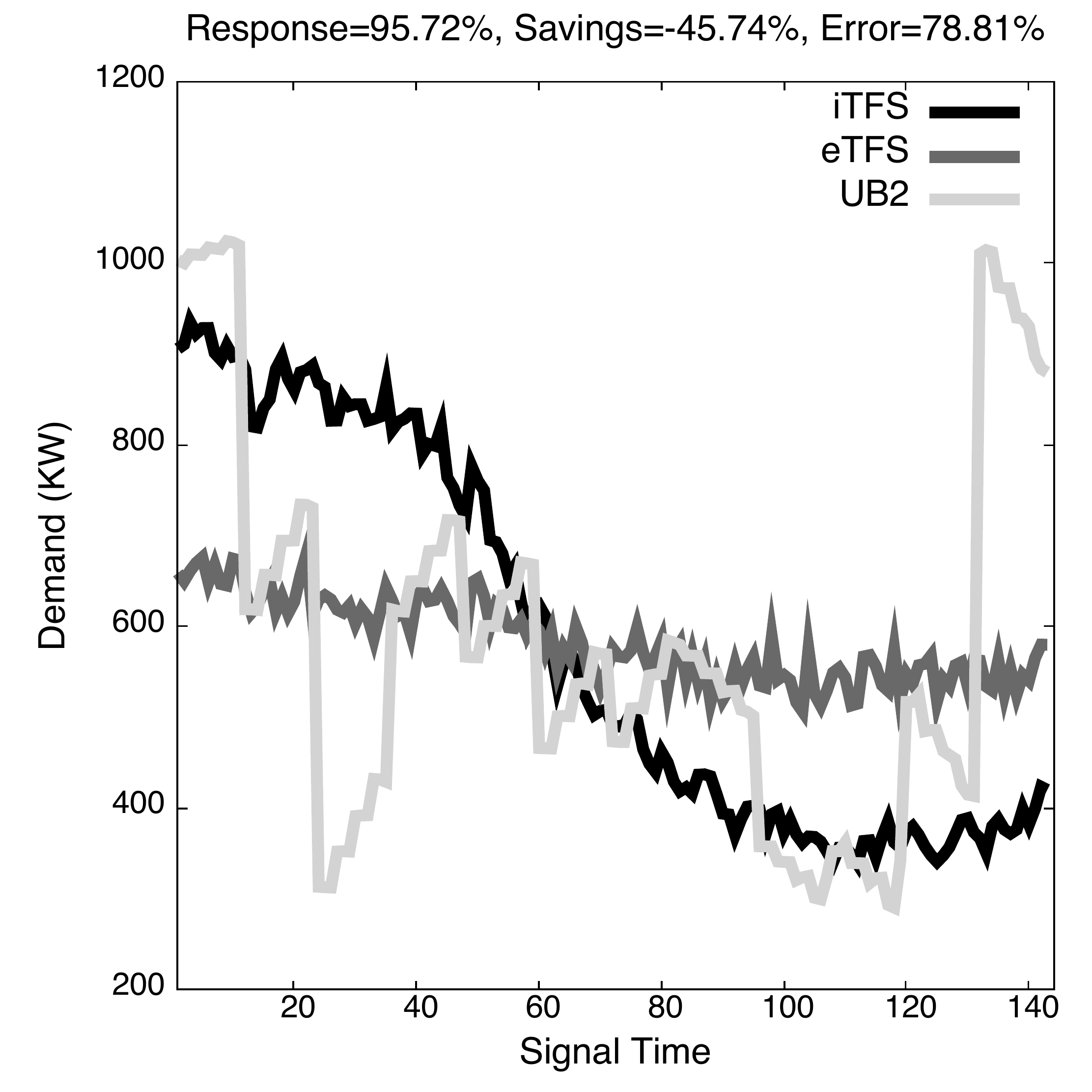}}
\subfigure[\PNWEvening, \shuffle, \minCost]{\includegraphics[width=0.49\columnwidth]{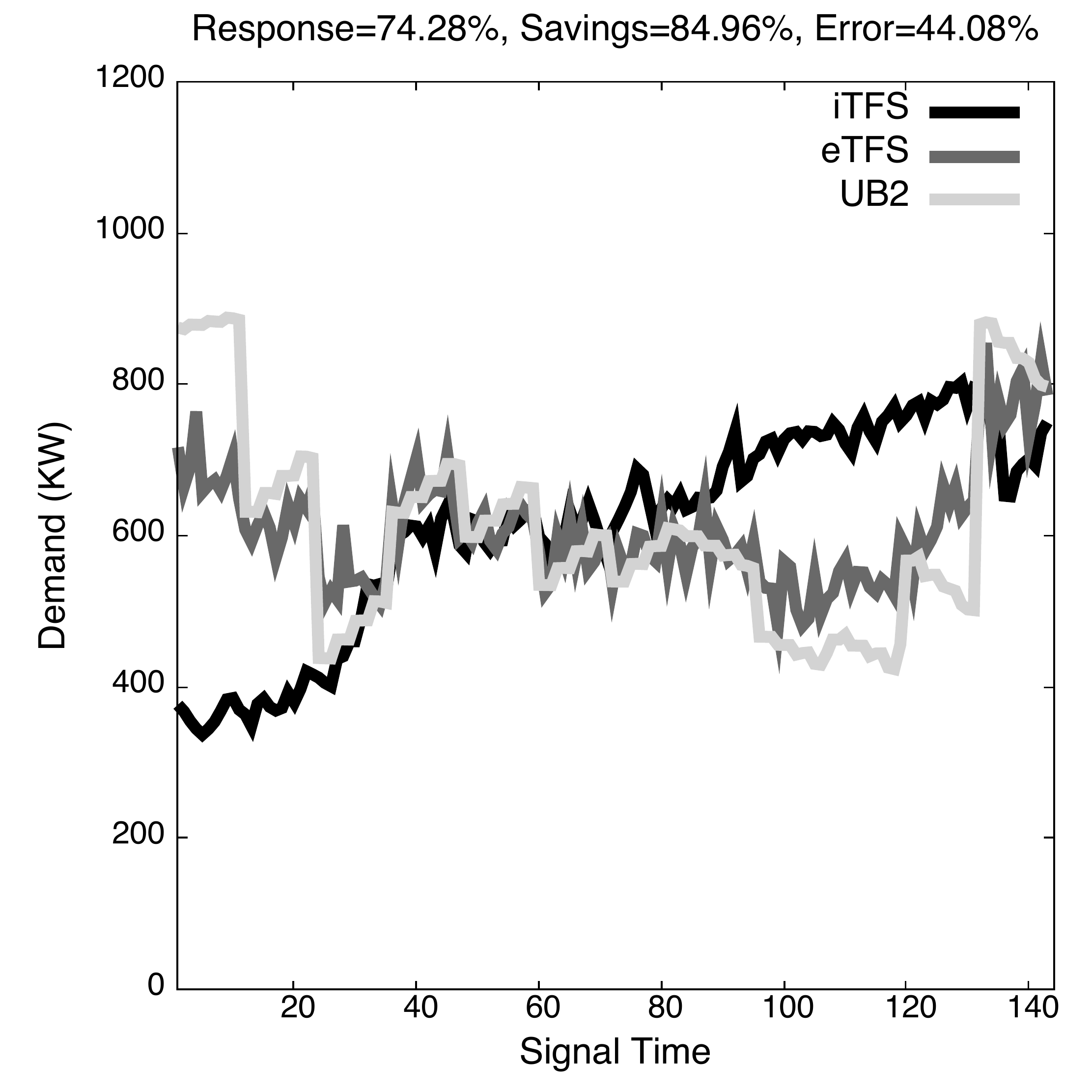}}
\caption{\eTFS signals with maximum response under the maximum entropy scenario. For comparison, the \iTFS and \upperBoundB signals are shown.}\label{fig:signals-maximum-entropy}
\end{figure}

Figure~\ref{fig:signals-maximum-entropy} illustrates the maximum entropy scenario. The maximum response is 65.49\%, 66.47\%, 95.72\% and 74.28\% for \SIM, \EDF, \PNWMorning and \PNWEvening respectively. A response with a higher demand at the beginning and at the end of the \eTFS signal is achieved during which generation cost is low. 

\begin{figure}[!htb]
\centering
\subfigure[\SIM, \shiftParam{20}, \minCost]{\includegraphics[width=0.49\columnwidth]{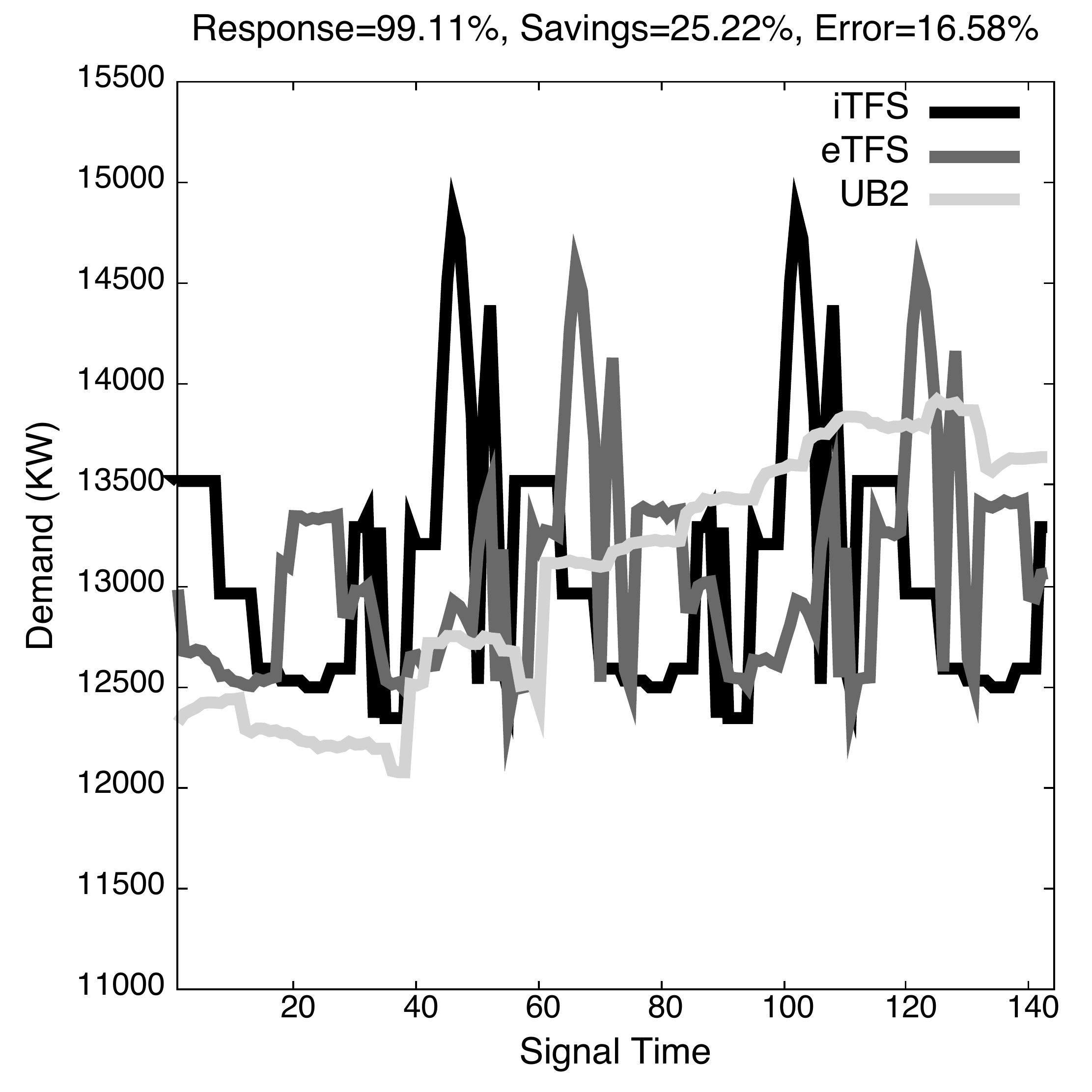}}
\subfigure[\EDF, \shuffle, \minRmseA]{\includegraphics[width=0.49\columnwidth]{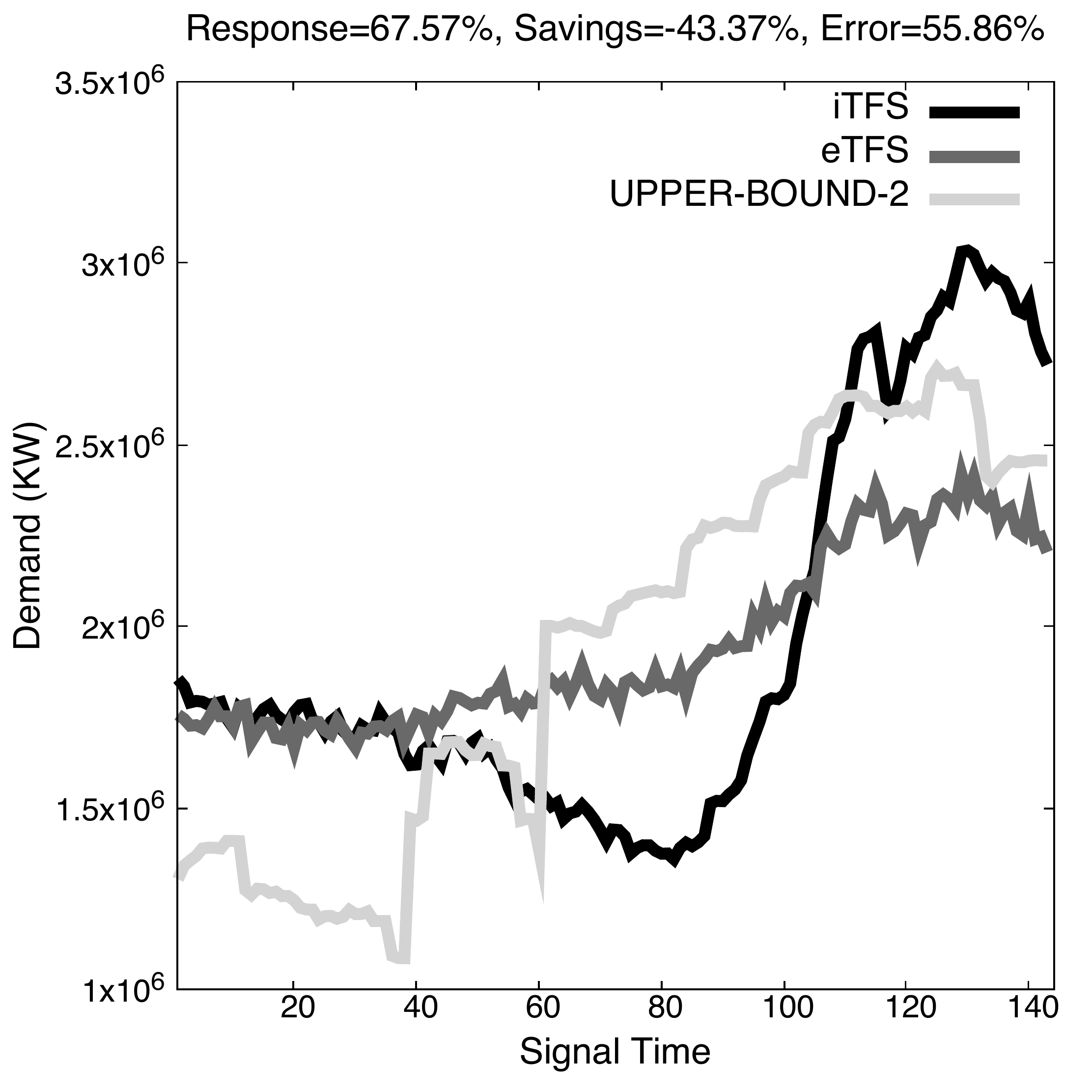}}
\subfigure[\PNWMorning, \shuffle, \minCost]{\includegraphics[width=0.49\columnwidth]{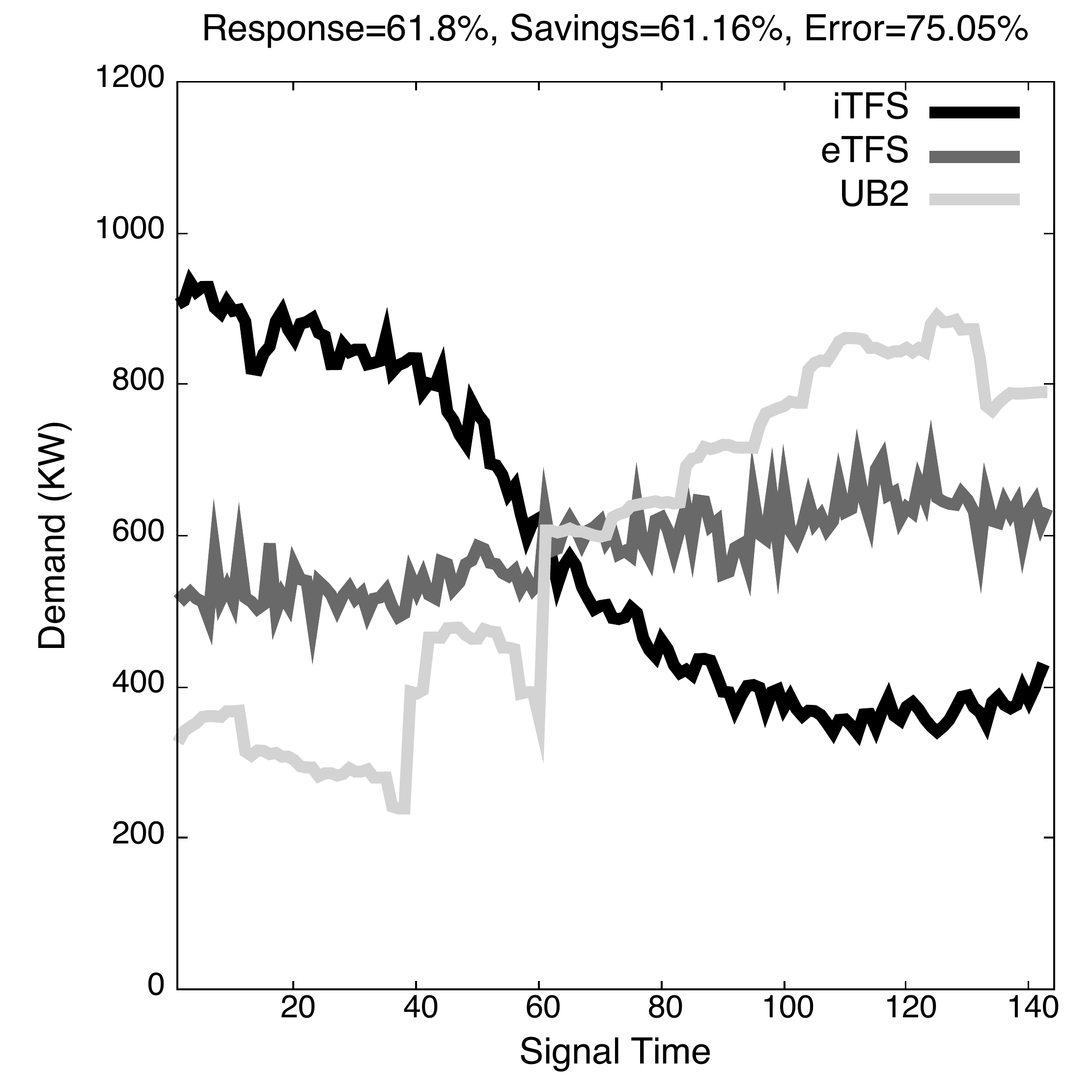}}
\subfigure[\PNWEvening, \shuffle, \minCost]{\includegraphics[width=0.49\columnwidth]{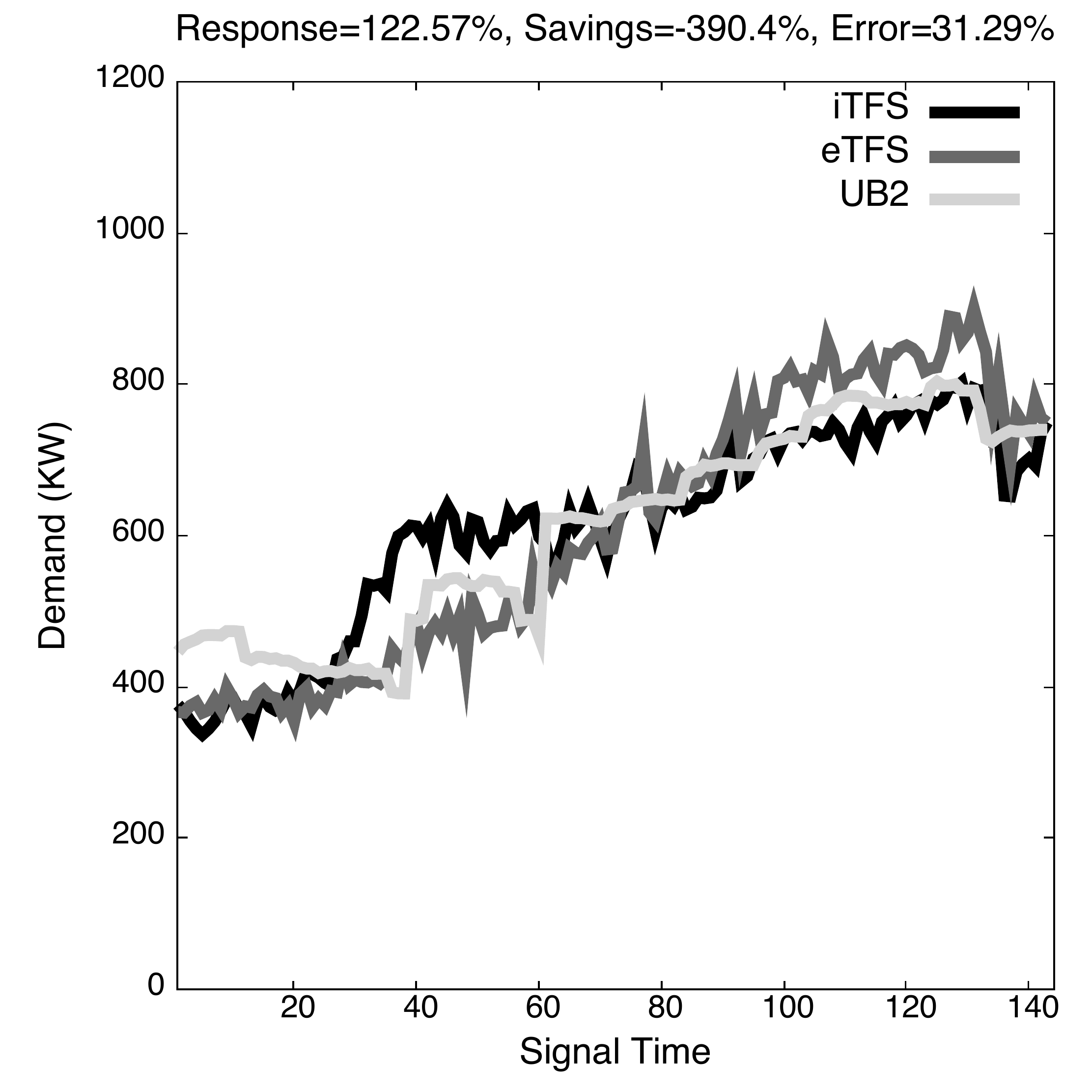}}
\caption{\eTFS signals with maximum response under the minimum entropy scenario. For comparison, the \iTFS and \upperBoundB signals are shown.}\label{fig:signals-minimum-entropy}
\end{figure}

Figure~\ref{fig:signals-minimum-entropy} illustrates the minimum entropy scenario. The maximum response is 99.11\%, 67.57\%, 61.8\% and 122.57\% for \SIM, \EDF, \PNWMorning and \PNWEvening respectively. The maximum response in \SIM is achieved with \minCost that results in a low volatility error of 16.58\% confirming the significant negative correlation between error-savings as shown in Figure~\ref{fig:correlation-coefficient}. Note that in Figure~\ref{fig:signals-minimum-entropy}d, the \iTFS signal and the \upperBoundB signal have a very similar trend with a volatility error of 31.29\%. Given that response and savings are relative measurements, low differences result in significant changes in the magnitude of response and savings. This explains the more extreme values computed with the \PNWEvening dataset. 

\section{Discussion and Conclusion}

This paper illustrates extensive experimental findings using data from real-world operational supply-demand systems that confirm the feasibility of a new type of bottom-up, online self-regulation using Internet of Things technologies. In contrast to earlier work, this paper contributes a generic self-regulatory framework shaped around standardized concepts and implemented Internet of Things technologies for an easier adoption and applicability in Smart City applications. 

An evaluation methodology, integrated within this framework, allows the systematic assessment of optimality and system constraints, resulting in more informative and meaningful comparisons of different self-regulatory settings with an impact on how system operators can design controllability strategies for self-regulating supply-demand systems. For example, the comparison of different agent selections by measuring response, savings and volatility error shows that purely cost-based metrics, such as savings, on which earlier work mainly relies on, cannot adequately quantify the dynamics and capabilities of regulatory mechanisms to adjust demand. Instead, comparisons between response and savings suggest trade-offs that open up new pathways to evaluate multiple dimensions and engineer multi-objective self-regulatory systems. The correlation results of the three metrics empirically validate the effectiveness of the selection functions for which they are designed, i.e. maximize savings vs. maximize response vs. minimize volatility error. Minimizing the volatility error comes in hand with maximizing savings, rather than maximizing response. 

A higher informational diversity in the options of agents, from which selections are made, results in higher performance as shown for the evaluated plan generation schemes. Acquiring informational diversity requires application-specific knowledge that can create new innovative opportunities for self-regulation in domains such as natural ecosystems~\cite{Burkhard2012}, sharing economies~\cite{Santi2014,Pournaras2017} or even participatory policies for science~\cite{Sarewitz2007}.

\bibliographystyle{elsarticle-num-names}
\bibliography{epos} 

\appendix

\section{Mathematical Symbols}\label{app:math}

Table~\ref{table:math-symbols} outlines all mathematical symbols used in this paper in the order they appear. 

\begin{table}[!htp]\footnotesize\centering
\caption{An overview of the mathematical symbols}
\begin{tabular}{l p{6cm}}
\hline
Symbol & Interpretation \\ \hline
\timePoint & Time index \\
\horizon & Planning horizon \\
\TS & Index indicating a type of transactive signal \\
\signal{\signalType} & A transactive signal of type \TS \\
\signalValue{\signalType}{\timePoint} & The value of a transactive signal \TS at time \timePoint \\
\TIS & Transactive Incentive Signal \\
\TFS & Transactive Feedback Signal \\
\iTFS & Inelastic \TFS signal \\
\eTFS & Elastic \TFS signal \\
\upperBound & An upper bound \eTFS signal \\
\response & The response metric \\
\savings & The savings payoff metric \\
\reverseFunct{\timePoint,\signal{\signalType}} & The value of the reflected signal \TS at time \timePoint  \\
\meanFunct{\signal{\signalType}} & The mean value of a transactive signal \TS\\
\stdevFunct{\signal{\signalType}} & The volatility value of a transactive signal \TS\\
\normalizeFunct{\timePoint,\signal{\signalType}} & The value of the normalized signal \TS at time \timePoint \\
$\epsilon_{\mu},\epsilon_{\sigma}$ & The mean and volatility error \\
\agent & Agent index \\
\plan & Plan index \\
\numOfPlans & Number of possible plans \\
\demandPlans{\agent} & The possible plans of an agent \agent \\
\demandPlan{\agent}{\plan} & The possible plan \plan of an agent \agent \\
\demand{\agent}{\plan}{\timePoint} & The value of a possible plan \plan at time \timePoint \\
\selectedPlan & The index of the selected plan \\
\numOfAgents & Number of agents \\
\newTimePoint=\diversityFunct{\timePoint} & Diversification function of a generation scheme \\
\combinationalDemandPlans{\agent} & The combinational plans of an agent \agent \\
\combinationalDemandPlan{\agent}{\plan} & The combinational plan \plan of an agent \agent \\
\combinationalDemand{\agent}{\plan}{\timePoint} & The value of a combinational plan \plan at time \timePoint \\
\aggregatePlans{\agent} & The aggregate plans of an agent \agent \\
\aggregatePlan{\agent}{\plan} & The aggregate plan \plan of an agent \agent \\
\aggregateDemand{\agent}{\plan}{\timePoint} & The value of an aggregate plan \plan at time \timePoint \\
\child & The first child of a parent agent \agent \\
\numOfChildren & The number of children that a parent agent has \\
\bruteForceFunct{\aggregatePlans{1},...,\aggregatePlans{\numOfChildren}} & Brute force over \numOfChildren sequences of aggregates plans \\
$\vartheta$ & Average price reduction \\
$\alpha$ & Average negative prices \\
\heterogeneity & Heterogeneity parameter for disaggregation \\
\minFunct{\signal{\signalType}} & The minimum value of a transactive signal \TS \\
\maxFunct{\signal{\signalType}} & The maximum value of a transactive signal \TS \\
\randNum & Random number in $[0,1)$\\
\hline
\end{tabular}\label{table:math-symbols}
\end{table}

\section{Normalization of a Reflection}\label{app:reflection}

Normalization aims at increasing the negative values to 0 so that $\signalValue{\TISIndex}{\timePoint} \in \mathbb{R}_{>0}$ and at the same time reducing the positive values so that the total reduction equals the total increase resulting in the same average price in the two signals. The reduction of the positive prices is distributed proportionally to their level so that none of the price reductions results in new negative prices. Mathematically, normalization of a \TIS reflection is expressed as follows:

\begin{equation}
\normalizeFunct{\timePoint,\signal{\TISIndex}} = \begin{cases}
\reverseFunct{\timePoint,\signal{\TISIndex}}+\frac{\reverseFunct{\timePoint,\signal{\TISIndex}}*\alpha}{\vartheta} &,\text{if $\reverseFunct{\timePoint,\signal{\TISIndex}}\geq 0$}\\
0 &,\text{if $\reverseFunct{\timePoint,\signal{\TISIndex}} < 0$}
\end{cases}
\end{equation}

\noindent where $\alpha$ is the average price reduction of the positive prices and $\vartheta$ is the average value of the negative prices. $\alpha$ is measured by summing the absolute values of the negative prices and dividing by the number of positive prices. It should hold that $\alpha>|\vartheta|$.

\begin{figure}[!htb]
\centering
\includegraphics[width=0.49\columnwidth]{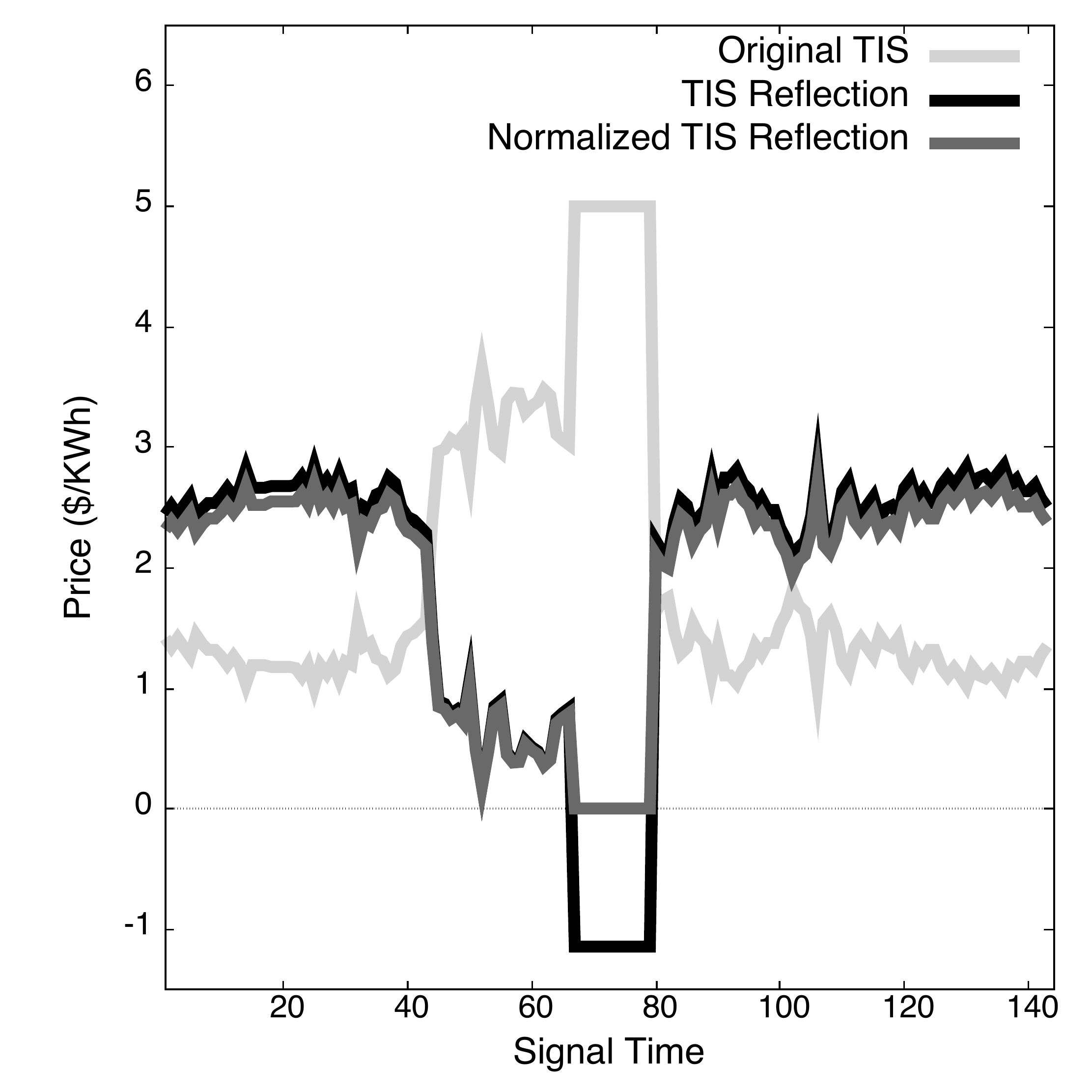}
\caption{Normalization of a TIS reflection with negative price values to a TIS signal with positive price values. The average price of the two signals remains the same.}\label{fig:tis-normalization}
\end{figure}

Figure~\ref{fig:tis-normalization} illustrates an example of the normalization performed in the TIS reflection for the generation failure scenario. This normalization is used for the computation of the \upperBoundA.

\section{Disaggregation}\label{app:disaggregation}

Given that the availability of real-time power demand data at the household level is still limited due to privacy and business concerns, disaggregation is performed in the \SIM and \EDF datasets for the evaluation illustrated in Section~\ref{sec:evaluation}. Two disaggregation approaches are followed in each case.

The \iTFS signals of the \SIM dataset are disaggregated as illustrated in Algorithm~\ref{alg:disaggregation}. The algorithm distributes the load uniformly among the agents (line 5 of Algorithm~\ref{alg:disaggregation}), however, it introduces some heterogeneity controlled by the parameter \heterogeneity (line 6 and 7 of Algorithm\ref{alg:disaggregation}).

\begin{algorithm}[!htb]
\centering
\small{
\begin{algorithmic}[1]
\REQUIRE \signal{\iTFSIndex}, \numOfAgents, \heterogeneity
\FORALL{\signalValue{\iTFSIndex}{t} $\in$ \signal{\iTFSIndex}}
\STATE{$x=\signalValue{\iTFSIndex}{t}$}
\FORALL{\agent $\in \{1,...,\numOfAgents\}$}
\IF{$\agent=\numOfAgents$ is \FALSE}
\STATE{$\avgLoad=x/(\numOfAgents-\agent+1))$}
\STATE{$\minAvgLoad=\avgLoad-\heterogeneity*\avgLoad$}
\STATE{$\maxAvgLoad=\avgLoad+\heterogeneity*\avgLoad$}
\STATE{$\randNum=\randFunct [0,1)$}
\STATE{$\demand{\agent}{1}{\timePoint}=\min(x,\avgLoad-\minAvgLoad*\randNum+\maxAvgLoad*\randNum)$}
\IF{$\demand{\agent}{1}{\timePoint}<0.0$}
\STATE{$\demand{\agent}{1}{\timePoint}=0.0$}
\ENDIF
\ELSE 
\STATE{$\demand{\agent}{1}{\timePoint}=x$}
\ENDIF
\STATE{$x=x-\demand{\agent}{1}{\timePoint}$} 
\STATE{$\demandPlan{\agent}{1}=\demandPlan{\agent}{1} \cup \demand{\agent}{1}{\timePoint}$} 
\ENDFOR
\ENDFOR
\ENSURE \demandPlan{\agent}{1}, $\forall \agent \in \{1,...,\numOfAgents\}$ 
\end{algorithmic}
}
\caption{Disaggregation of the \iTFS signal \signal{\iTFSIndex} to \numOfAgents agents.}\label{alg:disaggregation} 
\end{algorithm}

The \iTFS signals of the \EDF datasets are disaggregated by distributing each daily demand time series to a different agent. In other words, the daily variability of demand of a single household is transformed to variability of demand among different households. This approach results in disaggregated demand data for two groups of 724 agents that correspond to approximately 365 days for a period of 4 years.

\section{Performance Results in Detail}\label{app:response-savings-detail}

\begin{figure*}[!htb]
\centering
\subfigure[Ramp down]{\includegraphics[width=0.96\textwidth]{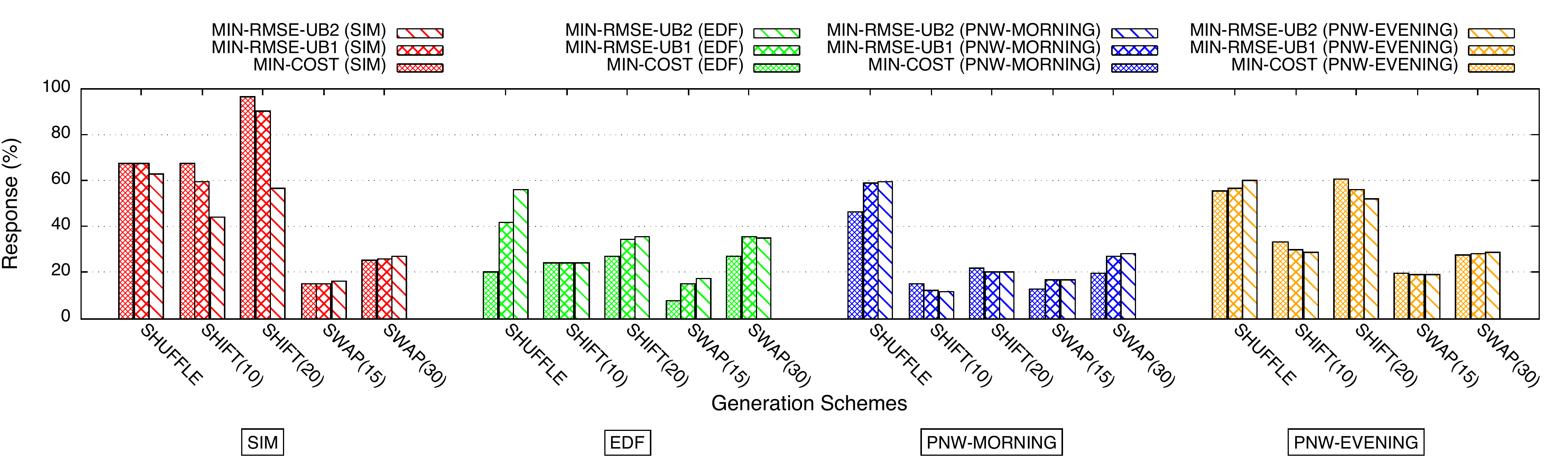}}
\subfigure[Generation failure]{\includegraphics[width=0.96\textwidth]{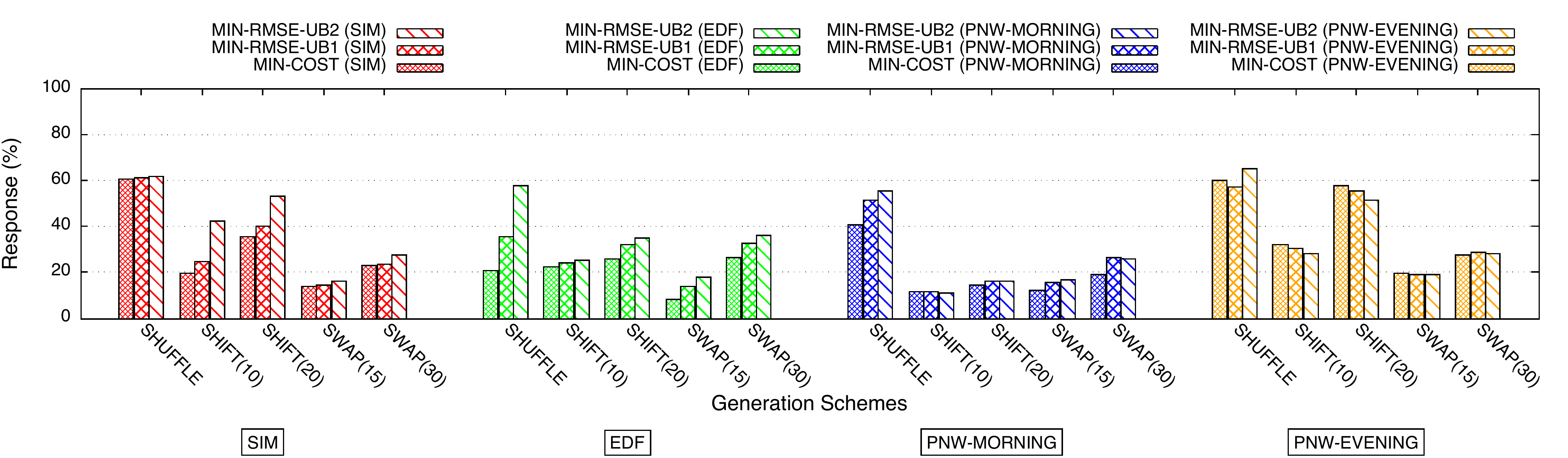}}
\subfigure[Maximum entropy]{\includegraphics[width=0.96\textwidth]{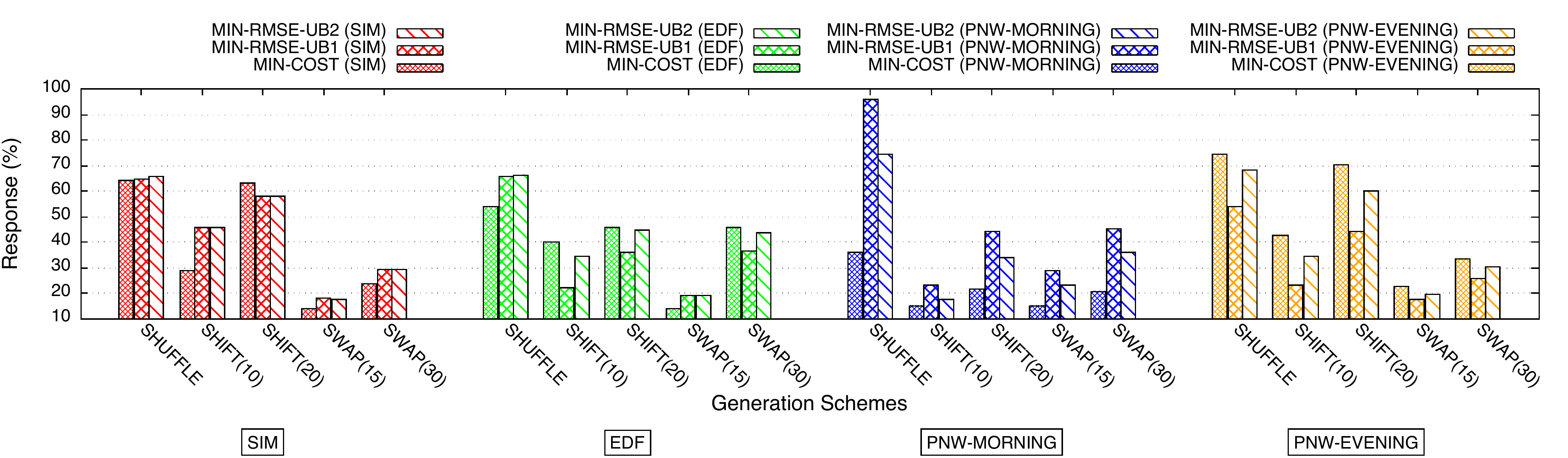}}
\subfigure[Minimum entropy]{\includegraphics[width=0.96\textwidth]{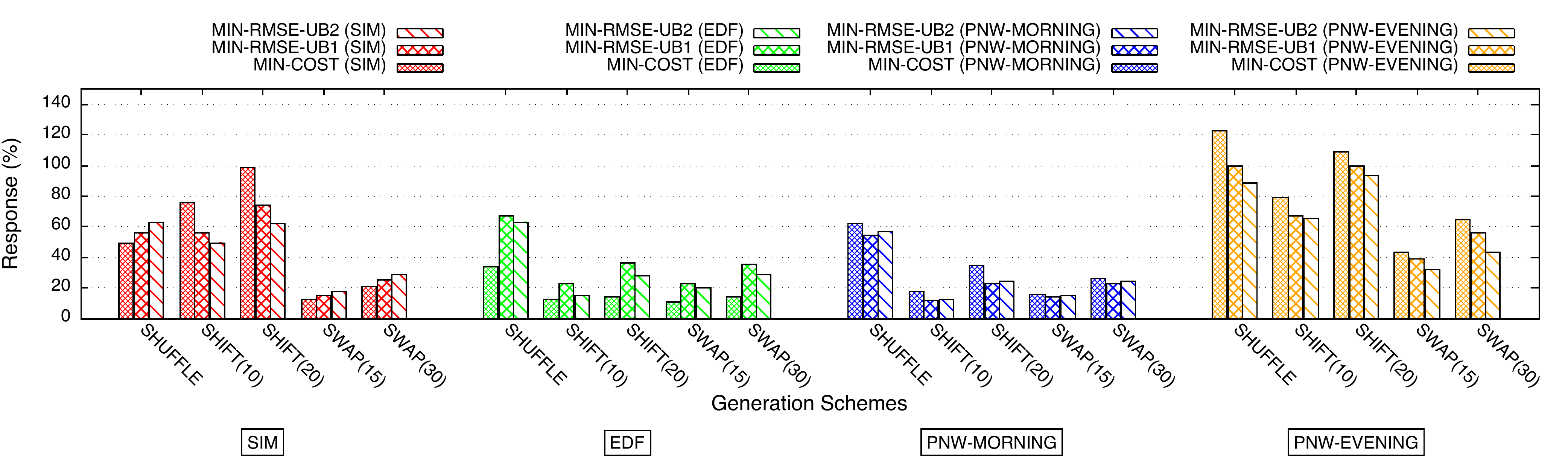}}
\caption{Response of EPOS under different generation schemes and selection functions.}\label{fig:response-detail}
\end{figure*}

\begin{figure*}[!htb]
\centering
\subfigure[Ramp down]{\includegraphics[width=0.96\textwidth]{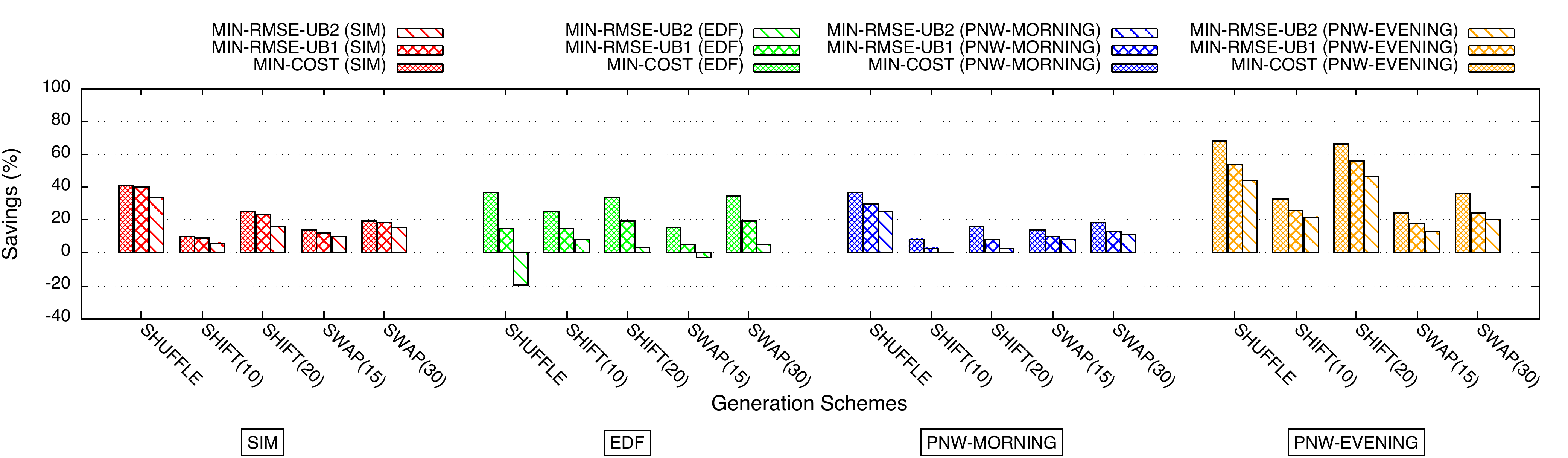}}
\subfigure[Generation failure]{\includegraphics[width=0.96\textwidth]{RAMP-DOWN-UB2-SAVINGS.pdf}}
\subfigure[Maximum entropy]{\includegraphics[width=0.96\textwidth]{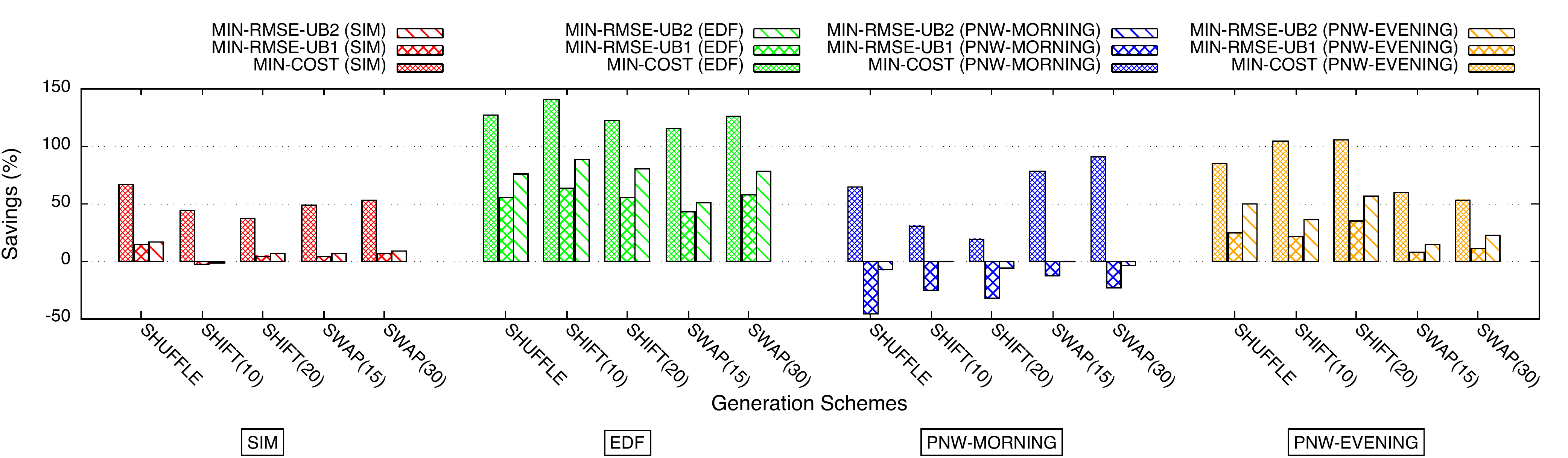}}
\subfigure[Minimum entropy]{\includegraphics[width=0.96\textwidth]{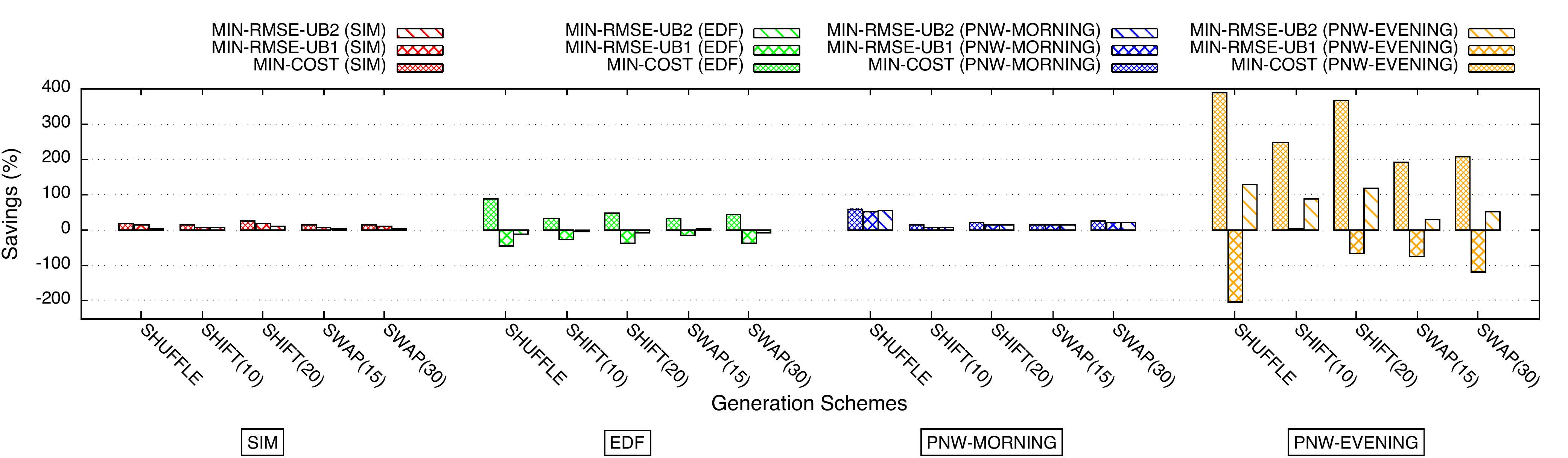}}
\caption{Savings of EPOS under different generation schemes and selection functions.}\label{fig:savings-detail}
\end{figure*}

Figure~\ref{fig:response-detail} and~\ref{fig:savings-detail} illustrate the measurements of response, savings and volatility error in detail. Results are illustrated for each of the studied aspects: generation scheme, selection function, dataset and regulatory scenario.  

\begin{figure*}[!htb]
\centering
\subfigure[Ramp down]{\includegraphics[width=0.96\textwidth]{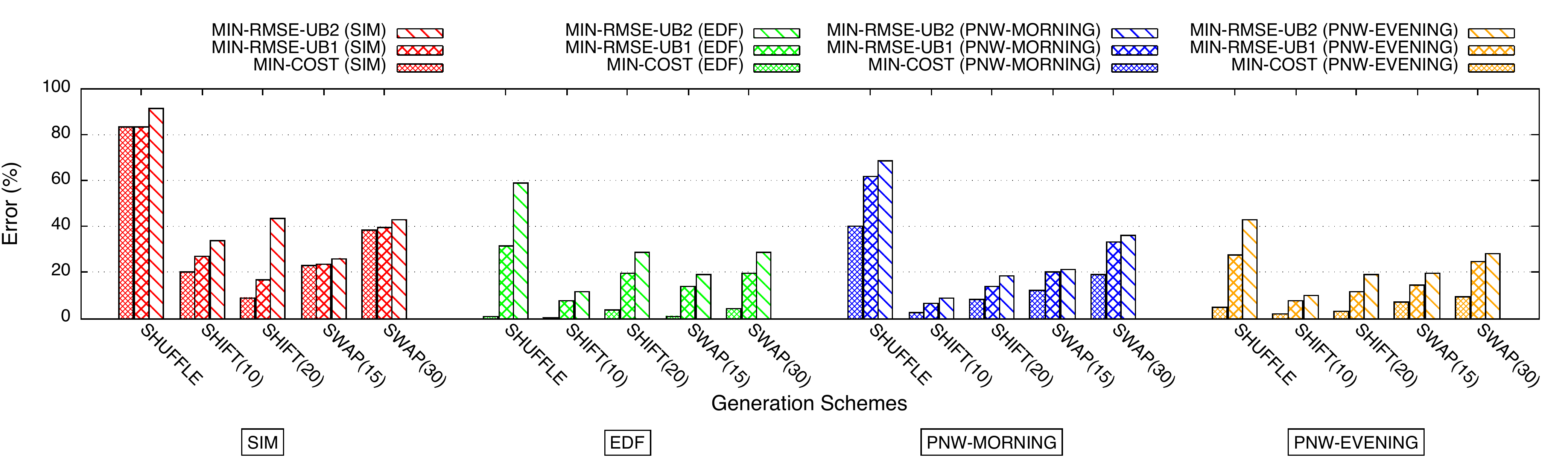}}
\subfigure[Generation failure]{\includegraphics[width=0.96\textwidth]{RAMP-DOWN-UB2-ERROR.pdf}}
\subfigure[Maximum entropy]{\includegraphics[width=0.96\textwidth]{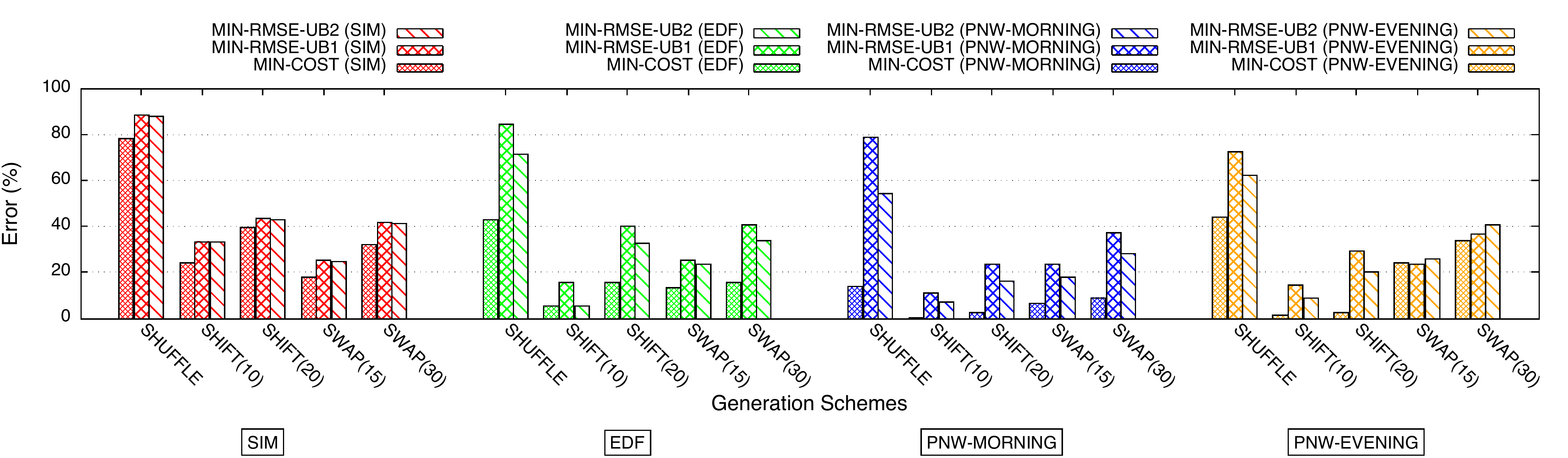}}
\subfigure[Minimum entropy]{\includegraphics[width=0.96\textwidth]{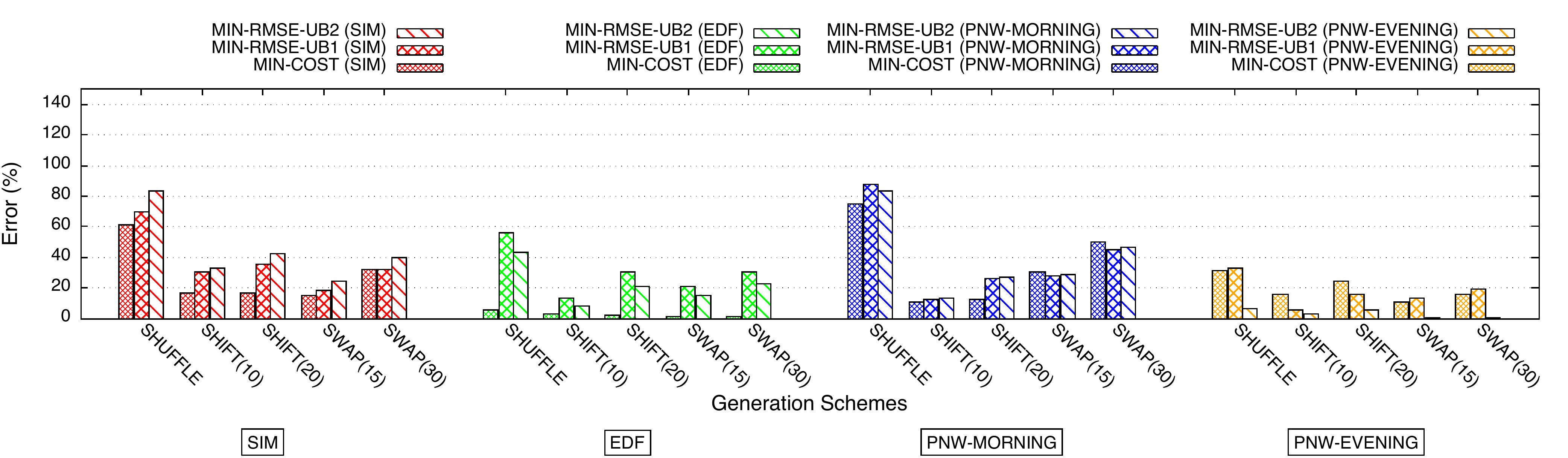}}
\caption{Volatility error of EPOS under different generation schemes and selection functions.}\label{fig:error-detail}
\end{figure*}

\end{document}